\documentclass[aps, reprint, superscriptaddress, twocolumn]{revtex4-2}
\usepackage{latexsym,amssymb,amsfonts,mathrsfs,amsmath,bm}
\usepackage{graphicx,newfloat,subfigure,lipsum}
\usepackage[usenames,dvipsnames]{xcolor}
\usepackage{siunitx} 
\usepackage{soul}
\usepackage{gensymb}
\usepackage[linktocpage,colorlinks=true,linkcolor=blue,citecolor=blue,breaklinks=true,urlcolor=blue]{hyperref}
\usepackage[utf8]{inputenc}
\usepackage{tikz}
\usepackage{scalerel}
\usepackage{chemformula}
\usetikzlibrary{svg.path}

\usepackage{makeidx} 
\makeindex

\renewcommand\vec{\bm}

\definecolor{orcidlogocol}{HTML}{A6CE39}
\tikzset{
  orcidlogo/.pic={
    \fill[orcidlogocol] svg{M256,128c0,70.7-57.3,128-128,128C57.3,256,0,198.7,0,128C0,57.3,57.3,0,128,0C198.7,0,256,57.3,256,128z};
    \fill[white] svg{M86.3,186.2H70.9V79.1h15.4v48.4V186.2z}
                 svg{M108.9,79.1h41.6c39.6,0,57,28.3,57,53.6c0,27.5-21.5,53.6-56.8,53.6h-41.8V79.1z M124.3,172.4h24.5c34.9,0,42.9-26.5,42.9-39.7c0-21.5-13.7-39.7-43.7-39.7h-23.7V172.4z}
                 svg{M88.7,56.8c0,5.5-4.5,10.1-10.1,10.1c-5.6,0-10.1-4.6-10.1-10.1c0-5.6,4.5-10.1,10.1-10.1C84.2,46.7,88.7,51.3,88.7,56.8z};
  }
}

\newcommand\orcid[1]{\href{https://orcid.org/#1}{\mbox{\scalerel*{
\begin{tikzpicture}[yscale=-1,transform shape]
\pic{orcidlogo};
\end{tikzpicture}
}{|}}}}

\begin{document}

\title{Culinary fluid mechanics and other currents in food science}

\author{Arnold J. T. M. Mathijssen\orcid{0000-0002-9577-8928}}
\email{amaths@upenn.edu}
\affiliation{Department of Physics \& Astronomy, University of Pennsylvania, 209 South 33rd Street, Philadelphia, PA 19104, USA}

\author{Maciej Lisicki\orcid{0000-0002-6976-0281}}
\email{mklis@fuw.edu.pl}
\affiliation{Institute of Theoretical Physics, Faculty of Physics, University of Warsaw, Pasteura 5, 02-093 Warsaw, Poland}

\author{Vivek N. Prakash\orcid{0000-0003-4569-6462}}
\email{vprakash@miami.edu}
\affiliation{Departments of Physics, Biology, Marine Biology and Ecology, University of Miami, 1320 Campo Sano Ave, Coral Gables, FL 33146, USA}

\author{Endre J. L. Mossige\orcid{0000-0002-6929-2932}}
\email{endrejm@uio.edu}
\affiliation{RITMO Centre for Interdisciplinary Studies in Rhythm, Time and Motion, University of Oslo (UiO), Forskningsveien 3A, 0373 Oslo, Norway}

\date{\today}

\begin{abstract}
\noindent
Innovations in fluid mechanics are leading to better food since ancient history, while creativity in cooking inspires applied and fundamental science. Here, we review how recent advances in hydrodynamics are changing food science, and we highlight how the surprising phenomena that arise in the kitchen lead to discoveries and technologies across the disciplines, including rheology, soft matter, biophysics and molecular gastronomy. This review is structured like a menu, where each course highlights different aspects of culinary fluid mechanics. Our main themes include multiphase flows, complex fluids, thermal convection, hydrodynamic instabilities, viscous flows, granular matter, porous media, percolation, chaotic advection, interfacial phenomena, and turbulence. For every topic, we first provide an introduction accessible to food professionals and scientists in neighbouring fields. We then assess the state-of-the-art knowledge, the open problems, and likely directions for future research. New gastronomic ideas grow rapidly as the scientific recipes keep improving too. 
\end{abstract}

\maketitle
\tableofcontents

\section{Introduction}

\begin{figure}[t]
        \includegraphics[width=\linewidth]{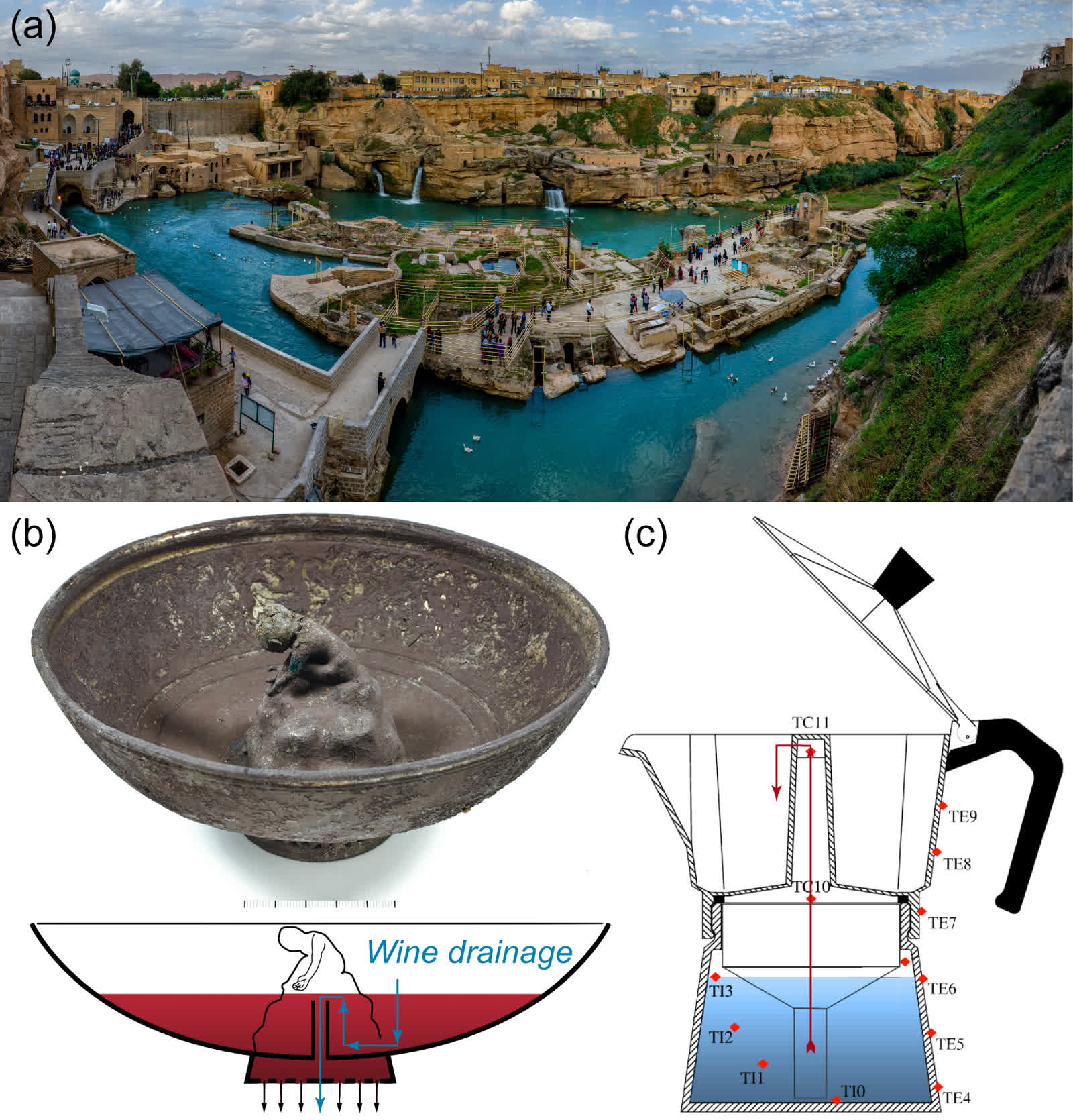}
        \caption{Hidden channels.
        \textbf{(a)} The Shushtar Historical Hydraulic System in Iran, listed by UNESCO as `a masterpiece of creative genius'. This irrigation system, dating back to the 5th century B.C., features canals, tunnels, dams, and water mills, which all work in unison. Image by Iman Yari, licenced under CC BY-SA 4.0.
        \textbf{(b)} The first discovered specimen of a Tantalus bowl (Pythagorean cup) from the late Roman period, catalogue no.\,9 of the Vinkovci treasure. This silver-gilt bowl empties itself when filled above a critical level by a hidden siphon, soaking a greedy drinker in wine. Image credit: Damir Dora{\v{c}}i{\'c}, Archaeological Museum in Zagreb, from \citet{vulic2017vinkovci}. 
        \textbf{(c)} The moka pot, a traditional stove-top coffee maker, where the boiling water percolates through the coffee by following the red arrows. From \citet{navarini2009experimental}. 
        }
        \label{fig:TantalusBowl}
    \end{figure}

The origins of fluid mechanics trace back to ancient water technologies [Fig.~\ref{fig:TantalusBowl}a], which supplied our earliest civilizations \index{ancient civilizations} with reliable food sources \cite{mays2010ancient}.
Subsequently, as soon as the water flows, surprising phenomena emerge beyond number. 
Their abundance naturally sparked the interest of the first inventors, since the kitchen can serve as a laboratory \cite{kurti1994kitchen} that is accessible to people of different backgrounds, ages, and interests.
As such, the scullery is a source of curiosity that has driven innovations throughout history \cite{drazin1987fluid}.
The problems that emerge while cooking have led to creative solutions that have not only improved food science, but also led to breakthroughs in modern engineering, medicine, and the natural sciences. 
In turn, fundamental research has improved gastronomy, and thus the cycle continues.
Hence, science and cooking are intrinsically connected across people and time.

Today, numerous chefs have written extensive cookbooks from a scientific perspective.
Well acclaimed is the work on molecular gastronomy \cite{this2006molecular}, which turned into a scientific discipline, as reviewed by \citet{barham2010molecular}.
Another recent movement, known for using advanced equipment including centrifuges and blow torches, is called modernist cuisine \cite{myhrvold2011modernist}. With its striking photography, sometimes tricked, it also connects science with art in the field of fine dining \cite{borkenhagen2017evidence}.
The book by \citet{mcgee2007food} is particularly influential too: Celebrity chef Heston Blumenthal stated it is ``the book that has had the greatest single impact on my cooking'', and then he wrote eight books himself.
Another excellent cookbook containing various experiments and scientific diagrams was written by \citet{lopez2015food}.
One of the first people to approach cooking systematically, a century earlier, was the `king of chefs and chef of kings' \citet{escoffier1903guide}, whose 943-page culinary guide still remains a golden standard in haute cuisine \cite{trubek2000haute}.

In the scientific community, a wave of excitement hit when \citet{kurti1988but} solicited recipes or essays on cooking from the members of the Royal Society.
Science can improve cooking, but they also showed that food can lead to better science, a notion that was not taken so seriously at the time.
Later, in her essay `Food for thought', Dame Athene Donald FRS pleads that the scientific challenges are as exciting in food as in any more conventional area. They should not be overlooked or, worse, sneered at \cite{donald2004food}.
Her early vision sparked scientists to regard food as an interdisciplinary research topic. 
Food science now spans across many fields, including materials science \cite[e.g.][]{mezzenga2005understanding}, food chemistry and physical chemistry \cite[e.g.][]{damodaran2017fennema}, nutrition genetics \cite[e.g.][]{capozzi2013foodomics}, food engineering \cite[e.g.][]{heldman2018handbook}, food microbiology \cite[e.g.][]{provost2016science, doyle2019food}, food rheology \cite[e.g.][]{ahmed2016advances}, soft condensed matter \cite[e.g.][]{vilgis2015soft, assenza2019soft, pedersen2019soft} and biophysics \cite[e.g.][]{foegeding2006food, nelson2020biological}.

However, to the best of our knowledge, and despite the overwhelming number of surprising hydrodynamic effects that emerge in the kitchen, there is no comprehensive review of fluid mechanics in gastronomy and food science. Therefore, we aim to address this topic here in a manner that first provides a broad overview and then highlights the frontier of modern research. As such, we aim to connect the following communities:

First, for chefs and gastronomy professionals, fluid mechanics can make or break their culinary creations. 
Hydrodynamic instabilities can ruin a layered cocktail [\S\ref{subsec:LayeredCocktails}], while the Leidenfrost effect helps with searing your steak [\S\ref{subsec:Leidenfrost}], and baristas learn about percolation to perfect their coffee [\S\ref{subsec:brewingCoffee}].
We will discuss these examples here, and quite a few more. 
Indeed, throughout this Review we aim to connect the science with food applications. 
We also point out some common mistakes in cooking and think of new ideas for recipes.

Second, for food scientists, it is important to unravel hydrodynamic effects in order to develop better food processing technologies \cite{knorr2011emerging}. For example, microfluidic techniques are now extensively used for edible foam generation and emulsification \cite{skurtys2008applications, gunes2018microfluidics}, but also bioactive compound extraction and the design of novel food microstructures \cite{he2020application}. 
More generally, fluid mechanics describes the transport of mass, momentum and energy, which is to be optimised in food processing \cite{welti2016transport} and food preservation \cite{gould2012new, amit2017review}. 
We will highlight a number of unexpected flow phenomena, and their relation with food science technologies.

Third, from the perspective of medicine and nutrition professionals, flow physics has led to novel health care solutions and provided insights on the physiology of digestion \cite{donald2004understanding}.
For example, flow devices can detect food-borne pathogens or toxins \cite{kant2018microfluidic}, which is essential for food safety \cite{bajpai2018prospects} and food quality control \cite{ozilgen2011handbook}. 
Similar technologies can equally be used for \textit{in vitro} fertilization for agricultural animal breeding, or other applications in animal health monitoring, vaccination and therapeutics \cite{neethirajan2011microfluidics}. 
Moreover, using next-generation DNA and protein sequencing with nanopore technology \cite{drndic202120}, the field of foodomics could help with improving human nutrition \cite{capozzi2013foodomics}.
In this article we will reflect on more of these food health innovations.

Fourth, for engineers and natural scientists, kitchen flows have led to breakthrough discoveries, and continue doing so. 
To name a few here, Agnes Pockels established the modern discipline of surface science after her observations of soap films while washing the dishes [\S\ref{subsec:SoapFilmDynamics}], and Pyotr Kapitza discovered the roll wave instability while under house arrest [\S\ref{subsec:RinsingFlows}]. 
Most universities and labs were closed during the COVID-19 pandemic, which again lead to an unasked-for wave of kitchen science \cite{american2020lab}. 
Moreover, culinary flows have given rise to engineering applications in completely different fields.
The piston-and-cylinder steam engine was inspired by Papin's pressure cooker [\S\ref{subsec:Eureka}], and inkjet printers rely on capillary breakups observed in the sink [\S\ref{subsec:RayleighPlateau}].
Indeed, because of the low activation barrier, the kitchen is a hotspot for curiosity-driven research where new ideas arise.

Fifth, for policy makers, the evolution of food science often lies at the heart of historical developments \cite{toussaint2009history, mays2010ancient} and it is key to the future of our planet \cite{foley2011solutions}. Our soil resources are under stress worldwide \cite{amundson2015soil}, the use of land has global consequences \cite{foley2005global}, and food from the sea is similarly limited~\cite{costello2020future}.
Solutions may come from food technology innovations, global policy reforms, and better science education.

Finally, for science educators, the kitchen can serve as an exceptional classroom \cite{rowat2014kitchen, benjamin1999rayleigh, vieyra2017kitchen} or indeed a lab \cite{kurti1994kitchen}.
Being a natural gateway to learning about fluid mechanics, food science demonstrations equally connect to numerous other disciplines.
Examples include teaching oceanography \cite{glessmer2020teach}, chemistry education \cite{schmidt2012using,piergiovanni2019modernist}, geology \cite{giles2020barriers}, soft matter physics \cite{ogborn2004soft}, and the science of cooking for non-science majors \cite{miles2009science}.
Recently, based on their successful edX (online) and Harvard University course, \citet{brenner2020science} connected {\it haute cuisine} with soft matter science in a textbook.
Indeed, a lot of science awaits to be discovered during our daily meals.

This Review is structured like a menu:
We begin with washing our hands in \S\ref{sec:kitchenSinkFundamentals} about kitchen sink fundamentals, where we provide a brief introduction to fluid mechanics that is accessible to scientists across the disciplines.
Then we are ready to pour ourselves a cocktail, which we discuss in \S\ref{sec:Drinks} concerning multiphase flows. 
The first course might be a consomm\'e, so in \S\ref{sec:SoupsSauces} we focus on complex fluids and food rheology.  
The main course is often hot, so we review thermal effects in cooking in \S\ref{sec:PotsPans}. 
Tempted by dessert with honey and ice cream, we consider Stokesian flows in \S\ref{sec:HoneyMicro}. 
We then brew a coffee after the lavish meal, the thought of which sparks interest in granular flows and porous media, as discussed in \S\ref{sec:CoffeeSugar}. 
Pouring another cup of tea, we discuss different aspects of non-linear flows and turbulence in \S\ref{sec:Turbulence}. 
Once the meal comes to an end, it is time to wash the dishes, which brings our attention to interfacial flows in \S\ref{sec:TheDishes}. 
We conclude the Review with an extensive discussion in \S\ref{sec:Discussion}.

\section{Kitchen Sink Fundamentals}
\label{sec:kitchenSinkFundamentals}

We begin this Review by introducing the basics of fluid mechanics in the context of food science. 
Flow experts may choose to skip this section, or to refresh their memory while enjoying the various anecdotes.
Starting with surprising aspects of hydrostatics, we quickly transition to the hydrodynamics of wine aeration, hydraulic jumps and satellite dishes, to name a few.
Some of these concepts are not as simple as appearance makes believe. In the words of \citet{drazin1987fluid},
    \begin{quote}
    \emph{A child can ask in an hour more questions about fluid dynamics than a Nobel Prize winner can answer in a lifetime.}
    \end{quote}
As things get more complicated in the later sections, we will often refer back to these kitchen sink fundamentals.

\subsection{Eureka! Surprising hydrostatics}
\label{subsec:Eureka}

In his work ``On Floating Bodies'', Archimedes of Syracuse (c.287–c.212 BC) described the principles of hydrostatics\index{Archimedes principle}. 
The buoyancy\index{buoyancy} force on an immersed object equals the weight of the fluid it displaces \cite{chalmers2017one}. 
To see this, we note that the hydrostatic pressure increases with depth.
Consider, for example, an immersed bottle of volume $V = L^3$ that experiences a pressure difference of $\Delta p = \rho g L$ between its top and bottom surfaces, where $\rho$ is the fluid density, $\rho_0$ is the object density, $\Delta \rho = \rho - \rho_0$ is the density contrast, and $g$ is the gravitational acceleration. 
The buoyancy force is this pressure difference multiplied by the surface area, $L^2$, giving $F_b = \rho g V$, which indeed is the weight of the displaced fluid. 
The total force, including buoyancy and gravity on the object, is $F = \Delta\rho g V$, which vanishes for neutrally buoyant objects. 
Note that buoyancy applies not to solid objects only, but also to fluids of different densities [see layered cocktails, \S\ref{subsec:LayeredCocktails}].

Greece being a sea-faring nation, we note the importance of Archimedes' principle to describe the stability of ships: 
If the center of gravity is above the metacentre (which is related to the centre of buoyancy), the boat will topple \cite{barrass2011ship,lautrup2011physics}.
We can test this by floating a cup upside-down in the kitchen sink.
Another classic experiment is throwing a stone out of a boat (or an upright floating cup). Does the water level rise?
As we will see throughout this Review, buoyancy is essential in many food science processes, including bubbly drinks [\S\ref{subsec:BubblyDrinks}], heat convection [\S\ref{subsec:HeatingBoilingRBC}], and latte art [\S\ref{subsec:brewingCoffee}].

The concept of pressure became more established in the 17th century. 
Evangelista Torricelli (1608-1647) understood that \textit{``We live submerged at the bottom of an ocean of the element air, which by unquestioned experiments is known to have weight''.}
Hence, he invented the barometer, by realizing that mercury in a top-sealed tube was supported by the pressure of the air \cite{west2013torricelli}.
Torricelli also wrote that air pressure might decrease with altitude, a prediction later demonstrated by Blaise Pascal (1623-1662).
This led to the discovery of Pascal's law\index{Pascal's law}, which states that a pressure change at any point in an enclosed fluid at rest is transmitted undiminished throughout the fluid \cite{batchelor2000introduction}.
This law lies at the heart of many applications with `communicating vessels'\index{communicating vessels} including water towers, modern plumbing, water gauges, and barometers \cite{middleton1964history}.

While hydrostatics may seem elementary compared to hydrodynamics, it can still be rather counter-intuitive.  
For example, gravity can readily be defied when turning a glass of water up-side down, aided with a special trick \cite{linden2020upside}.
Another puzzling effect appears in siphons \cite{potter1971siphon}. 
They are devices wherein fluids first flow upwards, over a hill, and then downwards, without a pump.
Siphons are used in modern washing machines, certain toilet flushing systems, and anti-colic baby bottles \cite{marshall2021assessing}.
While it is generally agreed upon that these flows are driven by gravity, it depends on the situation whether the fluid moves primarily because of pressure differences or intermolecular cohesion and capillary forces \cite{hughes2011secret, binder2011explicit, richert2011siphons, jumper2014towards} [see \S\ref{subsec:wettingCapillaryAction}].
Hence, siphons remain an active topic of research \cite{boatwright2015height, wang2022open}.

One of the oldest pranks in history that uses this siphon effect is the Tantalus bowl [Fig.~\ref{fig:TantalusBowl}b], also known as the Pythagoras cup\index{Pythagoras cup}\index{Tantalus bowl}. 
It is a drinking vessel that functions normally when filled moderately, but if the liquid level rises above a critical height, the siphon flow is initiated and the bowl will drain its entire contents.
Therefore, the bowl concretises a ``Tantalean punishment'', taking away pleasure from to those who get too greedy!
While described in ancient literature, the earliest specimen of a Tantalus bowl was discovered only very recently, in the Vinkovci treasure \cite{vulic2017vinkovci}.
Another culinary example of non-intuitive hydrostatics is Heron's fountain\index{Heron's fountain}, attributed to Heron of Alexandria (c.10-c.70 AD). It can spout water higher than its reservoirs without a pump, so it appears to be a perpetual motion machine, but in the end it is just very clever hydraulics. Furthermore, in his treatise on pneumatics, Heron describes a total of 78 different inventions and discoveries made by himself and earlier ancient philosophers \cite{greenwood1851pneumatics}.

The use of pressure in the kitchen expanded with the invention of the steam digester by Denis Papin (1647-1713).
His improved designs included a stream-release valve to prevent the machine from exploding, an essential feature in all modern pressure cookers\index{pressure cooker}, and coffee makers like the moka pot\index{moka pot} [see \S\ref{subsec:brewingCoffee}, Fig.~\ref{fig:TantalusBowl}c].
Besides kitchen appliances \cite{abufarah2022simulations}, the steam digester was the forerunner of the autoclave to disinfect medical instruments, and the piston-and-cylinder steam engine\index{steam engine} \cite{ferguson1964origins}.
Pascal's principle also underpins the hydraulic press\index{hydraulic press}, a force amplifier capable of uprooting trees, invented by Joseph Bramah (1748-1814).
Bramah invented and improved many other culinary technologies during the Industrial Revolution, including high-pressure public water mains\index{water mains} and the beer engine\index{beer engine} \cite{dickinson1941joseph}.

\subsection{Navier-Stokes equations and the Reynolds number}
\label{subsec:NavierStokes}


Moving on from hydrostatics, we shift our attention to fluids in motion. 
These are described by the famous equations named after Claude-Louis Navier (1785-1836) and George Gabriel Stokes (1819-1903). 
When the velocity of the fluid is much smaller than the speed of sound, which is true for typical kitchen flows, the Navier-Stokes equations\index{Navier-Stokes equations} can be written as
        \begin{subequations} \label{eq:NavierStokes}
        \begin{align}
        \label{eq:NavierStokes1}
        \rho \frac{D\vec{u}}{Dt} &= - \vec{\nabla} p + \mu \nabla^2 \vec{u}  + \vec{f},
        \\
        \label{eq:NavierStokes2}
        0 &= \vec{\nabla} \cdot \vec{u}.
        \end{align}
        \end{subequations}
Here $\vec{u}(\vec{x}, t)$ is the flow velocity at position $\vec{x}$ and time $t$, and $p(\vec{x}, t)$ is the pressure field, $\rho$ is the fluid density, $\mu$ is the dynamic viscosity\index{viscosity},  $\nu = \mu / \rho$ is the kinematic viscosity, $\vec{f}$ is a body force (usually gravity) acting on the fluid, and the operator $D/Dt = \partial/\partial t + (\vec{u} \cdot \vec{\nabla})$ is the material derivative, which includes both temporal and spatial variations experienced by a fluid element. 
Physically, Eq.~\eqref{eq:NavierStokes1} stems from the conservation of momentum, essentially Newton's second law applied to an infinitesimal fluid parcel, and Eq.~\eqref{eq:NavierStokes2} describes the conservation of mass.

\begin{figure}[t]
    \includegraphics[width=1\linewidth]{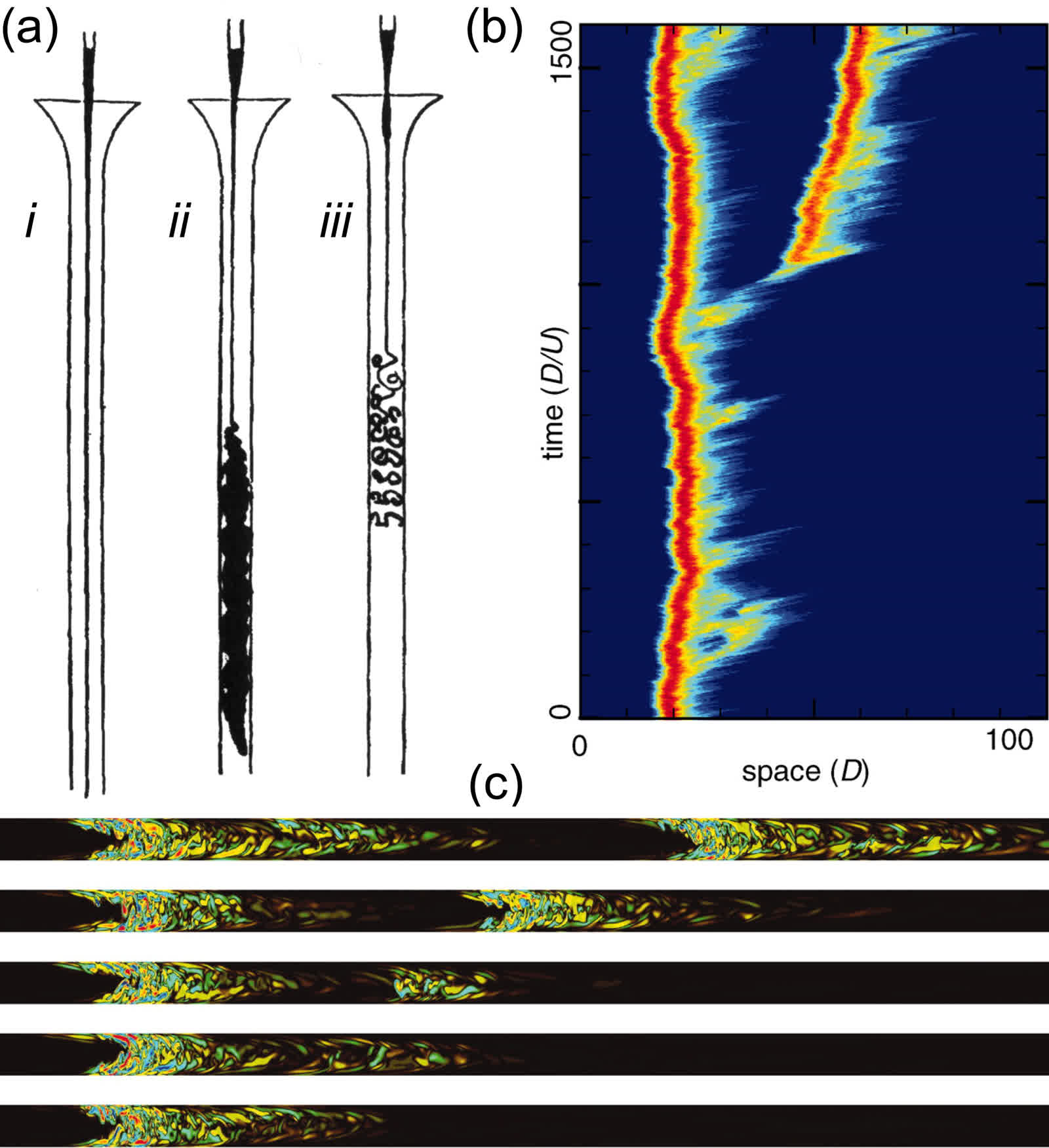}
    \caption{
    Turbulent pipe flow.
    \textbf{(a)} 
    Drawings by \citet{reynolds1883number}, showing (i) laminar pipe flow, (ii) turbulent flow, and (iii) turbulent flow observed by the light of an electric spark.
    \textbf{(b)} Space-time diagram from a numerical simulation at $\text{Re} = 2300$, showing the process of turbulent puffs splitting.
    \textbf{(c)} Visualization of puff splitting in a cross-section of the pipe. Time increases in the snapshots from bottom to top. Red is positive and blue is negative streamwise vorticity.
    (b,c) From \citet{avila2011onset}.
    }
    \label{fig:turbulentPipeFlow}
\end{figure}

The Navier-Stokes equations give us an insight into the character of flow.
In his seminal work, \citet{reynolds1883number} observed the shape of a streak of dye injected into a pipe flow. 
For low pumping speeds, the dye forms a straight line, parallel to the streamlines [Fig.~\ref{fig:turbulentPipeFlow}a.i].
Such flow is called laminar\index{laminar flow} because the fluid parcels move in lamellas parallel to each other.
Increasing the pumping speed, we see gradual distortion of this regular pattern, and the flow transits to turbulence\index{turbulent flow}.
Here, the streak of dye is quickly smeared all over the flow domain [Fig.~\ref{fig:turbulentPipeFlow}a.ii], because of increased mixing and chaotic streamlines [Fig.~\ref{fig:turbulentPipeFlow}a.iii].

To characterise the transition from laminar (ordered) to turbulent (disordered) flow \cite{mullin2011experimental}, we may use the Navier-Stokes equations and examine the relative magnitude of inertial to viscous forces which compete in shaping the flow. By considering a steady flow with a characteristic speed $U_0$, varying over a length scale $L_0$, we find that the relative magnitude of the convective term to the viscous term, called the Reynolds number, can be written as\index{Reynolds number}
    \begin{equation}
    \label{eq:ReynoldsNumber}
    \text{Re} \equiv \frac{\text{Inertial forces} }{\text{Viscous forces}} =  \frac{|\rho (\vec{u} \cdot \vec{\nabla})\vec{u}|}{|\mu \nabla^2 \vec{u}|} \sim \frac{\rho U_0 L_0}{\mu}.
    \end{equation}
The character of flow will crucially depend on this quantity. The low Reynolds number regime, $\text{Re}\ll 1$, is called Stokes flow [see \S\ref{subsec:LowReynolds}], where viscosity dominates inertia, while the other limit, $\text{Re}\to\infty$, corresponds to inviscid and typically turbulent flow [see \S\ref{sec:Turbulence}]. The critical Reynolds number, $\text{Re}_c$, will mark the value above which the turbulent solution dominates the flow character. In the next section, we will examine this transition in the particular example of a pipe flow.

Today, over 200 years after their formulation, the Navier-Stokes equations have still not been solved in general. 
This is mainly due to the non-linearity of the convective term, $(\vec{u} \cdot \vec{\nabla})\vec{u}$, in the material derivative. 
Even basic properties of their solutions have never been proved, particularly the `existence and smoothness problem'. 
One of the seven Millennium Prizes of \$1 million can be earned with a correct solution \cite{carlson2006millennium}.

\subsection{Drinking from a straw: Hagen–Poiseuille flow}
\label{subsec:PoiseuilleFlow}

Having in mind the notorious mathematical difficulty of fluid mechanics, solutions to the Navier-Stokes equations can be found in specific cases. An important example is laminar pipe flow, which occurs when we drink through a straw. This current is driven by a difference in pressure, $\Delta p$, where we consider a cylindrical tube of radius $R$ and length $L$ with negligible gravity. 
Then, the volumetric flow rate (the flux) going through the tube is described by the Hagen–Poiseuille equation\index{Hagen–Poiseuille flow}\index{Poiseuille flow|see {Hagen–Poiseuille flow}},
    \begin{equation}
    \label{eq:HagenPoiseuille}
        Q = \pi R^4\Delta p/(8 \mu L).
    \end{equation}
The expression was first deduced experimentally, independently by Gotthilf Hagen (1797-1884) and Jean Poiseuille (1797-1869), and soon after it was confirmed theoretically \cite{sutera1993history}.
Perhaps the best-known theory \cite{stokes1880theories} was derived by Sir George Stokes (1819-1903) in 1845, but he did not publish it until 1880, supposedly because he was not certain about the validity of the ``no-slip'' boundary condition\index{no-slip boundary condition} of vanishing velocity at the walls \cite{sutera1993history}.
Stokes also derived the exact flow velocity $\vec{u}$ everywhere in the pipe.
Starting from the Navier-Stokes equations \eqref{eq:NavierStokes} in cylindrical coordinates $(\rho, \theta, z)$, 
assuming that the flow is steady, axisymmetric, and that the radial and azimuthal components of the velocity are zero, one finds the parabolic flow profile,
    \begin{equation}
    \label{eq:PoiseuilleFlow}
        u_z = - \Delta p (R^2-\rho^2)/(4\mu L),
    \end{equation}
which, as expected, is strongest at the center line.
The consequences of the Hagen–Poiseuille equation \eqref{eq:HagenPoiseuille} can be substantial:   
It is 16 times harder to drink through a straw that is 2 times thinner, to achieve the same flux.
This fourth-power scaling is even more problematic for microscopic flow channels, in the field of microfluidics\index{microfluidics} [see \cite{tabeling2005introduction, bruus2008theoretical, squires2005microfluidics, kirby2010micro} and also \S\ref{subsec:microfluidicsInFoodScience}].
For large pipes or fast flows, conversely, the hydraulic resistance\index{hydraulic resistance} increases and the flow becomes turbulent \cite{avila2011onset}.

In the kitchen context, we can see this flow transition in a sink.  When the tap is opened a little, the water column is clear and can be used as an optical lens. The Reynolds number is low, and the flow is laminar.
Opening the tap further, the image begins to fluctuate. Then bubbles appear and begin to jump around vigorously, blurring the water.
When the tap is opened all the way, the column turns completely opaque and white. This is due to the entrainment of air bubbles, causing Mie scattering\index{Mie scattering}, which is roughly independent of the wavelength of light~\cite{hulst1981light}, as opposed to Rayleigh scattering that turns the sky blue, as explained independently by Smoluchowski~\cite{smoluchowski1908molekular} and Einstein~\cite{einstein1910theorie}. 

The critical Reynolds number, $\text{Re}_c$, can be measured directly from this faucet experiment: 
The characteristic length scale $L_0$ is often chosen to be the diameter of the faucet nozzle, $d\sim \SI{1}{\centi\metre}$, and the velocity scale $U_0$ can be determined easily by holding a cup under the faucet at the onset of turbulence \cite{thomsen1993estimating}.
One should find a volumetric flow rate of about $Q \sim \SI{1.8}{\centi\metre^3 \per\second}$, which corresponds to  $\text{Re}_c \approx 2300$ in pipe flow \cite{heavers1990laminar, schlichting2016boundary}.
Near this critical value, one observes puff splitting events [Fig.~\ref{fig:turbulentPipeFlow}b,c], showing that spatial proliferation of chaotic domains is critical to fluid turbulence \cite{avila2011onset}.

Interestingly, one can also determine the diameter of the valve inside the faucet.
Without seeing it, the onset of turbulence can still be heard, as a hissing sound. 
Using the relation $\text{Re} = 4Q/(\nu \pi d)$, with known Re, we can find the critical $Q$ at which the sound emerges, and thus compute the valve diameter. 
Because the valve is usually smaller than the nozzle, this happens at a lower flow rate.
In medicine, this listening technique called auscultation \cite{chizner2008cardiac}\index{auscultation} can be used to detect narrowing of blood vessels, sounds referred to as bruit or vascular murmurs \cite{stein1976turbulent, marsden2014optimization, seo2017method} and, similarly, obstructions of the airways in respiratory conditions \cite{grotberg2001respiratory,  kleinstreuer2010airflow, bohadana2014fundamentals}\index{respiratory flow}\index{cardiovascular flow}.
In \S\ref{subsec:SoundGeneration} we will talk more about hydrodynamic sound generation.

\subsection{Wine aeration: Bernoulli principle}
\label{subsec:BernoulliPrinciple}

In his book ``Hydrodynamica'', Bernoulli (1700-1782) found that pressure decreases when the flow speed increases. More generally, Bernoulli's principle\index{Bernoulli principle} is a statement about the conservation of energy along a streamline. Swiss mathematician Leonhard Euler (1707-1783) used this principle to derive the modern form of the Bernoulli equation,
    \begin{equation}
    \label{eq:BernoulliEquation}
    \tfrac{1}{2}u^2 + \Psi + w = \text{constant along a streamline},
    \end{equation}
where $\Psi=gz$ is the  force potential due to gravity, $w$ is the enthalpy of the fluid per unit mass, and where it is assumed that the flow is steady and that friction due to viscosity is negligible. For an incompressible fluid this is $w=p/\rho$, which is often appropriate for water. For a compressible fluid additional information is required about its thermodynamic energy \cite{batchelor2000introduction}.

Despite its apparent simplicity, the Bernoulli principle is a powerful tool in many applications. 
On the one hand, it can be used to compute flow rates by measuring a pressure difference, for example to determine the speed of an aircraft with a Pitot tube \cite{anderson2010fundamentals}.
On the other hand, it can be used to compute pressure differences by measuring flow rates, for example to determine the pressure distribution around a plane wing (airfoil) and thus the lift force \cite{anderson2010fundamentals}\index{airfoil}. 

\begin{figure}[t]
    \includegraphics[width=1\linewidth]{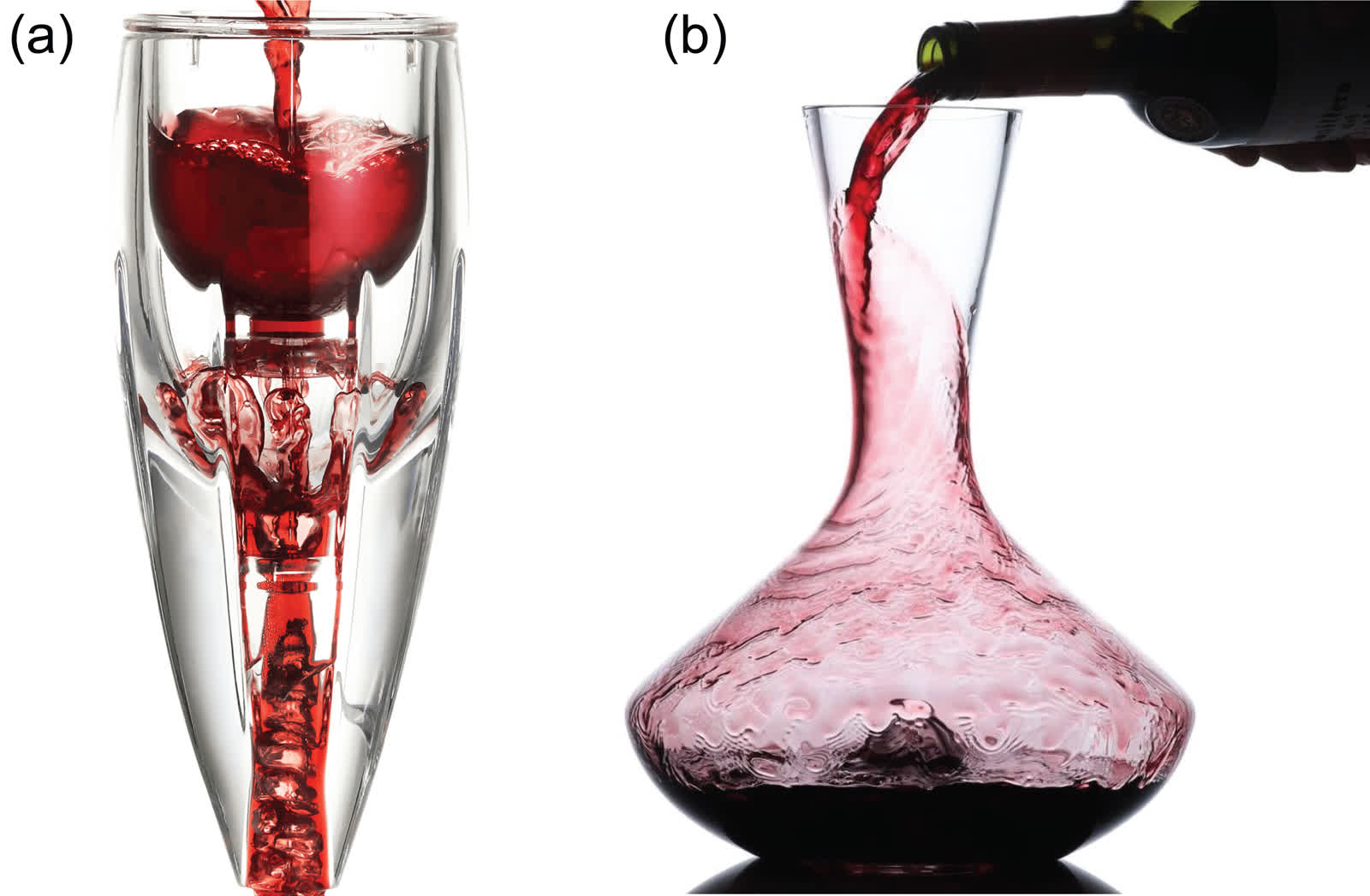}
    \caption{Wine aeration.
    \textbf{(a)} Oxygen injection using the Venturi effect: The wine moves down into a narrow funnel by gravity. In the funnel the liquid accelerates, which lowers the pressure compared to the surrounding atmosphere, as described by the Bernoulli principle [Eq.~\eqref{eq:BernoulliEquation}]. Hence, air bubbles are drawn in, which aerate the wine. 
    \textbf{(b)} Wine decanter: By pouring and swirling the liquid around, ripples form that mix in oxygen efficiently. See \S\ref{subsec:RinsingFlows} about thin film instabilities.
    (a,b) From Vintorio Wine Accessories, with permission. 
    }
    \label{fig:wineAeration}
\end{figure}

Together with the principle of mass conservation, the Bernoulli equation also explains the Venturi effect\index{Venturi effect}:
In a pipe constriction, the pressure decreases as the flow speed increases.
In the kitchen, this Venturi effect is exploited in wine aerators\index{wine} [Fig.~\ref{fig:wineAeration}a]:
The wine moves through a main tube with a constriction, where the lower pressure is used to draw in bubbles from a side tube. 
These air bubbles can improve the wine flavour \cite{ribereau2006handbookv2, balboa2011sensory}. 
Indeed, this was already known to Louis Pasteur, who famously wrote \textit{``C'est l'oxyg\`ene qui fait le vin''} \cite{pasteur1873etudes}, or ``It's the oxygen that makes the wine''.
Note, Fig.~\ref{fig:wineAeration}b shows aeration based on mixing with thin film ripples, which we describe in \S\ref{subsec:RinsingFlows}.
The Venturi effect is also used in gas stoves and grills, where inspirators mix air with flammable gas (instead of wine) to enhance combustion efficiency.
One can read more about the use of fire in the kitchen in \S\ref{subsec:flamesandfire}.
Finally, the Bernoulli principle can be used to regulate pressures in hydraulic devices such as food grippers, for instance to handle sliced fruits and vegetables \cite{davis2008end,petterson2010bernoulli}.

\subsection{Pendant droplets: Surface tension}
\label{subsec:HangingDrop}

The Bernoulli and Navier-Stokes equations describe the motion of fluids, but they don't say much about the interesting phenomena that occur at interfaces: How can droplets hang upside-down from a tap? 
Water molecules are attracted to each other by cohesion forces, particularly by hydrogen bonds \cite{rowlinson2013molecular, de2004capillarity}.
This cohesion\index{cohesion} leads to a surface tension\index{surface tension}, $\gamma$, a force per unit length, which acts as if there was a taut elastic sheet covering the liquid interface \cite{marchand2011surface}.
Indeed, surface tension acts to minimise the surface area, so free droplets tend to be spherical.
This inward force is balanced by a higher pressure inside the droplet. The pressure difference across the water-air interface, called the Laplace pressure\index{Laplace pressure}, is given by the Young-Laplace equation\index{Young-Laplace equation}, 
    \begin{equation}
    \label{eq:YoungLaplace}
    \Delta p = - \gamma H,
    \end{equation}
where $H$ is the mean curvature of the interface. For spherical droplets of radius $R$, we have $H=2/R$. The expression is named after Thomas Young (1773-1829) and Pierre-Simon Laplace (1749-1827).

In the kitchen, the surface tension of water-air interfaces can be measured with a dripping tap experiment: 
A drop of radius $R$ hanging from a faucet balances surface tension $(F_\gamma \sim 2 \pi R \gamma)$ with the force of gravity $(F_g = \frac{4}{3}\pi R^3 \Delta \rho g)$.
Therefore, the pendant drop\index{pendant drop} will fall if it grows larger than $R^* \sim \lambda_c$, the capillary length\index{capillary length} 
    \begin{equation}
    \label{eq:CapillaryLength}
    \lambda_c = \sqrt{\frac{\gamma}{\Delta \rho g}}.
    \end{equation}
Then, by measuring this critical droplet radius, the surface tension can be estimated using the known values for gravity $g$ and the density difference $\Delta \rho$ between water and air.
For a more accurate result, we should take the exact shape of the nozzle and the pendant drop into account \cite{berry2015measurement}\index{tensiometry}.
For water-air interfaces, we then find the surface tension $\gamma \approx 0.07$ N/m.
This value (double-O-seven) is easy to remember because the non-dimensional parameter that characterises the importance of gravity compared to surface tension is called the Bond number\index{Bond number}, 
    \begin{equation}
    \label{eq:BondNumber}
    \text{Bo} \equiv \frac{\text{Gravitational forces} }{\text{Capillary forces}}
     = \frac{\Delta \rho g L_0^2}{\gamma} = \left( \frac{L_0}{\lambda_c} \right)^2,
    \end{equation}
named, however, after the English physicist Wilfrid Noel Bond (1897-1937). Here, $L_0$ is a characteristic length scale, such as the drop radius $R$. Note that the Bond number is also called the E\"otv\"os number (Eo)\index{E\"otv\"os number}, named after the Hungarian physicist Lor\'and E\"otv\"os (1848–1919).
    
Surface tension can lead to a vast number of counter-intuitive phenomena, and new effects are still discovered every day.
This is particularly important for nanotechnology and miniaturization in microfluidics, where the Bond numbers are inevitably small.
Also in microgravity experiments, the effects of surface tension are amplified \cite{kundan2015thermocapillary}, as highlighted by wringing out a wet cloth in the International Space Station \cite{youtube2017wringing}. 
An alternative simple method, accessible for students, is to use a smartphone camera to measure the shape of a hanging droplet \cite{goy2017surface}.
At the end of our experiment, when the pendant drop falls, it makes a distinct `plink' sound, which we explain later in \S\ref{subsec:SoundGeneration}. 
Pliking droplets are also an inherent part of drip coffee, discussed in \S\ref{subsec:brewingCoffee}.

\subsection{Rising liquids: Wetting and capillary action}
\label{subsec:wettingCapillaryAction}

Besides cohesion forces that lead to surface tension, liquid molecules are also subject to adhesion forces\index{adhesion}, when they are attracted to other molecules \cite{rowlinson2013molecular}. 
This can be observed directly when looking at a water droplet sitting on a kitchen benchtop \cite{de2004capillarity}. 
The line where the liquid, gas and the solid meet is called the contact line or the triple line.
The contact angle\index{contact angle}, $\theta_c$, is defined as the angle between the liquid-gas interface and the surface. 
If the benchtop is hydrophilic (attracts water), the droplet spreads out and wets the surface \cite{bonn2009wetting}\index{wetting}, leading to a small contact angle of $0 \leq \theta_c < 90^\circ$.
But if the adhesion forces are weak, such as on waxed surfaces, the surface is called hydrophobic (fears water\index{hydrophobic surface}). Then the cohesion forces prevent spreading by pulling the drop together, leading to a large contact angle of $90^\circ < \theta_c \leq 180^\circ$. 
The equilibrium contact angle is found by minimising the total free energy, for example using calculus of variations.
This leads to the Young equation\index{Young equation},
    \begin{equation}
    \label{eq:YoungEquation}
    \cos \theta_c = \frac{\gamma_\text{SG} - \gamma_\text{SL} }{ \gamma_\text{LG} },
    \end{equation}
where the interfacial energies\index{interfacial energy} $\gamma_{ij}$ encode the relative strengths of the cohesion and adhesion forces between the three phases $i,j \in ($Liquid, Gas, Solid$)$.
Note that without subscript we imply $\gamma = \gamma_\text{LG}$, and the gas phase is sometimes also called the vapour phase.
This equation may be improved by accounting for advanced effects like interface curvature and contact line dynamics \cite{tadmor2004line, jasper2019generalized}.

The degree to which a drop will wet a substrate can be estimated with the spreading parameter\index{spreading parameter} \cite{de2004capillarity}, given by
    \begin{equation}
    \label{eq:SpreadingParameter}
    S = \gamma_\text{SG} - \left(\gamma_\text{LG} + \gamma_\text{SL}\right).
    \end{equation}
When $S>0$, the drop spreads indefinitely towards a zero equilibrium contact angle, as is the case of silicone oil spreading on water. When $S<0$, the drop instead forms a finite puddle, as when a drop of cooking oil is placed on a bath of water in a simple kitchen experiment. Due to its weight, the drop deforms the water surface. The amount of distortion depends on the relative magnitude of buoyancy to surface tension forces, as characterized by the Bond number [Eq.~\eqref{eq:BondNumber}].
The spreading dynamics for partially wetting droplets were recently studied by \citet{durian2022spatters}.

Leonardo da Vinci (1452-1519) first recorded that liquids tend to flow spontaneously into confined spaces, an effect now called ``capillary action'' or ``capillarity'' \cite{de2004capillarity}\index{capillary action}.
When dipping a narrow glass tube or a straw into water, the liquid will rise up until it reaches a constant height $h$. The narrower the capillary, the more the liquid ascends. 
This is somewhat unexpected because the Hagen–Poiseuille equation \eqref{eq:HagenPoiseuille} suggests that the flow rate should decrease for smaller tube radii, if the pressure difference were constant.
Moreover, the flow even moves against the hydrostatic pressure imposed by gravity.
This effect is also explained by intermolecular adhesion and cohesion forces, as the liquid is attracted up by the surfaces.

Quantitatively, the height to which the water rises in a capillary can be calculated by balancing the hydrostatic pressure, $\Delta p_g = \rho g h$, with the Laplace pressure as in Eq.~\eqref{eq:YoungLaplace}.
If the capillary is cylindrical with inner radius $R_i$, and the meniscus has a spherical shape, its radius of curvature is $R = R_i/ \cos \theta_c$ in terms of the contact angle. Combining these ingredients yields Jurin's law \cite{jurin1718account}\index{Jurin's law},
    \begin{equation}
    \label{eq:JurinLaw}
    h = 2 \gamma_\text{LG} \cos \theta_c / (\rho g R_i).
    \end{equation}
For a typical glass microchannel of $R_i=\SI{50}{\micro\metre}$, the water can rise up to $\sim30$ cm and much more for thinner tubes. 
Hence, capillary action has many applications \cite{de2004capillarity}.
For example, you may like growing your own basil for cooking. 
Plants use capillarity for transporting water from the soil to their leaves, together with other mechanisms including osmosis and evaporation \cite{jensen2016sap, katifori2018transport}.

Conversely, if the contact angle $\theta_c > 90^\circ$ for hydrophobic surfaces, $h$ turns negative in Jurin's law [Eq.~\eqref{eq:JurinLaw}], so liquid is expelled.
Then one can observe dewetting\index{dewetting}, the process of a liquid spontaneously retracting from a surface \cite{redon1991dynamics, reiter1992dewetting, herminghaus1998spinodal}.
Consequently, thin liquid films are metastable or unstable on these surfaces, as it breaks up into droplets. 
Therefore, dewetting is often not desirable in industrial applications because it can peel off protective coatings or paint \cite{palacios2019pollock}, and in machinery it can inhibit lubrication\index{lubrication} [\S\ref{subsec:lubrication}]. Dewetting also has important implications for human health. For example, dewetting of lung surfactant layers can inhibit breathing  \cite{hermans2015lung}, and dewetting of the tear film caused by e.g. dry eye disease  \cite{madl2022mucin} or by wearing contact lenses  \cite{suja2022dewetting} can cause ocular  discomfort and vision loss.
Dewetting is also important in solid-state physics because it can damage thin solid films.
Sometimes, this effect can be turned into an advantage for making photonic devices and for catalyzing the growth of nanotubes and nanowires \cite{thompson2012solid}.
An example of a fine-dining accessory exploiting the effects of surface tension is the `floral pipette' \cite{burton2013biomimicry}, which presents a novel means of serving small volumes of fluid in an elegant fashion. 
Not least, wetting properties are crucial for making cooking equipment with non-stick coatings [\S\ref{subsec:nonStickCoatings}].

\subsection{Break-up of jets: Plateau-Rayleigh instability}
\label{subsec:RayleighPlateau}

Dripping kitchen taps offer a direct example of an important hydrodynamic instability [Fig.~\ref{fig:hydraulicjumps1}a]:
When a vertical stream of water leaves a worn tap, it narrows down, stretched by gravity.
Once the liquid cylinder is sufficiently thin, we observe its breakdown into droplets before hitting the sink. 
\citet{plateau1873statique} was the first one to describe this instability systematically, which then enticed \citet{Rayleigh1879} to provide a theoretical description and a stability analysis of an inviscid jet. 
He showed that a cylindrical fluid column is unstable to disturbances whose wavelengths $\lambda$ exceed the circumference of the cylinder. 
The most unstable mode for a jet of radius $R$ has the wave number $k = 2\pi/\lambda \approx 0.697/R$ and from the growth rate of this mode, a typical jet breakup time can be estimated as $\tau_b \sim 3 \sqrt{\rho R^3/\gamma}$, which in a kitchen sink is typically a fraction of a second. 
Kitchen jets exhibit also cross-sectional shape oscillations attributed to capillary waves which can easily be seen in home-made experiments \cite{wheeler2012tap}. By measuring these variations of jet eccentricity, occurring at a frequency proportional to $\tau_b^{-1}$, one can measure the surface tension, following an experimental protocol by \citet{bohr1909determination}.

The physics and stability of liquid jets are fundamentally important for a number of applications, as reviewed by \citet{eggers2008physics}. The break-up of jets has some universal features, which result in a number of self-similar solutions. The scale invariance is manifested both in the conical shape of the tip of a French baguette, and in a bimodal distribution of droplets produced, independently of the initial conditions, in the pinch-off process. The latter is an important technological problem in ink-jet printing \cite{eggers1997nonlinear, martin2008inkjet}. 
In microfluidic systems, the surface-tension assisted breakup is also used to create mono-disperse droplets \cite{anna2016droplets}, and drinking (lapping) mammals including cats and dogs may adjust their jaw-closing times to the pinch-off dynamics to maximise water intake \cite{jung2021pinch}.
Citrus fruits can eject high-speed microjects from bursting oil gland reservoirs, up to 5000 g-forces, comparable to the acceleration of a bullet leaving a rifle \cite{smith2018high}.

\subsection{Hydraulic jumps in the kitchen sink}
\label{subsec:hydraulicJumps}

\begin{figure}[t]
    \centering
    \includegraphics[width=1.0\linewidth]{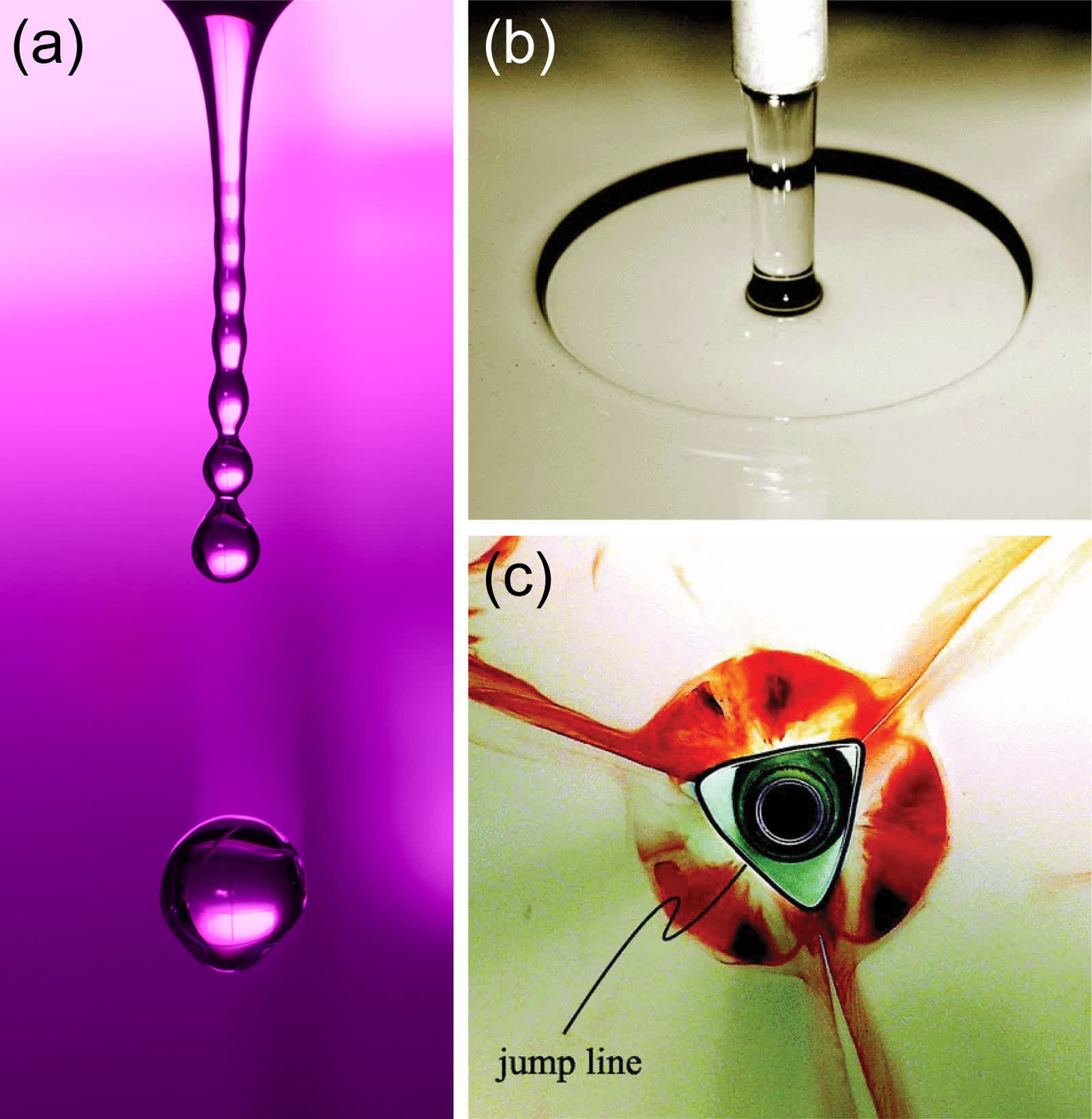}
        \caption{Jets in the kitchen sink.
        \textbf{(a)} Example of the Rayleigh-Plateau instability, where a thin jet from a faucet breaks into droplets. Image by Niklas Morberg, licenced under CC BY-NC 2.0.
        \textbf{(b)} A circular hydraulic jump forms when a thicker liquid jet impinges on a planar surface.
        \textbf{(c)} A triangular hydraulic jump, seen from below through a glass plate. The impinging jet is the black center region, the jump line is the black line surrounding it, and an additional roller vortex is marked with a red dye.
        (b,c) From \citet{martens2012model}.
        }
    \label{fig:hydraulicjumps1}
\end{figure}

When a thicker water jet (that does not break up) impinges on the kitchen sink, it will first spread out in a thin disk at high velocity. 
Surprisingly, at some distance from its origin, the thickness of the film suddenly increases; a hydraulic jump \index{hydraulic jump} is formed [Fig.~\ref{fig:hydraulicjumps1}b]. Again, Leonardo da Vinci was the first known to study hydraulic jumps \cite{hager2013energy}, and \citet{rayleigh1914theory} first described it mathematically. He postulated that a balance between inertia and gravity forces lead to the jump. In other words, a jump is expected when the\index{Froude number} Froude number, 
     \begin{equation}
     \label{eq:FroudeNumber}
     \text{Fr} 
     \equiv \frac{\text{Inertial forces} }{\text{Gravitational forces}}
     = \frac{U_0 }{ \sqrt{g L_0}},
     \end{equation}
transitions through unity, since the flow in the thin film continually loses momentum as is spreads out radially.
Here, $U_0$ is the velocity at the free surface of the film, $g$ is gravity, and $L_0$ is the film thickness. Rayleigh did not include effects of surface tension [\S\ref{subsec:HangingDrop}] in his analysis. However, he wrote that ``On the smallest scale surface-tension doubtless plays a considerable part, but this may be minimised by increasing the stream, and correspondingly the depth of the water over the plate, so far as may be convenient''. \citet{watson1964radial} included effects of viscosity in his description of circular jumps, and two years later, in 1966,  \citet{olsson1966radial} performed complementary experiments to validate Watson's theory and hypothesized that surface tension contributed to the loss of kinetic energy at the jump. \citet{bush2003influence} expanded Watson's theory by exploring the role of surface tension on the formation of circular hydraulic jumps, and \citet{mathur2007gravity} found surface tension to dominate over gravity for films on the micrometer scale. However, all of the above-mentioned authors regarded capillary pressures (due to surface tension, see Eq.~\eqref{eq:YoungLaplace}) as being negligible compared to hydrostatic pressures (due to gravity) on the scale of kitchen sink hydraulic jumps.
\citet{bhagat2018origin} observed that when a strong jet impinges on a planar surface, the radius of the disk (inner region before the jump) is independent on the orientation of the surface. They concluded that gravity is not causing the hydraulic jump when the film is sufficiently thin, as it is when produced by a strong jet. Instead, a capillary pressure competes with the transport of momentum, and this balance is characterised by the Weber number\index{Weber number},
     \begin{equation}
     \label{eq:WeberNumber}
     \text{We}
     \equiv \frac{\text{Inertial forces} }{\text{Cohesion forces}}
     = \frac{\rho U_0^2 L_0}{\gamma},
     \end{equation}
where $\gamma$ denotes surface tension and $\rho$ denotes the density of the impinging liquid.
For weaker jets, however, gravity can no longer be neglected. By balancing the energy at the jump and adopting the approach by \citet{watson1964radial} by assuming a boundary layer flow inside the film, \citet{bhagat2018origin} found the following more general criterion for a circular jump to occur:
    \begin{equation}
    \text{We}^{-1} + \text{Fr}^{-2} = 1,
    \end{equation}
where the first term is associated with capillary waves and the second is associated with gravity waves. 

As we have seen, the rich physics characterizing circular hydraulic jumps has attracted researchers for centuries, and the degree to which surface tension controls these jumps remains an active research topic. \citet{duchesne2019surface} and \citet{bohr2020surface} consider a static control volume and argue that surface tension has a negligible influence as it is fully contained in the Laplace pressure, while \citet{bhagat2020circular, bhagat2022circular} come to a different conclusion by an energy-based analysis. Another aspect of hydraulic jumps concerns the influence of different surface coatings on the jump radius and shape, and \citet{walker2012shearthinning} showed that when a water jet impinges on a shear thinning liquid [see \S\ref{subsec:FoodRheology}], the radius becomes time dependent. Later, the same group used viscoelastic liquids to enhance the degree of particle removal through inducing normal stresses [see \S\ref{subsec:FoodRheology}] that `lift' the particles away from the substrate \cite{walker2014enhancedremoval,hsu2014instabilities}.
\citet{abdelaziz2022non} discuss non-circular jumps for inclined jets.
Finally, it is also possible to create polygonal jumps, either by leveraging hydrodynamic instabilities in viscous liquids \cite{ellegaard1998creating, ellegaard1999cover, bush2006experimental, martens2012model, nichols2020geometry}, as displayed in Fig.~\ref{fig:hydraulicjumps1}c, or by utilizing micro-patterned surfaces \cite{dressaire2009thin}.

\subsection{How to cook a satellite dish}
\label{subsec:RotatingLiquid}

\begin{figure}[t]
    \includegraphics[width=1\linewidth]{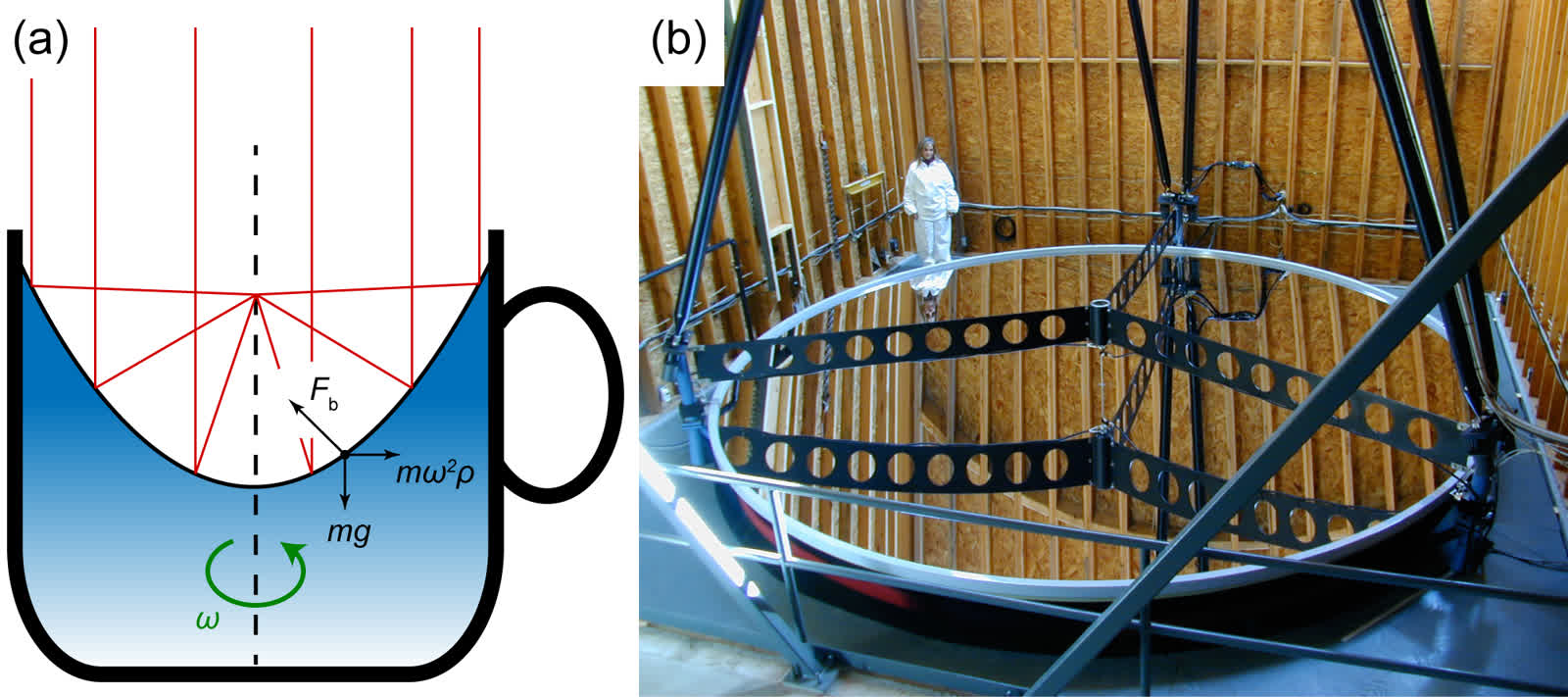}
    \caption{Liquid mirrors.
    \textbf{(a)} Diagram of a rotating liquid that forms a parabolic reflector by the principle of hydrostatic equilibrium. The black arrows denote the force balance between gravity, rotation and buoyancy.
    \textbf{(b)} The Large Zenith Telescope uses this principle. It was one of the largest optical telescopes in the world. Diameter: $\SI{6.0}{\metre}$, rotation period: $\SI{8.5}{\second}$, mercury thickness: $\SI{1.2}{\milli\metre}$, accessible area of sky: 64.2\,deg$^2$. Person for scale. From \citet{hickson2007large}. 
    }
    \label{fig:liquidMirror}
\end{figure}

The importance of parabolas to focus light rays was already known since classical antiquity: Diocles described it in his book \textit{On Burning Mirrors}, and legend has it that Archimedes of Syracuse (c.287–c.212 BC) used these to burn down the Roman fleet \cite{knorr1983geometry}.
The latter is probably fictional, but Archimedes did write that the surface of a rotating liquid forms a paraboloid [Ibid].
At hydrostatic equilibrium, the gravitational force on a fluid element is balanced by the centripetal force and buoyancy [Fig.~\ref{fig:liquidMirror}a], such that the liquid height profile is given by
\begin{equation}
    \label{eq:ParaboloidMirror}
    h(\rho) - h(0) = \omega^2 \rho^2 / (2g), 
\end{equation}
where $\rho$ is the radial distance from the rotation axis, $\omega$ is the angular velocity, and the corresponding focal distance is $f = g/2\omega^2$ [Fig.~\ref{fig:liquidMirror}a].
Liquid-mirror telescopes use exactly this concept: the Large Zenith Telescope [Fig.~\ref{fig:liquidMirror}b] is made of a 6-meter pool of liquid mercury, which is rotated such that the camera sits at the focal point \cite{hickson2007large}.
Instead of a parabolic mirror, the earliest known functional reflecting telescope, which was made by Isaac Newton (1642–1727), used a spherical mirror, because paraboloids are hard to fabricate \cite{wilson2007reflecting}.
For modern large telescopes, the parabolic mirror is sometimes made by spinning molten glass in a rotating furnace.
You can try to do this yourself in the kitchen, by melting some wax (or gelatin) and letting it cool on a record turntable.
Once it has solidified, you could even coat it with reflective paint.
Parabolic reflectors are also widely used in solar cookers and large-scale solar engineering \cite{price2002advances}, opening up interesting avenues for future research in renewable energy technologies \cite{duffie2020solar}.

Equation \eqref{eq:ParaboloidMirror} holds under the assumptions that the whole fluid rotates as a rigid body and that there is no local rotation of neighbouring fluid elements. 
Then the flow is called irrotational \cite{acheson}. 
In closed containers, one must also account for surface tension and potential dewetting when the bottom becomes dry in the middle of the vessel \cite{lubarda2013shape}. 
For higher rotation speeds, however, this static description is no longer valid as the flow becomes rotational. 
In fact, this transition is also highlighted by a symmetry breaking, leading to the formation of polygonal rotating structures \cite{jansson2006polygons, bergmann2011polygon} before all symmetry is lost in turbulence at even higher rotation speeds.

This local rotation of fluid elements is quantified by the fluid vorticity \cite{acheson},\index{vorticity} defined as $\vec{\omega} = \nabla\times \vec{u}$.
The Navier-Stokes equation \eqref{eq:NavierStokes} for an incompressible fluid ($\nabla\cdot\vec{u}=0$) can be recast upon taking a curl of both sides, giving
\begin{equation}
    \frac{D\vec{\omega}}{Dt} = (\vec{\omega}\cdot\nabla)\vec{u} + \nu \nabla^2\vec{\omega}.
\end{equation}
Here the first term of the right-hand side accounts for the stretching or tilting of vorticity due to the flow gradients, while the last term describes the diffusion of vorticity in the fluid. 
Vortices formed across all scales, from atmospheric to molecular processes, are described by their velocity or vorticity distribution. The simplest models assume an axisymmetric velocity field in which the fluid circles around the vortex axis. However, the flow field is more complex in most practical cases.
For instance, secondary flows can arise due to this circling motion and friction with surfaces \cite{okulov2022influence}.
Interestingly, these secondary flows give rise to the tea leaf effect [see \S\ref{subsec:teaLeafParadox}].

\subsection{Washing and drying hands, skincare}
\label{subsec:dryingHands}

In Shakespeare’s Scottish play, Lady Macbeth repeatedly washes her hands `for a quarter of an hour' to cleanse away her murderous guilty conscience. At the start of the COVID-19 pandemic\index{COVID-19}, her troubled soliloquy was used to satirise WHO posters that offer personal hygiene instructions in public restrooms \cite{smith2020macbeth}. 
Jokes aside, washing hands with soap is ``a modest measure with big effects'' to combat pathogen dispersal \cite{group1999hand}, which is of particular importance for the food industry \cite{todd2010outbreaks}.
Also coronaviruses can be cleaned off the skin with soap, or with hand sanitisers that contain sufficiently high concentrations of agents such as ethanol or isopropanol \cite{chin2020stability, poon2020soft, bar2020science}.
\citet{dancer2020revising} also reminds us that besides our hands, we should not forget to clean the surfaces that we touch, following the legacy of Florence Nightingale (1820-1910), often called the founder of modern nursing.
Despite the importance of proper sanitation, its hydrodynamics is not so well explored.
\citet{mittal2020flow} recently wrote \textit{``Amazingly, despite the 170+ year history of hand washing in medical hygiene \cite{rotter1997years}, we were unable to find a single published research article on the flow physics of hand washing.''} 
There is of course a large body of literature about micelle formation and multi-phase flows [\S\ref{sec:Drinks}], foaming [\S\ref{subsec:Foam}] and the physics of micro-organisms [\S\ref{subsec:MicroSwimmers}], but connecting this network of knowledge in the context of personal hygiene is only just starting. 

Motivated by this gap in the literature and without access to a lab due to stay-at-home orders during the pandemic, \citet{hammond2021will} conducted a theoretical assessment of hand washing. Using a lubrication approximation, he came a long way in describing how and when a virus particle is released from our hands when we rub them together.
Hammond found that the rubbing speed needs to exceed a certain value set by the depth of the surface undulations of the skin, which in his model are represented by sinusoidal waves. 
Surprisingly, he found that multiple rubbing cycles are needed to remove a particle. 
More generally, the study of washing biological surfaces could widen our understanding of hydrodynamic interactions between particles, rough surfaces and fluid flows, which is a largely unexplored research field with important implications in the food industry, for example. A natural extension of this work is to include viscoelastic effects, which are likely to enhance the particle removal \cite{walker2014enhancedremoval} [see \S\ref{subsec:hydraulicJumps}], and which also might more realistically represent the material properties of the soap film.  
Moreover, two recent review papers that discuss the biological physics and soft matter aspects of COVID-19 were written by \citet{poon2020soft} and \citet{bar2020science}.

After washing our hands, it is essential to dry them properly \cite{todd2010outbreaks, gammon2020covid}. 
When we use a towel, the water gets pulled into the fabric by capillary action [see \S\ref{subsec:wettingCapillaryAction}].
This only works well if the towel is more hydrophilic (water-loving) than the surface of our hands.
Paper and cotton cloth are especially hydrophilic, aided further by the large surface area of the fibres. 
Another method is to dry hands by evaporation.
Whereas evaporation has been studied extensively on idealised surfaces [see \S\ref{subsec:CoffeeRingEffect}], not so much is known about wetting and evaporation on soft materials like the skin \cite{lopes2012evaporation, gerber2019wetting}.
An ongoing debate is whether the dispersal of viruses and bacteria can be stopped more efficiently by warm air dryers, or jet dryers, which on the one hand may avoid having to touch surfaces but on the other hand could cause pathogen aerosolization \cite{huang2012hygienic, mittal2020flow, reynolds2020comparison, best2014microbiological}, which is especially problematic in food processing plants \cite{kang1989biological}.

A common medical condition that comes with washing and drying hands frequently is xeroderma, or dry skin\index{skin care} \cite{walters2012cleansing}.
This can lead to symptoms including itching, scaling, fissure, or wrinkling \cite{cerda2003geometry, aharoni2017smectic}.
These problems can often be alleviated with moisturisers or emollients, but in more severe cases an effective treatment requires understanding the underlying biophysical mechanisms \cite{proksch2020dry}.
Liquid transport has been studied in the networked microchannels of the skin surface \cite{dussaud2003liquid}, as well as the physics of stratum corneum lipid membranes \cite{das2016physics}, and more generally soft interfacial materials \cite{brooks2016soft}.
Connecting the disciplines of physics and medicine will become increasingly important in future research.

\section{Drinks \& Cocktails: Multiphase Flows}
\label{sec:Drinks}

After washing our hands, it is time to start dinner with a beverage of choice.
In this section we review a wealth of hydrodynamic phenomena that can emerge inside your drink.
Surely referring to this, the Roman emperor Marcus Aurelius (121--180 AD) once said:
\begin{quote}
    \emph{Look within. Within is the fountain of good, and it will ever bubble up, if thou wilt ever dig.}
\end{quote}
Examples of surprising effects happening inside beverages are shock waves in the tears of wine, effervescence in Champagne, or `awakening the serpent' during whisky tasting.
These multi-phase flows \cite{brennen2005fundamentals, michaelides2016multiphase} have seen rapid scientific advances recently, and they are applied extensively in industrial processes.
Perfect to contemplate while waiting for the main course to arrive, or to impress at a cocktail party.
But, as the Dutch proverb says, \textit{``Don't look too deep into your glass''}.

\subsection{Layered cocktails} 
\label{subsec:LayeredCocktails}

A classic example of a culinary multi-phase fluid is a layered cocktail [Fig.~\ref{fig:LayeredCocktail}a]. For instance, an Irish flag cocktail is made by first pouring cr\`eme de menthe, then a layer of Irish cream, topped off with orange liqueur. 
This beverage is called stably stratified, because each layer is less dense than the one below it, so buoyancy keeps the layers separate [see Archimedes' principle, \S\ref{subsec:Eureka}].
Note that density of liquids can be measured with high accuracy using a hydrometer \cite{lorefice2006calibration}, which can also used in breweries for assessing the strength of alcohol (alcoholometer) or in the dairy industry for measuring the fat content of milk (lactometer).
Then, multiple coloured cocktail layers can be formed using a density chart for the different liquid ingredients, also called a specific gravity chart \cite{cocktaildensitychart}. 
Stratification is essential for life on Earth, both in the atmosphere \cite{mahrt2014stably}, where sharp cloud layers can be observed, and in the ocean \cite{li2020increasing}, where water layers can be characterised by large gradients in density (pycnocline\index{pycnocline}), but also gradients in temperature (thermocline\index{thermocline}) or salinity (halocline\index{halocline}), immediately impacting environmental stratified flows \cite{grimshaw2002environmental} or phytoplankton migration \cite{sengupta2017phytoplankton}, and more generally geophysical fluid mechanics \cite{pedlosky1987geophysical}.

\subsubsection{Inverted fountains}

The cocktail layers will separate readily if the ingredients are immiscible\index{immiscible fluids}, such as, say, lemon water and rose oil.
However, if the liquids are miscible\index{miscible fluids}, it is recommended to pour the layers slowly (ideally along the side of the glass with the help of a spoon) because otherwise the layers will mix.
We can understand this turbulent and miscible mixing process as an `inverted fountain' \cite{turner1966jets, hunt2015fountains}, where the lighter fluid is forced down into the heavier fluid, opposed by buoyancy \cite{xue2019fountain}. 
The (inverted) height of the fountain, $z_{f}$, and thus the mixing volume, depends strongly on the Reynolds number [Eq.~\eqref{eq:ReynoldsNumber}], where $U_0$ is the pouring velocity and $L_0$ is the radius of the injected jet, but also the densimetric Froude number, where $g$ in Eq.~\eqref{eq:FroudeNumber} is replaced by $|g'|$, the reduced gravity due to buoyancy, given by
     \begin{equation}
     \label{eq:FountainBuoyancy}
     g' = g (\rho_{i}-\rho_{a})/\rho_{a},
     \end{equation}     
in terms of $\rho_{i}$ and $\rho_{a}$, the densities of the injected and the ambient fluid.
Conventionally, $g'$ is negative for inverted fountains.
For large Froude numbers, \citet{turner1966jets} showed that the fountain height is given by
    \begin{equation}
    \label{eq:TurnerFountain}
    z_f \approx 2.46 L_0 \text{Fr}. 
    \end{equation}   
This classical result predicts a linear relation, which agrees well with modern experiments. 
Note that different scaling laws have been derived for weaker fountains with a smaller Froude number, as reviewed by \citet{hunt2015fountains}.


\subsubsection{Internal waves}

\begin{figure}[t]
    \centering
    \includegraphics[width=\linewidth]{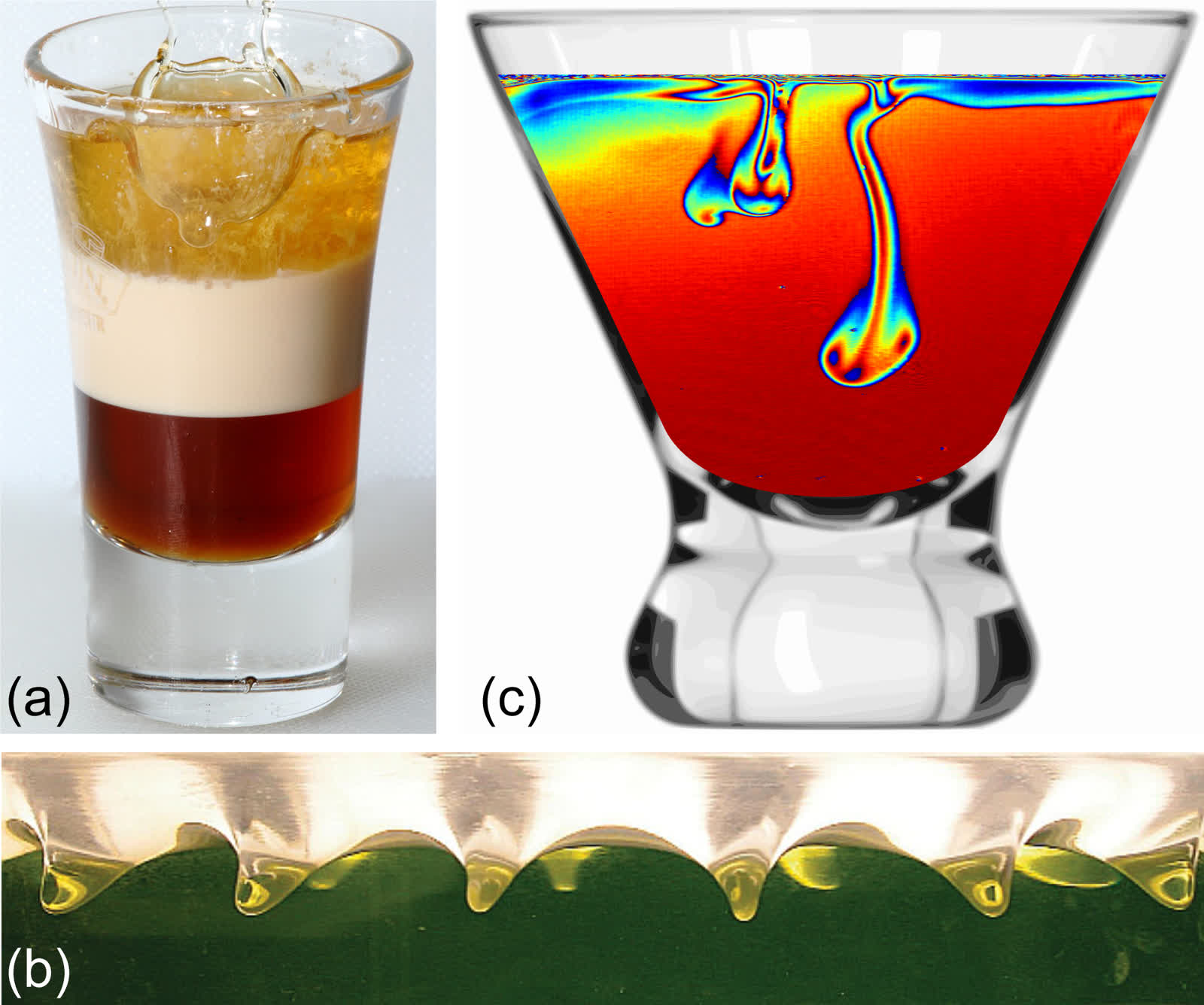}
    \caption{Multi-phase cocktails.
    \textbf{(a)} B-52 shot made by layering Kahlua, Bailey's Irish Cream and Grand Marnier, with a splash on top. Image from J. D. Baskin on Flickr, licenced under CC BY 2.0.
    \textbf{(b)} Kelvin-Helmholtz instability waves formed at an oscillated water-oil interface. From \citet{yoshikawa2011oscillatory}. 
    \textbf{(c)} Evaporation-induced Rayleigh-Taylor instability. Colours indicate the ethanol concentration, from blue (low) to red (high), measured with Mach–Zehnder interferometry. Image courtesy of Sam Dehaeck. 
    }
    \label{fig:LayeredCocktail}
\end{figure}

Once the cocktail layers are established, more interesting flow phenomena can be observed.
For example, when the glass is slightly disturbed, gravity waves can be seen (not to be confused with the gravitational waves in general relativity), and these waves propagate inside the fluid instead of on its surface \cite{benjamin1967internal, helfrich2006long}.
Specifically, they are called interfacial (internal) waves \index{internal waves} when they propagate horizontally along an interface characterised by a density gradient, $d\rho/dz<0$.
Consider a fluid parcel in a continuously stratified fluid, with a smooth density profile $\rho(z)$, that is in hydrostatic equilibrium at $z_0$.
If the parcel of density $\rho_0=\rho(z_0)$ is displaced a vertical distance $\Delta z=z-z_0$ where the surrounding fluid has a different density $\rho(z_0+\Delta z)$, it will feel a gravitational restoring force [\S\ref{subsec:Eureka}]. To first order, that leads to simple harmonic motion \cite{vallis2017atmospheric} with an oscillation frequency given by 
    \begin{equation}
    \label{eq:InternalWaveFrequency}
    f = \frac{1}{2\pi} \sqrt{-\frac{g}{\rho_0}\frac{d\rho}{dz}}.
    \end{equation} 
This expression was independently co-discovered by \citet{vaisala1925uber} and \citet{brunt1927period}, so it is called the Brunt–V\"ais\"al\"a frequency.
These internal oscillations are typically slow compared to surface waves at the liquid-air interface, because the density gradient between the liquid layers is much smaller [see \S\ref{subsec:Ripples}].
Internal waves are common in oceanography \cite{garrett1979internal} and atmospheric science \cite{pedlosky1987geophysical}, where they lead to beautiful patterns such as rippled clouds or lenticular clouds \cite{lamb2011physics}.

\subsubsection{Kelvin–Helmholtz instability}
\label{subsubsec:KelvinHelmholtzInstability}

Another spectacular atmospheric phenomenon is the formation of fluctus clouds, which look like breaking ocean waves in the sky. 
They are caused by the shear-induced Kelvin–Helmholtz (KH) instability\index{Kelvin–Helmholtz instability} \cite{kelvin1871hydrokinetic, drazin2004hydrodynamic}, when two fluid layers move alongside each other.
Indeed, the same can be observed when a layered cocktail is sheared. 
By rotating the glass, a velocity gradient $\partial u/\partial z$ is created between the stratified layers.
This shear is especially pronounced if the liquid spins up by friction with the bottom wall instead of the side walls [see \S\ref{subsec:teaLeafParadox}].
The velocity gradient drives the KH instability, while it is opposed by buoyancy, quantified by the density gradient $\partial \rho/\partial z$. 
The ratio of these two forces is encoded by a dimensionless quantity named the (densimetric) Richardson number, 
     \begin{equation}
     \label{eq:RichardsonNumber}
     \text{Ri}
     \equiv \frac{\text{Buoyancy forces} }{\text{Shear forces}}
     =
     \frac{g^{\prime}}{\rho} \frac{\partial \rho / \partial z}{(\partial u / \partial z)^{2}}.
     \end{equation}
The fluid layers are unstable when the shear is large enough, when $\text{Ri} \lesssim 1$, depending on the system configuration. 
In a setup resembling our cocktail, the instability was characterised recently in a spin-up rotating cylindrical vessel by \citet{yan2017profile}, and in an oscillatory cylindrical setup by \citet{yoshikawa2011oscillatory}, as shown in Fig.~\ref{fig:LayeredCocktail}b.
Naturally, the KH instability will cause the stratified layers to mix with one another, as reviewed by Peltier and Caulfield \cite{peltier2003mixing}, or cause emulsion formation if the layers are immiscible [\S\ref{subsec:Emulsions}]. 
In the latter case, surface tension stabilises the short wavelength instability on top of buoyancy, which strongly affects emulsion formation \cite{thorpe1969experiments, drazin1970kelvin}.
Therefore, the KH instability is important for many processes in industry and food science. 
Think about making mayonnaise with a blender, for example. 
From a fundamental point of view, understanding these flows is intrinsically connected with the heart of theoretical physics: symmetries. 
Only recently, \citet{qin2019kelvin} described that the KH instability results from parity-time symmetry breaking.
Moreover, the Kelvin-Helmholtz instability also features in the magnetohydrodynamics of the sun \cite{foullon2011magnetic}, ocean mixing \cite{pedlosky1987geophysical}, relativistic fluids \cite{bodo2004kelvin} and superfluids \cite{blaauwgeers2002shear}.

\subsubsection{Rayleigh-Taylor instability}
\label{subsubsec:RayleighTaylorInstability}
Until now we have discussed stably stratified cocktails. 
Yet, when a heavier fluid sits on top of a lighter fluid, the latter pushes into the former by gravity, so the mechanical equilibrium is unstable.
Any small perturbation will lead to a familiar pattern of finger-like structures with a mushroom cap, as seen in Fig.~\ref{fig:LayeredCocktail}c.
This phenomenon is explained by the Rayleigh-Taylor (RT) instability\index{Rayleigh-Taylor instability}, which was first discovered by \citet{Rayleigh1883}, and later described mathematically by \citet{Taylor1950} together with systematic experiments by \citet{lewis1950instability}. 
Many developments followed, and \citet{chandrasekhar2013hydrodynamic} extended the theoretical description in his famous book. 
Like the KH instability, the RT instability is relevant across the disciplines, from the astrophysics of supernovae \cite{kuranz2018high, abarzhi2019supernova} to numerous technological applications \cite{drazin2004hydrodynamic}. 

The RT instability arises because the system seeks to minimise its overall potential energy.
Its onset is primarily governed by the Atwood number, 
    \begin{equation}
        \label{AtwoodNumber}
        \text{At} = \frac{\rho_{h} - \rho_{l}}{\rho_{h} + \rho_{l}},
    \end{equation} 
the non-dimensional difference between the densities of the heavier and the lighter fluid, $\rho_{h}$ and $\rho_{l}$. 
This  number is most likely named after George Atwood FRS (1745-1807), who also invented the Atwood machine, but we could not find the original source.
To describe the RT instability more generally, one must account for the fluid viscosities and surface tension \cite{andrews2010small}, and potential effects due to fluid compressibility \cite{boffetta2017incompressible}.
Moreover, the dynamics depend strongly on the initial conditions: They begin with linear growth from perturbations, which transitions to a non-linear growth phase involving characteristic structures of rising `plumes' and falling `spikes'. Subsequently, these plumes and spikes interact with each other through merging and competition, and roll up into vortices. The final stages are characterized by turbulent mixing [see \S\ref{sec:Turbulence}]. 

In the words of \citet{benjamin1999rayleigh}, the RT instability is a fascinating gateway to the study of fluid dynamics [see also \S\ref{subsec:learning}].
It can readily be observed in kitchen experiments, but it also occurs spontaneously without us even noticing [Fig.~\ref{fig:LayeredCocktail}c]:
In a well-mixed (non-layered) cocktail, the alcohol evaporates faster than water. 
Hence, at the air interface, a water-rich layer develops that is denser than the bulk mixture, which gives rise to the RT instability \cite{dehaeck2009evaporating}. 
The plumes of such `evaporating cocktails' are observed using a Mach–Zehnder interferometer. By demodulating the fringe patterns using a Fourier transform method \cite{kreis1986digital}, it is possible to compute the refractive index field, and hence the local ethanol concentration.
Evaporation-induced {R}ayleigh-{T}aylor instabilities also occur in polymer solutions \cite{mossige2020evaporation}, so your cocktail need not necessarily be alcoholic.

Interestingly, the RT instability is often inseparable from the Kelvin-Helmholtz instability. RT flows create velocity gradients that trigger KH billows, while KH flows create density inversions that trigger RT fingers.
Moreover, the RT instability is closely related to the Richtmyer–Meshkov (RM) instability, when two fluids of different density are impulsively accelerated \cite{abarzhi2019supernova}, and to the Rayleigh-B\'enard convection, where an instability occurs due to heating a single liquid from below or cooling it from above, which we describe in more detail in \S\ref{subsec:HeatingBoilingRBC}.

\subsection{Tears of wine}
\label{subsec:Tears}

\begin{figure}[t]
    \includegraphics[width=1\linewidth]{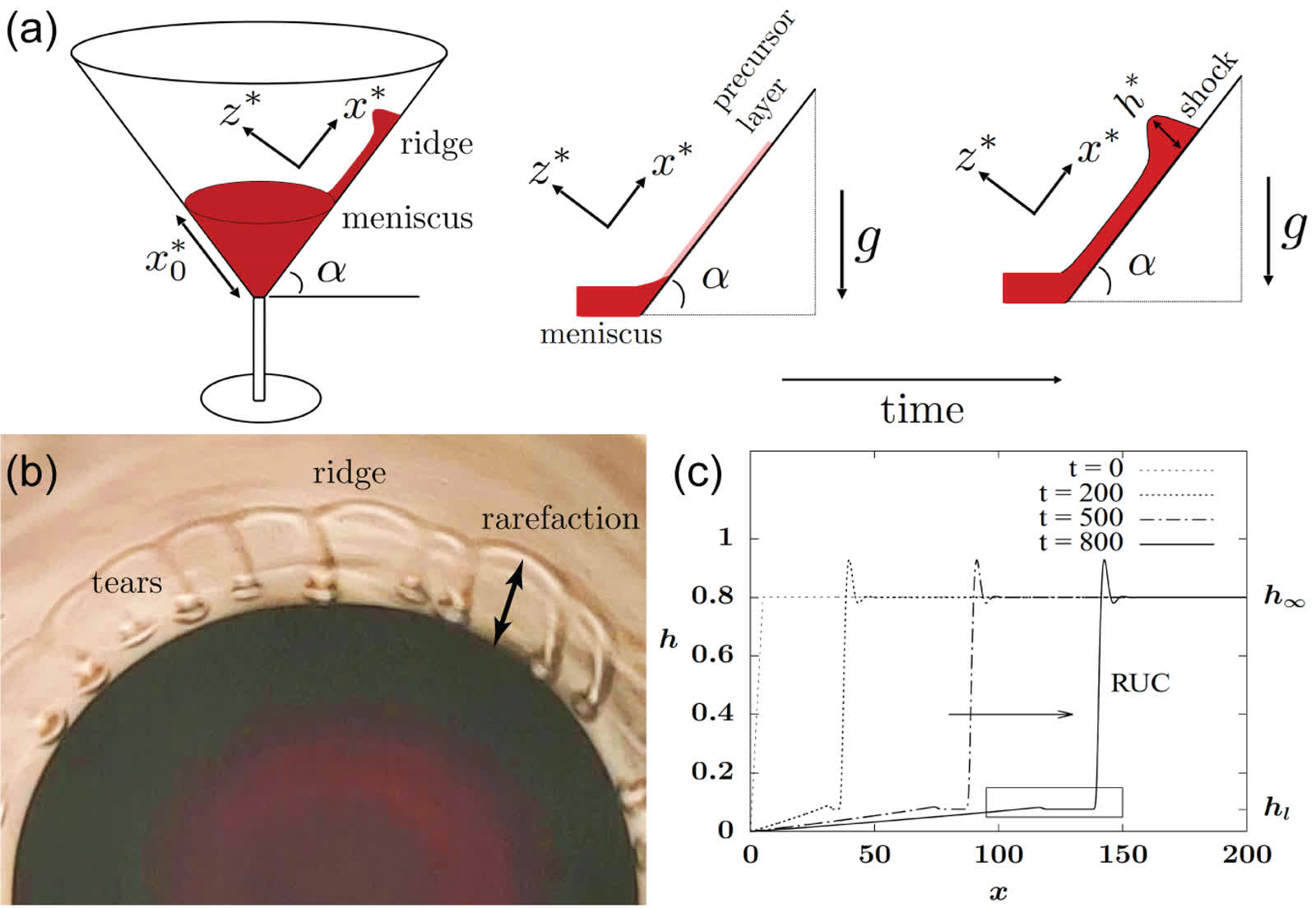}
    \caption{Shock waves in tears of wine. 
    \textbf{(a)} Schematic of a conical-shaped glass of inclination angle $\alpha$, showing a one-dimensional thin wine film traveling up an inclined flat glass surface. The film height $h^*$ is exaggerated for clarity.
    \textbf{(b)} Experiment using 18\% ABV port wine and $\alpha=\SI{65}{\degree}$. Swirling the wine around the glass creates a front that forms out of the meniscus. The draining film advances up the glass and destabilises into wine tears.
    \textbf{(c)} The formation of a reverse-undercompressive (RUC) shock.
    (a-c) From \citet{dukler2020theory}. 
    }
    \label{fig:tearsOfWine}
\end{figure}

One of the most surprising phenomena in multi-phase flows is the Marangoni effect\index{Marangoni effect}, named after Carlo Marangoni (1825-1940) \cite{marangoni1871ueber}.
You may have seen this effect already in the kitchen, when a droplet of dish soap falls into a bowl of water sprinkled with pepper: Within the blink of an eye, the pepper moves to the edges by an outward flow along the liquid-air interface. 
Another striking example is adding food colouring drops to a bowl of milk, where poking it with a soap-covered cotton bud generates beautiful flow patterns (try it!). 
These Marangoni flows arise because the surfactant molecules in the soap lower the surface surface tension \cite{levich1969surface}, leading to a difference in surface tension along the interface, of $\Delta \gamma = f \gamma$, where the factor $f \sim 10^{-1}$ for most soaps.
Consequently, the water without soap pulls more strongly on the water with soap, generating a current from regions of lower to higher surface tensions. 
The flow strength can be estimated roughly as $u \approx \Delta \gamma / \mu$, so even small fractions $f$ give rise to fairly strong flows, using $\gamma \sim \SI{0.07}{\newton\per\metre}$ and viscosity $\mu \sim \SI{0.9}{\milli\pascal\second}$ for water.
Of course, more detailed calculations must take other influences into account, including solubility, surface contamination and system geometry \cite{levich1969surface, halpern1992dynamics, lee2007spreading, roche2014marangoni, kim2017solutal}. 
The ratio between advective and diffusive transport over a characteristic length scale $L_0$ is given by the Marangoni number\index{Marangoni number},
     \begin{equation}
     \label{eq:MarangoniNumber}
     \text{Ma}
     \equiv \frac{\text{Advective transport rate} }{\text{Diffusive transport rate}}
     =
     \frac{\Delta \gamma L_0}{\mu D},
     \end{equation}
where $D$ is the diffusivity of the surfactants or any additive that changes the surface tension.

Fortunately, the Marangoni effect does not only occur with soap, but also with edible ingredients.  
In fact, the phenomenon was first identified by James Thomson (1822-1892) in the characteristic ``tears'' or ``legs'' of wine\index{tears of wine}\index{legs of wine} \cite{thomson1855curious}, and indeed other alcoholic drinks including liquors and whisky\index{whisky} [see \S\ref{subsec:WhiskyTasting}].
These tears are formed because the alcohol is more volatile than water, and it has a lower surface tension \cite{fournier1992tears}.
To see this, pour yourself a glass of wine [see Fig.~\ref{fig:tearsOfWine}a].
In the thin meniscus that the wine forms with the glass surface, the alcohol evaporates faster than the water, so the surface tension here is higher than in the bulk.
The wine is then pulled up the meniscus, forming a thin film that starts climbing up along the side of the glass.
After a few seconds, the film forms a ridge approximately $\SI{1}{\centi\metre}$ above the meniscus.
This ridge becomes unstable under its own weight as more wine climbs up, so it collapses into ``tears'' that fall down towards the meniscus [Fig.~\ref{fig:tearsOfWine}b].
Large tears can fall back into the bulk, but small tears can also be pulled up again by the continuously climbing film that replenishes the ridge.
This can cause the tears to bounce up and down, especially at the meniscus. The effects are beautifully imaged using the Schlieren or shadow projection techniques \cite{settles2001schlieren}.

So far, so good, but the mechanism by which the droplets form and collapse is more complex. 
As the wine film climbs to its terminal height, when the Marangoni stresses are balanced by gravity, 
this transient stationary state is subject to various hydrodynamic instabilities \cite{vuilleumier1995tears, hosoi2001evaporative, fanton1998spreading}.
Besides alcohol concentration differences, the evaporation also induces temperature gradients that lead to additional Marangoni stresses \cite{venerus2015tears}.
The ridge instability that triggers the formation of wine tears was also studied and analysed with a Plateau-Rayleigh-Taylor theory \cite{nikolov2018tears}.
Yet, the dynamic formation of the ridge itself is still not well understood. 

Until now we have discussed a wine film that spontaneously climbs up a dry wine glass, but it is common practice among connoisseurs to swirl the wine around.
This often creates a wet coating film much higher than the terminal climbing height, which can give rise to rather different behaviours.
\citet{dukler2020theory} showed that such swirled films can feature a `shock wave' that climbs out of the meniscus [Fig.~\ref{fig:tearsOfWine}c], again driven by Marangoni stresses due to evaporation.
This wave can be observed as a ridge that propagates upwards, where the wine film above the shock front is thicker than below it.
Specifically, the dynamics can be described as a reverse undercompressive (RUC) shock. 
This type of shock wave is unstable: 
Small inhomogeneities in the wine film are amplified into thick drops, which then fall down as tears. 
As described previously for rising films driven by thermal gradients \cite{reverse2003sur, munch2003pinch}, different shock morphologies can occur in other circumstances. This is of great scientific and technical interest, including dip-coating and painting applications.
Moreover, it would be interesting if people's dining experience could be improved by developing a new dish that uses the shock wave as a surprise effect.

\subsection{Whisky tasting}
\label{subsec:WhiskyTasting}

\begin{figure}[t]
    \includegraphics[width=1\linewidth]{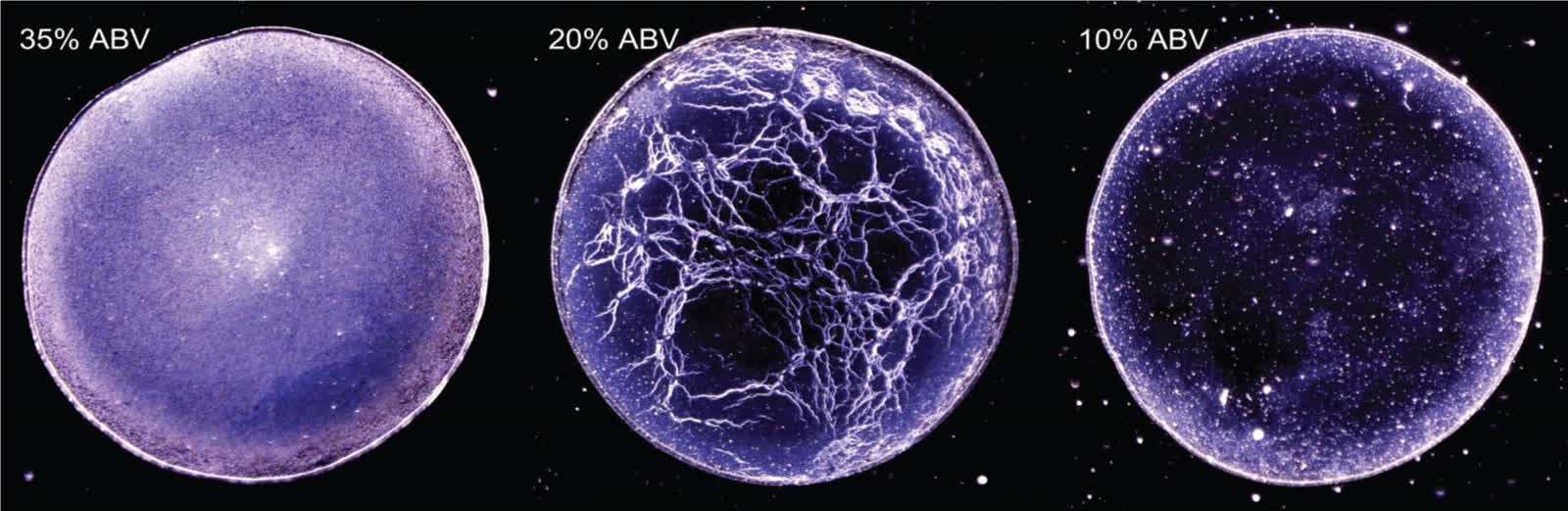}
    \caption{Whiskey webs. Different patterns emerge after letting whiskey droplets of different alcohol percentages evaporate on a glass surface. At 35 $\%$ ABV \textbf{(left)}, the deposits are evenly distributed,  while at 10 $\%$ ABV \textbf{(right)}, the deposits are distributed preferentially near the rim of the drop. At intermediate (20 $\%$ ABV) concentration  \textbf{(middle)}, the deposits form complex `whiskey web' patterns.  
    From \citet{williams2019whiskey}. 
    }
    \label{fig:whiskeyWebs}
\end{figure}

We use all of our senses when we taste whisky or whiskey  \cite{velasco2013assessing}.
The spelling whiskey is common in Ireland and the United States, while the term whisky is common for produce from the UK and most other countries.
While complete volumes have been written about the art of whisky tasting, see e.g. \cite{maclean2020malt}, we would like to highlight some of its hydrodynamic aspects, which are often clearly visible.
That is, a good deal of information may be gained by assessing the appearance and dynamics of spirits \cite{miller2019whiskyscience, russell2014whisky}.
The following tests can give clues about the whisky quality and vintage before any smelling or tasting.

As a first examination, it is customary to inspect the tears that we described in \S\ref{subsec:Tears}.
This gives an indication of the whisky's alcohol content, its viscosity and its surface tension, which in turn depend on the exact chemical composition.
When the tears run down slowly, with thick legs, it indicates that it will give more mouthfeel  \cite{maclean2010viscimetry}. 
Conversely, if the legs are thin and run quickly, the whisky is likely to be younger and of a lighter body.
This is because the texture changes during the aging process, as viscous natural oils and other compounds are released from the wooden casks \cite{mosedale1995effects}, which inevitably also influence the whisky colour.
The rheological and thermophysical properties are also affected by storage and temperature \cite{hlavac2013influence}.
However, a whisky with more pronounced tears is not  automatically sweeter or better in quality, since the tear formation is a purely physical phenomenon. 
Indeed, the tears vanish when the glass is covered, since the evaporation-induced Marangoni stresses disappear.

A second experiment is the `beading' test \cite{miller2019whiskyscience, davidson1981foam, maclean2018beading}.
When a whisky bottle is shaken vigorously, a foam can appear on the liquid interface if the alcohol concentration is higher than approximately $50\%$ alcohol by volume (ABV).
The beading is not necessarily more pronounced at higher concentrations, but it is not observed below a certain percentage.
Beading can also say something about the age of the whisky: The bubbles tend to last longer in older vintages because the compounds released from the wooden casks can stabilise the foam.
Read more about foam stability in \S\ref{subsec:Foam}.

A third inspection method is called whisky viscimetry\index{viscimetry}  \cite{smith2009azwhisky, maclean2010viscimetry}.
When adding a little water to the whisky, small vortices called `viscimetric whirls' appear when the liquids of different viscosities mix with one another.
Connoisseurs sometimes refer to this phenomenon as `awakening the serpent' \cite{smith2009azwhisky}.
These vortices only last for a few seconds, but again they tell us something about the texture of the whisky.
The more persistent the whirls, the thicker the mouthfeel and the higher the alcohol concentration.
To the best of our knowledge, this effect in whisky has not been quantified systematically in the scientific literature, but it is related to the miscible droplet dynamics discussed in \S\ref{subsec:miscibledropshape}.

Depending on the distillation method, spirits reach an initial strength of $\sim70\%$ ABV (pot still) or even higher  (column still).
Most whiskies are then diluted down to $\sim60\%$ ABV prior to storage in casks. After maturation, they are often mixed with more water to $\sim40\%$ ABV, the minimum in most countries. 
There are several reasons for this dilution:
First, it can enhance the flavour because many of the taste-carrying molecules, such as guaiacol, are thermodynamically driven up to the liquid-air interface at low ethanol concentrations \cite{karlsson2017dilution}.
Second, the ethanol concentration influences the sensory perception \cite{velasco2013assessing, harwood2020influence}, where lower strengths are more palatable by most consumers. 
Third, besides enhancing the flavour of spirits, dilution can lead to a better mouthfeel.
At $\SI{20}{\celsius}$, the viscosity of pure ethanol and water is $\mu_1 \approx \SI{1.2}{\milli\pascal \second}$ and $\mu_2 \approx \SI{1}{\milli\pascal \second}$, respectively, but the viscosity of an ethanol-water mixture features a maximum of $\mu_{12} \approx \SI{3}{\milli\pascal \second}$ at 40-50\% ABV \cite{dizechi1982viscosity}.
This surprising non-linear effect of binary mixture viscosities is described in more detail in \S\ref{subsec:Blending}.

Remarkably, diluting your drink can also help with distinguishing whether it is whisky or whiskey:
An evaporating bourbon droplet of 40\% ABV tends to leave a uniform surface deposition [Fig.~\ref{fig:whiskeyWebs}], while a diluted droplet at 20\% ABV leaves distinctive patterns called whiskey webs \cite{williams2019whiskey, carrithers2020multiscale, williams2021whiskey}.
Apparently, Scotch whisky and other distillates do not feature these web patterns and they are unique across different samples of American whiskey, so they could act like fingerprints.
Indeed, the flavour profile results from the intricate chemical composition, which also affect the web patterns through the interplay of bulk chemistry with surfactants and polymers.
Similarly, the webs do not form in droplets below 10\% ABV, where instead the coffee-ring effect\index{coffee-ring effect} is observed [see \S\ref{subsec:CoffeeRingEffect}].
In general, this rich variety of surface depositions results from a combination of intrinsic (chemical composition) and extrinsic factors (temperature, humidity) that lead to an interplay of Marangoni flows and macromolecular surface adsorbtion. 
The non-uniform residues are often undesired in many industrial applications including 3D printing, so whisky experiments could help us with understanding and controlling uniform coatings \cite{kim2016controlled}.
Drying drop technologies may also be used for wine and hard drinks quality control \cite{yakhno2018drying}.

\subsection{Marangoni cocktails}
\label{subsec:MarangoniCocktails}

\begin{figure}[t]
    \includegraphics[width=1\linewidth]{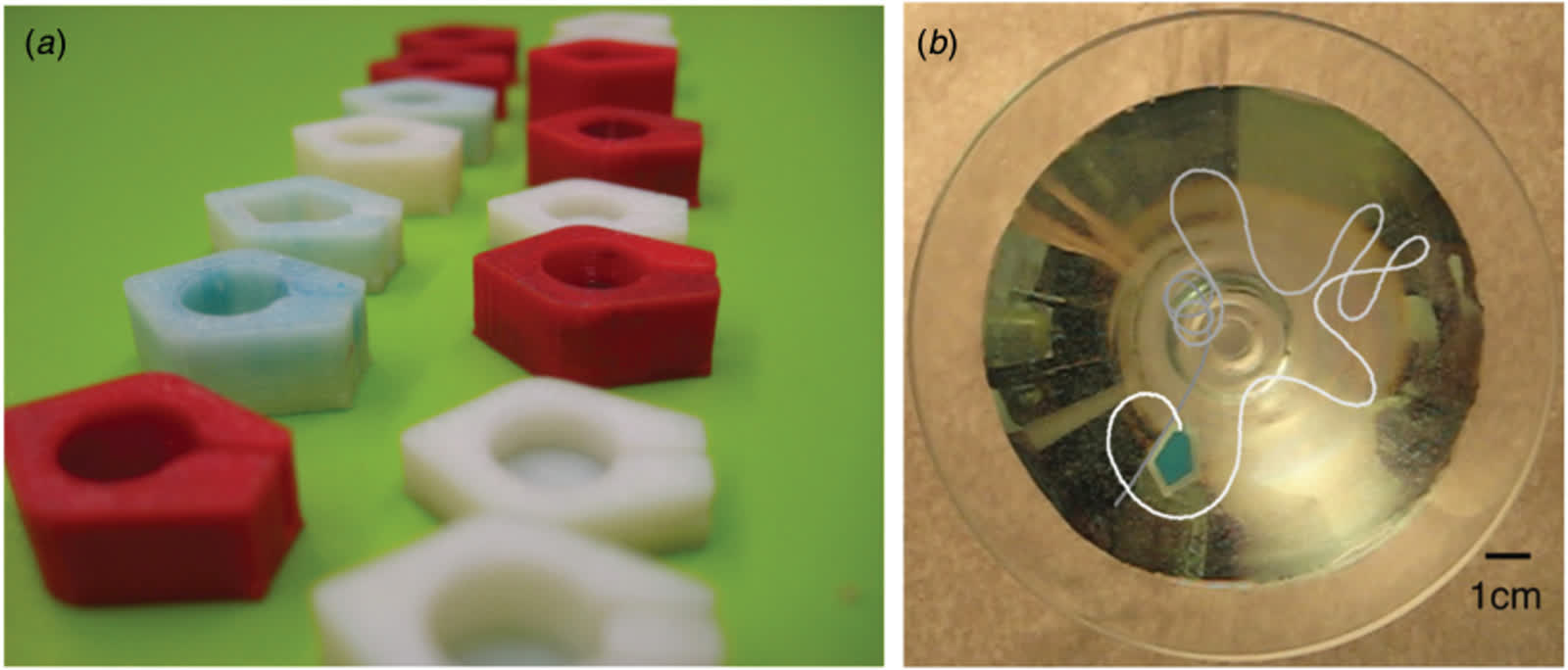}
    \caption{Marangoni-stress powered cocktail boats.
    \textbf{(a)} A fleet of cocktail boats with different designs. The fuel (any liquor) is stored in the central cavity. The thin slit at the rear slowly releases the fuel into the glass, which establishes a surface tension gradient that drives the boat forward.
    \textbf{(b)} Trajectory in a cocktail glass. This boat is $\sim \SI{1.5}{\centi\metre}$ long and fueled by Bacardi 151 (75\% ABV).
    (a,b) From \citet{burton2013biomimicry}. 
    }
    \label{fig:CocktailBoat}
\end{figure}

Another well-known demonstration of the Marangoni effect is the `soap boat'\index{soap boat} \cite{keller1998surface, youtube2014marangoniboat}. 
These boats propel themselves in the kitchen sink by releasing surfactant molecules from the back: 
The surface tension of water is then higher at the front, so the boat is effectively pulled forwards.
The same effect is also used by water-walking insects as a quick escape mechanism \cite{bush2006walking}.
However, this propulsion is short lived when using soap as fuel in a closed geometry, because the interface becomes saturated with surfactants.
A prolonged motion can be achieved by using other commonly available fuels \cite{renney2013easy}.
Moreover, continuously moving boats can be made with camphor, a volatile surfactant that evaporates before the interface can saturate, allowing for persistent propulsion \cite{kohira2001synchronized}.

Recently, this technology was extended to create alcohol-powered `cocktail boats' that move around in your glass \cite{burton2013biomimicry, burton2014cocktail}. We depict them in Fig.~\ref{fig:CocktailBoat}. A typical commercial spirit can provide a surface tension difference up to $\Delta \gamma \sim 50$mN/m compared to pure water, but sugar and other cocktail ingredients tend to reduce this value somewhat.
By collaborating with chefs, various materials were tested to make the boats edible.
Gelatin boats were found to be capable of sustained motion and suitable for a wide range of flavourings, but they are susceptible to dissolving and sticking to the glass walls.
Wax boats performed the best, with speeds up to 11cm/s and travel times up to 2 minutes, but unfortunately they are not well digestible \cite{schmidt1997animal}.
It would be interesting if future research could improve or discover new edible materials.

Marangoni propulsion does not only lend itself to appetizing divertissements.
The same mechanism can be used to create microscopic swimming droplets \cite{maass2016swimming}, which can be used as cargo carriers that move deep inside complex flow networks \cite{jin2021collective}.
Recently, \citet{dietrich2020microscale} developed very fast Marangoni surfers that can swim over ten thousand body lengths per second, and \citet{timm2021remotely} developed Marangoni surfers that can be remotely controlled.
More generally, similar \textit{phoretic}\index{phoresis} effects \cite{anderson1989colloid}, where interfacial flows are driven by gradients in concentration, electric fields, temperature etc., can be exploited to make a broad range of self-propelled colloids that are of extraordinary interest to understand collective dynamics and emergent phenomena out of equilibrium \cite{howse2007selfmotile, koch2011collective, marchetti2013hydrodynamics, bechinger2016active, cates2015motility, elgeti2015physics, zottl2016emergent}.
The same mechanisms are also at play in active emulsions \cite{weber2019physics}, the transport of molecules in biological systems \cite{anderson1986transport, needleman2017active}, and the fragmentation of binary mixtures into  many tiny droplets, a process called Marangoni bursting \cite{keiser2017bursting}.

\subsection{Bubbly drinks} 
\label{subsec:BubblyDrinks}

\begin{figure}[t]
    \includegraphics[width=1\linewidth]{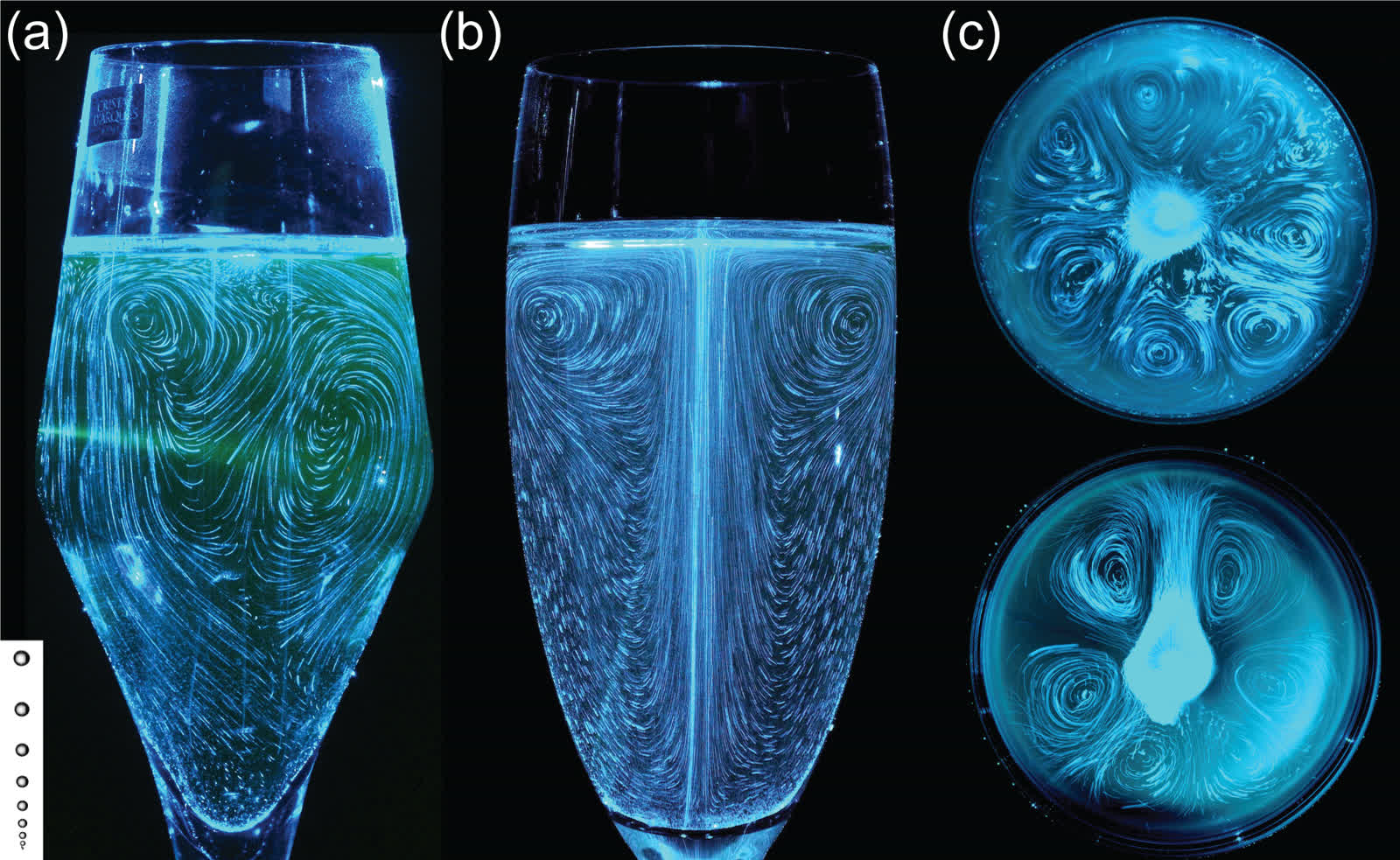}
    \caption{Laser tomography of champagne glasses. 
    \textbf{(a)} Natural, random effervescence in an untreated glass. Inset: Growth of bubbles as they rise.
    \textbf{(b)} Stabilized eddies in a surface-treated glass.
    (a,b) Courtesy of G\'erard Liger-Belair.
    \textbf{(c)} Counter-rotating convection cells self-organise at the air-champagne interface.
    From \citet{beaumont2016unveiling}. 
    }
    \label{fig:ChampagneCells}
\end{figure}

Go ahead and pour yourself some nice sparkling wine into a glass, and observe the beautiful sight of rising bubbles and their effervescence [Fig.~\ref{fig:ChampagneCells}(a)]. Champagne and sparkling wines are supersaturated with dissolved CO\textsubscript{2} gas, which, along with ethanol, is a product of the wine fermentation process \cite{liger2008}. When the bottle is uncorked, there is a continuous release of this dissolved  CO\textsubscript{2} gas in the form of bubbles. Hence, this physicochemical system provides a great opportunity to study several fundamental fluid mechanics phenomena involving bubbles: their nucleation, rise, and bursting dynamics, which in turn affect the taste of carbonated drinks \cite{planinsic2004fizziology, liger2008, polidori2009bubbles, chandrashekar2009taste, zenit2018fluid, mathai2020bubbly, rageZenit2020bubblesNature}. 

Before savoring the wine or bubbles, you need of course to first open the bottle. We are all familiar with the curious ``pop'' sound, and of course the dangers of uncontrolled corks flying out! This uncorking process is also accompanied by the formation of a small fog cloud just above the bottle opening. It has been shown recently that uncorking champagne creates supersonic $\text{CO}_2$ freezing jets \cite{liger2019under}. What is the underlying physical principle behind these interesting phenomena? It turns out that there is a sudden gas expansion when the bottle is uncorked (pressure drop from about 5 atm to 1 atm). This leads to a sudden drop in the temperature (about \SI{90}{\celsius}), resulting in condensation of water vapour in the form of a fog cloud. 

The uncorking also leads to a drop in the CO\textsubscript{2} partial pressure above the champagne surface. Hence, the dissolved CO\textsubscript{2} in the champagne is no longer in equilibrium with its partial pressure in the vapour phase. In fact, just after the uncorking, it turns out that the champagne is supersaturated with CO\textsubscript{2}. As described by \citet{lubetkin1988nucleation}, this is quantified by the supersaturation ratio,
\begin{equation}
   S = (c_{L}/c_{0}) - 1,
\end{equation}
where $c_{L}$ is the CO\textsubscript{2} concentration in bulk liquid and $c_{0}$ is the equilibrium CO\textsubscript{2} concentration corresponding to partial pressure of CO\textsubscript{2} of 1 atm. Just after uncorking, $c_{L}/c_{0} \approx 5$, so $S\approx4$, and the champagne must degas in order to achieve stable thermodynamic equilibrium. The gas loss occurs through two mechanisms, by diffusion through the liquid surface (invisible to us), and by the vigorous bubbling (effervescence) that we can readily observe and also hear \cite{poujol2021sound} [see \S\ref{subsec:SoundGeneration}].  

How do these bubbles form in the first place? The bubbles do not just pop out of nothing, the CO\textsubscript{2}-dissolved gas molecules need to cluster and push their way to overcome attraction forces that hold together the liquid molecules. Hence, the bubble formation process is controlled by a nucleation energy barrier\index{nucleation energy barrier} \cite{ford2004statistical}. As described by \citet{jones1999cycle}, the critical radius of curvature $r^{*}$ that is necessary for gas pockets to overcome this barrier is 
    \begin{equation}
       r^{*} \approx 2\gamma/(p_\text{atm}S),
    \end{equation}
where the surface tension of champagne is $\gamma \approx \SI{50}{\milli \newton \per \metre}$, the atmospheric pressure is $p_\text{atm} \approx \SI{e5}{\pascal}$, and $S\approx$ 4 at the time of uncorking. Using these values, we find that this critical radius for bubble formation is $r^{*}\approx\SI{0.25}{\micro\metre}$.
Such large bubbles are not likely to form spontaneously.
Indeed, instead of this homogeneous nucleation in the bulk of the liquid, bubble formation is more likely to occur at prenucleation sites on glass surfaces, typically at the bottom.

As such, the pleasing effervescence (bubbling) that we observe in champagne can arise from either natural or artificial sources \cite{liger2008, polidori2008artificial}, as shown in Fig.~\ref{fig:ChampagneCells}(a,b) respectively. 
On the one hand, natural effervescence refers to bubbling from a glass which has not received any specific surface treatment. When champagne is poured into a glass, a majority of bubble nucleation sites are found on small ($\SI{100}{\micro\metre}$ long) hollow, cylindrical cellulose-fibre structures which contain trapped gas cavities (lumen). Such hollow tubes are typically adsorbed on the wall of a glass
poured with champagne. Another source of natural effervescence are gas pockets trapped in tartrate crystals precipitated on the glass wall. Hence, natural effervescence can vary significantly depending on how the glasses are cleaned, dried, and stored. 
On the other hand, artificial effervescence refers to bubbling from a glass surface where precise imperfections have been engraved by the glass manufacturer. The typical imperfections introduced on the glass are micro-scale scratches to produce a specific pattern, which give rise to bubbling phenomena that are markedly different from natural effervescence \cite{liger2008}. 

The bubble release mechanism from a fibre's lumen has been well studied \cite{liger2008}. After a champagne bottle is uncorked, the supersaturation of carbon dioxide implies that such molecules will escape to the vapour phase using every available gas/liquid interface. The trapped tiny air pockets on fibre lumens offer gas/liquid  interfaces to the dissolved carbon dioxide molecules enabling them to cross the interface to gas pockets. The CO\textsubscript{2} gas pockets grow in size, and when it reaches the fibre tip, it is ejected as a bubble. However, a portion of the gas packet is left trapped behind in the lumen, and the bubble ejection cycle continues until the dissolved CO\textsubscript{2} supply is depleted.
Interestingly, the detachment of bubbles from the nucleation site is analogous to a dripping faucet [\S\ref{subsec:HangingDrop}].
The size of the bubble at the moment of detachment is approximately the radius of the mouth of the cellulose fibre, $R_0 \approx \SI{10}{\micro \metre}$.

After the bubbles detach, they rise towards the liquid surface due to their buoyancy, and they also grow in size since they absorb more dissolved CO\textsubscript{2} molecules. The repetitive production of bubbles from the nucleation sites has been captured in a model by \citet{liger2002kinetics}, and it has been found that the bubble radius $R$ increases linearly with time $t$ as: 
\begin{equation}
   R(t) = R_0 + kt ,
\end{equation}
where $R_0$ is the initial bubble radius, $k = dR/dt$ is the growth rate. Bubble rise experiments conducted with champagne and sparkling wines revealed $k$ values around $\SI{400}{\micro\metre \per \second}$ and experiments in beer revealed growth rates of around $\SI{150}{\micro\metre \per \second}$, indicating that the physicochemical properties of the liquids influenced the bubble growth rate \cite{liger2002kinetics}. 

According to wine tasters, the smaller the better, which in this context means that small bubbles make a better wine. Hence, plenty of attention has been focused on modeling the average size of the rising bubbles [Fig.~\ref{fig:ChampagneCells}(a), inset], which is a resultant of their growth rate and velocity of ascent. As discussed in detail by \citet{liger2006modeling}, the average bubble radius is
\begin{equation}
   R \approx (2.7 \times 10^{-3}) T^{5/9} \left( \frac {1}{\rho g} \right)^{2/9} \left( \frac {c_{L} - k_{H} p_\text{atm}} {p_\text{atm}} \right)^{1/3} h^{1/3},
\end{equation}
where $T$ is the liquid temperature, $\rho$ is the liquid density, $g$ is gravity, $k_{H}$ is Henry's law constant and $h$ is the distance travelled by the bubble from the nucleation site. It is interesting to note that so many factors can influence the average bubble size, which is typically in the sub-millimeter length-scale in bubbly drinks. The bubble size in beer is significantly smaller than in champagne, and the reason for this is that the amount of dissolved CO\textsubscript{2} in champagne is about two times higher. 

Bubble growth is beautifully visualised by the oscillating dynamics of a raisin in a glass of Champagne, also known as the ``fizz-ball'' effect \cite{moinester2012fizz, cordry1998finicky}. Initially the raisin sinks, but the textured surface of the raisin provides nucleation sites for bubbles to grow on the surface. As they grow larger, these attached bubbles pull up the raisin by buoyancy. However, when the raisin reaches the air interface, some of the bubbles pop, so the raisin sinks again. This cycle of rising and sinking continues for a long time, until the CO\textsubscript{2} is depleted. The fizz-ball effect can be observed in many carbonated drinks (beer, soda) and for many different objects (berries, seeds, chocolate chips).


In addition to the visual beauty and fascination, the bubbles actually play an important role in the drink -- the bubbles have been shown to generate large-scale time-varying convection currents and eddies inside the glass \cite{liger2008, beaumont2015flow, beaumont2016unveiling, dijkstra1992structure, beaumont2020computational}, often with surprising self-organised flow patterns [Fig.~\ref{fig:ChampagneCells}(b,c)]. Since they cause a continuous mixing of the liquid, bubbles are thought to play a key role in the flavor and aromatic gas release from the wine-air interface. These release rates are dependent on the fluid velocity field close to the surface, which is in turn significantly influenced by the ascending bubbles. As the bubbles collapse at the air interface, they radiate a multitude of tiny droplets into aerosols \cite{liger2009unraveling}, which evaporate and release a distinct olfactory fingerprint \cite{ghabache2016evaporation}.

In the future, we can look forward to several innovations in bubbly drinks, where numerous factors must be taken into consideration -- different types of glass shapes, natural versus artificial effervescence, engraving conditions, kinetics of flavor and CO\textsubscript{2} release under various conditions, and sensory analysis.


\subsection{Foams}
\label{subsec:Foam}

\begin{figure}[t]
\centering
\includegraphics[width=1\linewidth]{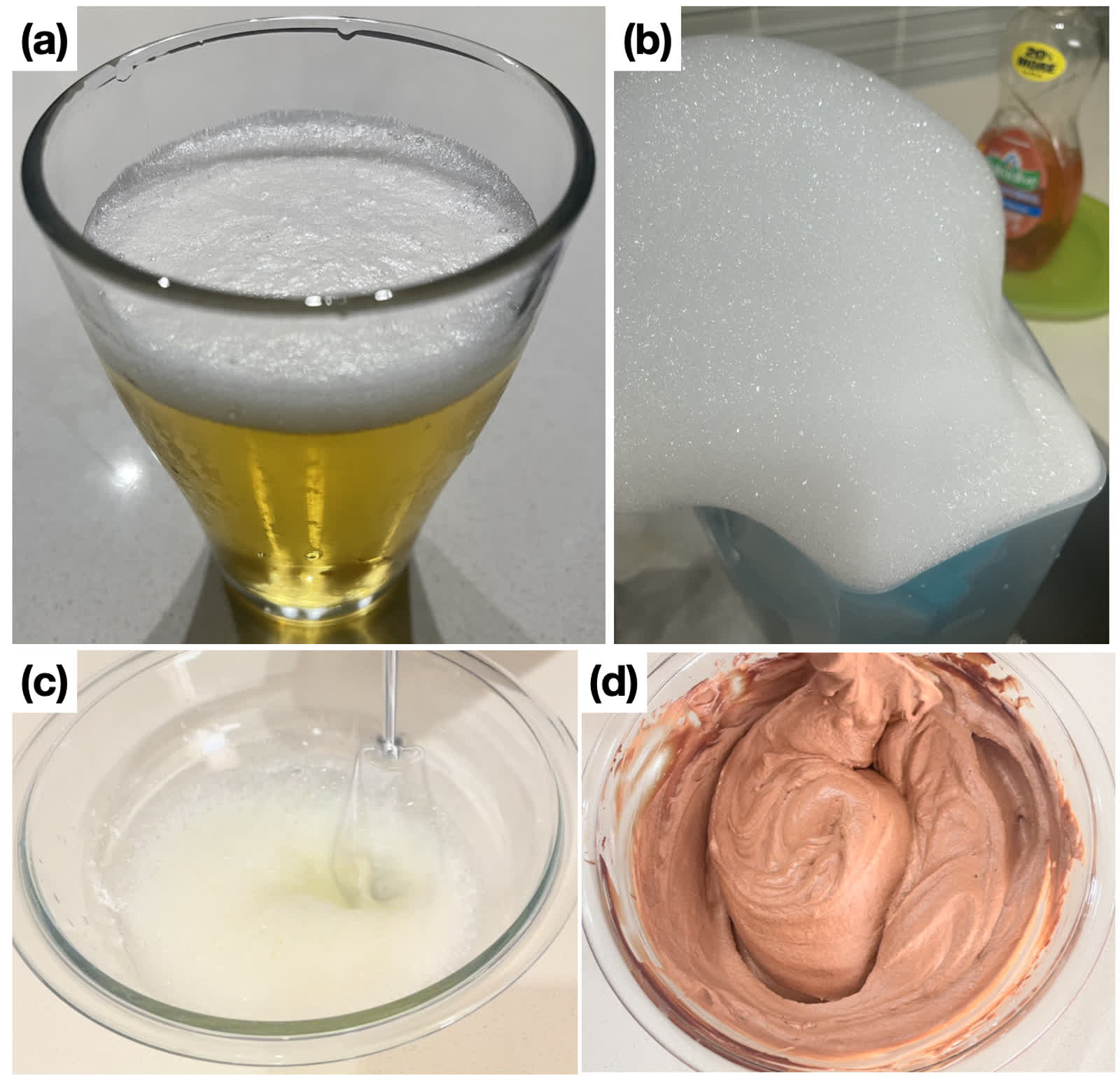}
\caption{Examples of foams in the kitchen: \textbf{(a)} Beer, \textbf{(b)}Dish washing, \textbf{(c)} Egg beating, and  \textbf{(d)} Chocolate mousse.}
\label{fig:foam}
\end{figure}

Bubbles [\S\ref{subsec:BubblyDrinks}] can burst when they reach the surface, but there is a finite lifetime associated with this process \cite{liger2008}. Thus, when the bubble production rate is very fast, greatly exceeding the surface bursting time-scales -- then the bubbles start accumulating on the surface to create layers of bubbles called ``foams'' \cite{cantat2013foams, weaire2001physics, kraynik1988foam, Weaire_2002FluidMechanicsFoams}. In many beers, these foams last long since they tend to be stabilized by proteins. They add to the visual appeal, and provide a creamy texture enhancing the mouthfeel [Fig.~\ref{fig:foam}]. However, in champagne, the foam is more fragile and less stable due to the lack of proteins. Foams are formed in many other fizzy drinks and also in specially prepared coffees such as cappuccino, where the foam layer lasts for a long time since it is stabilized by milk proteins. These observations naturally bring up questions concerning the mechanisms behind the formation, stability, age and drainage of foams -- we will discuss these aspects below.

A foam is essentially a dispersion of gas in liquid, and gas bubbles tightly occupy most of the volume. The liquid phase in the form of films and junctions is continuous unlike the gas phase. Foams are also characterized by the presence of surface-active molecules called surfactants, which stabilise the bubbles at the interfaces of gas and liquid \cite{manikantan2020surfactant} (the same type of molecule can also stabilise an oil/water interface in an emulsion [see \S\ref{subsec:Emulsions}] or on individual droplets [see \S\ref{subsec:immiscibledropsdynamics}]). Foams consist of significant quantities of gas, hence being less dense than the liquid it contains, and this is why a foam floats on the surface of a  liquid. Another interesting property of foam is the large surface area per unit volume, since the foam contains a large number of interfaces. Hence, foams enhance the possibilities for molecular transfer and find applications in foods for flavor enhancement (e.g. chocolate or spices) and also reduce the need for high sugar or salt content  \cite{cantat2013foams}. Foams also have special mechanical properties -- they exhibit both solid-like and liquid-like behavior \cite{janiaud2005foam}. If the deformation is not too high, foams can show weak visco-elastic solid properties and can return to its original shape. However, if the foam is subjected to a high deformation, it can behave like a visco-plastic solid that can be sculpted. Foams can flow like liquids and seep through pores and cavities, so they can be poured into containers and tubes of various shapes. The foam's viscous resistance increases less quickly with flow rate compared to a normal fluid enabling it to reduce frictional losses, hence foams behave as `yield stress fluids' [see \S\ref{subsec:FoodRheology}] with intermittent flow via avalanche-like topological bubble rearrangements \cite{durian1995foam}. Hence, given their unique properties, in addition to the food industry, foams find applications in many other areas of science and technology -- e.g. cosmetics \cite{durian1991scaling}, cleaning, reducing pollution, surface treatment, fire-fighting, army, and building materials \cite{cantat2013foams}. In food science, foams on beer, cream, and egg white have been imaged and characterised using magnetic resonance imaging (MRI)  to noninvasively estimate densities, drainage rates, and collapse throughout their structure \cite{german_stability_1989}.

We will now examine the physical properties that allow a foam to exist in equilibrium. There are four relevant length-scales to consider: (i) the meter scale, where the foam appears to be a soft and opaque solid, (ii) the millimeter scale, where individual bubbles can be distinguished in the foam, (iii) the micron scale, which reveals liquid distribution between bubbles, and (iv) the nanometer scale, where molecules (e.g. soap molecules) at the interfaces (air/water) are relevant. 
The physics of foams is hence a very broad subject covering so many length-scales \cite{cantat2013foams}, and here we will only touch upon a few aspects. 

At the scale of the gas/liquid interface, the surface tension [\S\ref{subsec:HangingDrop}] and the Young-Laplace law, Eq. \eqref{eq:YoungLaplace}, determine the shape of the interface. An interface is flat if geometric constraints allow it, while the surface of an interface that is completely surrounded by some fluid becomes spherical. The high pressure on the concave side  tries to curve the surface while surface tension tries to flatten it.

Foams are prepared using additives that chemically consist of a polar head and a tail with a long carbon chain. The head is hydrophilic and the tail is hydrophobic [see \S\ref{subsec:wettingCapillaryAction}]. The combination of these properties results in an {\textit{amphiphilic}} molecule (both water-loving and fat-loving). Such a molecule, when dissolved in water, tends to adsorb at the air/water interface. This forms a monolayer, which greatly affects the interfacial surface tension properties. Hence, these molecules are called surface-active molecules or {\textit{surfactants}}. In our everyday experience, there are several examples where we find many small bubbles that burst quickly (few seconds), e.g. in sparkling wine or champagne. Here, the small volume of gas in the bubble encloses a thin film which is unstable due to van der Waals forces, and hence breaks. However, the presence of surfactants such as soap molecules [Fig.~\ref{fig:foam}b] carry a small charge, giving rise to an electrostatic repulsion, which cancels out the van der Waals forces and stabilises the thin film -- helping the foam last for a longer period.

 A bubble is essentially a small volume of gas enclosed by a film of water. The bubble assumes the smallest possible surface area to contain the gas, which typically results in a spherical shape for an isolated bubble. The pressure inside most foam bubbles is only slightly greater than the atmospheric pressure and is not sufficient to compress the gas appreciably, so the volume of gas can be considered to be fixed. If the bubble has a large size, then the interface becomes deformed and is no longer spherical under external forces such as gravity \cite{prakash2012gravity}. When two bubbles come in contact, they share an interface and hence change shape to reduce the total interfacial area, and can no longer remain spherical. 
 

So far, we mainly discussed foams in the context of air bubbles in liquid. 
However, edible foams \cite{briceno2022role} extend to soft solids or even hard solids, such as meringues, bread, and chocolate mousse [Fig.~\ref{fig:foam}c,d]. They are a pleasure to eat because of their lightness and texture, which is determined by their complex mechanical properties \cite{robin2010extrusion, cantat2013foams, haedelt2007bubble, kraynik1999foam, janssen2020role}. Edible foams are often prepared by solidification through refrigeration, cooking or baking, such as bubbles in a pizza crust \cite{avallone2022rheology}. Since these foams are solidified before their collapse, they often do not need a stabilizing agent. Another interesting point is that air is an important raw material in these edible foams: air contributes greatly to increasing the volume of the  product, but it's practically free. On a final note, many other popular deserts are edible foams, these include favorites such as ice cream, marshmallows, many types of cakes, baked Alaska, etc. 

\begin{figure}[t]
     \centering
     \includegraphics[width=0.75\linewidth]{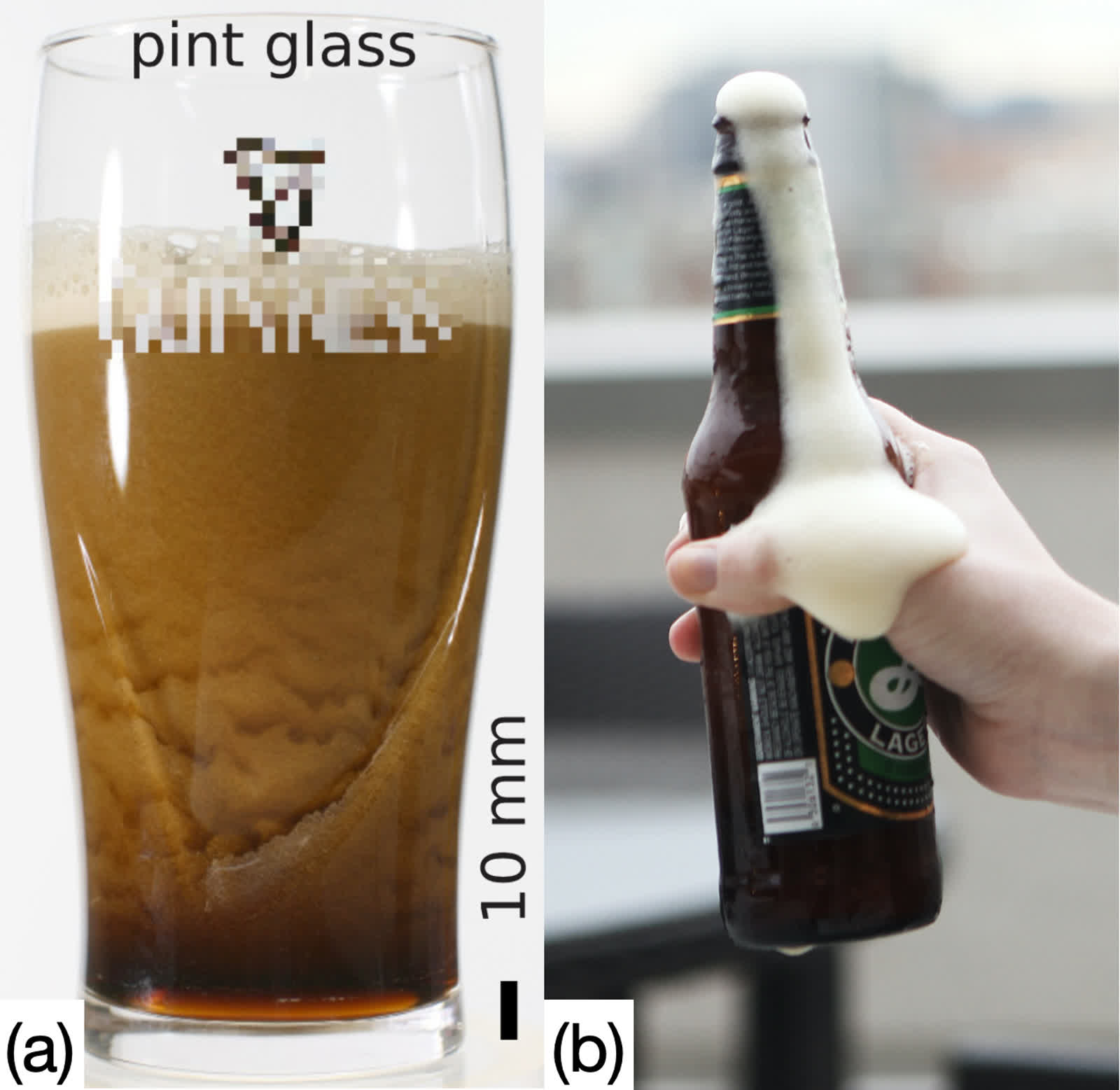}
     \caption{Examples of dynamic and unstable foaming phenomena: 
         \textbf{(a)} Unique foamy textures in stout with nitrogen bubbles. From \citet{watamura2021bubble}. 
         \textbf{(b)} Foam overflow `volcano' due to tapping on a beer bottle. Public domain image.
     }
     \label{fig:foamtapping}
 \end{figure}

We have mainly discussed foams which are stable for a finite duration of time. There are several interesting examples of dynamic and unstable foams, we show two popular examples in Fig.~\ref{fig:foamtapping}. We know in general that bubbles rise due to their buoyancy in a liquid, but in Guinness beer there is a collective downward movement of bubbles, creating a `cascade of bubble textures'. It has been demonstrated that this bubble texture cascade motion [Fig.~\ref{fig:foamtapping}a] arises due to a roll-wave instability of gravity currents \cite{watamura2019bubble}, a phenomenon that is analogous to the roll-wave instability in liquid films that cause water films to slide downhill on rainy days [\S\ref{subsec:RinsingFlows}]. Furthermore, it has been theoretically shown that these bubble cascades can occur in systems other than the Guinness beer~\cite{watamura2021bubble}. 

Another interesting phenomenon is `beer tapping' -- a beer bottle foams up resulting in an overflow when it is tapped from the top [Fig.~\ref{fig:foamtapping}b]. The fascinating fluid physics underlying this phenomenon was explained recently \cite{rodriguez2014beertapping}. It turns out that when the beer bottle is first hit at the top, a compression wave travels through the bottle. This wave gets rebounded through the liquid as an expansion wave. At the base of the bottle, the compression and expansion waves interact to cause `mother' bubbles to break up. This is a rapid cavitation process resulting in the formation of smaller `daughter' bubbles, which expand rapidly to create foam that starts to overflow \cite{rodriguez2014beertapping}.

\section{Soup Starter: Complex Fluids} \label{sec:SoupsSauces}

Most foods are neither purely liquid nor solid, but rather something in between: they are called viscoelastic or complex materials. 
The fractalist Benoit Mandelbrot (1924-2010) once said:
\begin{quote}
    \emph{A formula can be very simple, and create a universe of bottomless complexity.}
\end{quote}
Similarly, you can make excellent sauces following one simple recipe, where small variations in the ingredients can completely change the sauce flavour and consistency, because changes in the molecular interactions lead to very different food properties.
These complex properties strongly affect how we perceive taste, since they are directly linked with mouthfeel, the oral processing and texture of foods \cite{stokes2013oral, sahin2007physical}. The rheology of complex fluids is also of extreme importance in the food industry because they affect transport phenomena, production processes, storage, and processing techniques that need to be adapted to the properties of materials at hand \cite{borwankar1992rheology, brummer2006rheology, fischer2011rheology, ahmed2016advances}.
Complex fluids are a bit of a soup sandwich, so we begin this section with an introduction to food rheology.
Then we get into the thick of it, reviewing the science of food suspensions, emulsions, and the mixing of sauces. 
Shall we board the gravy train?

\subsection{Food rheology}
\label{subsec:FoodRheology}


A large amount of syrup can be collected by pulling a knife vertically out of the jar. 
First, the syrup is lifted up by tangential forces, `dragging along' neighbouring fluid layers. 
Then, as the syrup slowly drips down, the stream is stretched by tensile forces.
Thus, a \emph{viscous} fluid has two important properties: it resists both tangential and tensile stresses, because of internal friction between the molecules. 
However, the same experiment done with oobleck (a solution of cornstarch in water) will show you an even richer phenomenology:
Depending on the speed of lifting the knife, the solution flows easily, with difficulty, or even cracks like a solid! 
This is because the cornstarch solution has a complex internal microstructure which responds to external agitation with a certain delay. 
Viscosity\index{viscosity} alone is often not enough to determine the character of flow, but rather it is the interplay between the viscosity, elasticity, and inertia that gives rise to a wealth of different food properties. 

To understand the rheology of the fluid is to know the stresses that arise in the effect of actuation, and {\it vice versa}.
Consider a rheometer where a simple incompressible fluid is sheared between two parallel plates that translate in opposite directions. 
The fluid moves steadily along $x$, with a velocity $u_x(y)$ varying along $y$.
Then, the force (per unit area) required to shear the plates is determined by the shear stress, $\vec{\sigma}$. This tensor has only one component in our example, $\sigma_{xy}$, which is given by Newton's law of viscosity, 
    \begin{equation}
        \label{eq:NewtonViscosity}
        \sigma_{xy} = \mu \frac{\partial u_x}{\partial y},
    \end{equation}
where the constant of proportionality is the dynamic viscosity, $\mu$, and the second term is called the shear rate, $\dot{\gamma}$. 
So, a \textit{Newtonian fluid}\index{Newtonian fluid} is defined by the linear relation between stress and shear rate.
Most simple liquids are indeed Newtonian, including water, alcohol, and most thin oils. 
However, many kitchen fluids, such as milkshakes, emulsion dressings, and even chapati and bread dough, deviate from this linear relationship because of their complex internal structure. For example, \citet{louhichi2022flow} studied dough by characterizing the linear and nonlinear viscoelastic properties of an aqueous gluten solutions. Such \textit{non-Newtonian fluids}\index{non-Newtonian fluid} can feature many different types of behaviour.

\begin{figure}[t]
         \includegraphics[width=1\linewidth]{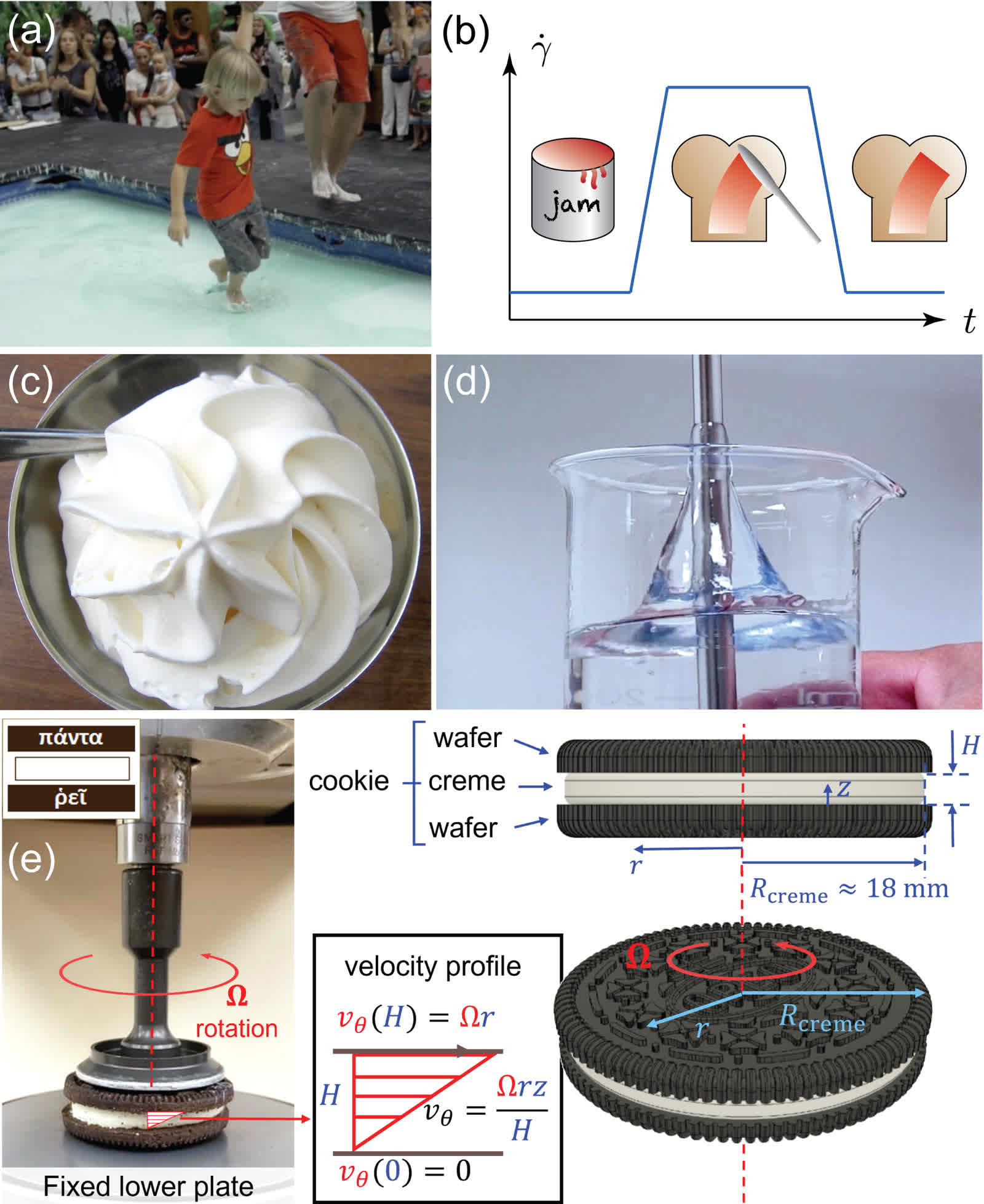}
         \caption{Examples of complex rheological behaviour of fluids: 
         \textbf{(a)} People walking over a swimming pool full of oobleck, a mixture of cornstarch and water. Image courtesy of Ion Furjanic, director of \citet{kix}.
         \textbf{(b)} Thixotropic fluids become thinner with time when they are sheared, and solidify again at rest. Classic examples are paint or sandwich spread.
         \textbf{(c)} Whipped cream is an example of a Bingham plastic, which can be squeezed out like a fluid, but then turns solid in the absence of stresses. From Wikimedia Commons, licensed under CC BY 2.0.
         \textbf{(d)} The Weissenberg rod climbing effect seen in a 2\% solution of high molecular weight polyacrylamide. From Wikimedia Commons, licensed under CC BY 4.0. 
         \textbf{(e)} Oreology. A sandwich cookie is mounted on a rheometer, where one wafer is rotated relative to the other. Hence, the properties of the creme between the wafers are measured. Image courtesy of Crystal E. Owens.
         }
         \label{fig:nonnewtonian}
     \end{figure}

In {\it shear-thickening}\index{shear-thickening} liquids, the viscosity increases with the shear rate. When agitated, they seem to harden and increase their resistance to motion. This is commonly seen in oobleck, mentioned above, and other starch solutions \cite{dintzis1996shear}. In some cases, they can effectively behave as solids, as seen in the famous demonstrations of people walking on a swimming pool filled with an aqueous mixture of cornstarch [Fig.~\ref{fig:nonnewtonian}a]. However, if a person stops mid-way through the pool, they would start to sink, which demonstrates stress relaxation as the shear rate is reduced. But one needs to be careful, because the material will solidify again when people try to resist the sinking.
Fluids that gradually become more viscous with the duration of stress are called {\it rheopectic} fluids.
A good example is the beating of egg whites, which slowly stiffen, because the proteins unravel and form large networks.
Rheopecty can also result from heat -- think of pancake batter in a frying pan [see \S\ref{subsec:CrepeMaking}]. 
 
The contrary behaviour is seen in {\it shear-thinning}\index{shear-thinning} fluids, whose viscosity decreases with increasing shear rate. Such fluids appear less viscous when set in motion. Think of yoghurt, mustard, clotted cream, or the well-known example of ketchup, reluctant to leave the bottle at first but spilling generously when shaken. Another example is paint, which should be easy to spread on a wall and stay there when the brush is removed. A particular sub-class of shear-thinning substances are {\it thixotropic} fluids \index{thixotropy}\cite{larson2019review}, that become less viscous over time when agitated, so their response is also time-dependent. Thixotropy is often a consequence of their fibrous or polymeric internal composition. Ketchup and yoghurt belong to this category, together with margarine and even honey at large strain \cite{munro1943viscosity}. This quality is desirable in spreads and jams, which should be easy to spread on toast, but stay solid once applied, as shown in Fig.~\ref{fig:nonnewtonian}b. The shear rate is almost a step function in this case, and the viscosity of jam would show an inverted relationship, with decreased viscosity when spreading at increased stress.

Some materials, called {\it Bingham plastics}\index{Bingham plastic} or {\it yield-stress materials}, behave as solids at low stresses but start flowing above a critical stresses.
A prime example is our daily experience with toothpaste, which only flows when enough force is applied. 
Yield stress is also seen in numerous food products \cite{griebler2022nonlinear, wilson2022spray, wilson2022rise}.
Think about mayonnaise, where ridges and peaks on the surface show the existence of a critical stress above which it flows \cite{figoni1983time, goshawk1998rheological, balmforth2014yielding}. 
Similarly, Fig.~\ref{fig:nonnewtonian}c shows that cream, once whipped, can be squeezed out from a cone, but will stay rigid on a piece of cake in the absence of forcing.
Moreover, many emulsions [\S\ref{subsec:Emulsions}] and foams [\S\ref{subsec:Foam}] behave as Bingham plastics, because a certain minimal amount of force is required for the bubbles to rearrange within the material before it can flow.
 
In most non-Newtonian fluids subject to small or slowly varying deformations, it suffices to assume that stress and strain are linearly related, and a plethora of phenomenological models have been proposed to quantify this relationship. Many non-Newtonian fluids are called viscoelastic because they exhibit both viscous (creep) and elastic (relaxation) effects. In fact, they are something in between fluids and elastic solids, and can display both behaviors depending on the circumstances. To investigate them, two typical tests can be run. The creep\index{creep test} test consists of measuring a time-dependent strain upon the application of a steady stress. The material starts to flow with a certain delay, which is measured by this experiment. In the linear approximation, doubling the stress doubles the strain. This is widely observed, for example, in processed fruit tissues \cite{alzamora2008exploring}, or in dynamic rheology measurements of honey \cite{yoo2004effect}. A complementary experiment, the stress relaxation\index{stress relaxation test} test, measures the time-dependent stress resulting from a steady strain. In the range of small deformations, many food products can be aptly described by linear models. Rheology measurements of frankfurters of various compositions show a good linear response for strains of up to about 3.8\% \cite{skinner1986linear}. Stress relaxation and creep recovery tests on oat grains also show linear behaviour for a range of temperatures and moisture content \cite{zhao2020viscoelastic}. Linear models have been successfully used to describe stress relaxation behaviour for a variety of semi-solid food products, such as agar gel, meat, mozzarella cheese, ripened cheese, and white pan bread \cite{delnobile2007use}. Even systems with finer microstructure, such as protein-stabilised oil-in-water emulsions, can also exhibit linear viscoelastic behaviour \cite{ruiz2013linear}.

To assess whether the deformations are large or rapid enough for a fluid to exhibit linear response, we can compare the typical observation time (or the process under consideration) $\tau_0$ to the typical stress relaxation time measured in the experiments above to yield the dimensionless Deborah number\index{Deborah number},
\begin{equation}
    \text{De} 
    \equiv \frac{\text{time scale of relaxation} }{\text{time scale of process}}
    = \frac{\tau}{\tau_0},
\end{equation}
which indicates whether the material should behave like a fluid (at low $\text{De}$) or exhibit non-Newtonian properties, with an increasingly manifested elasticity at high $\text{De}$. For example, viscoelastic ice cream \cite{bolliger2000relationships} should preferably be consumed at high $\text{De}$ for practical reasons [see \S\ref{subsec:IceCream}].
 
However, some biological fluids are non-Newtonian at any $\text{De}$.\index{non-Newtonian fluid} Rheology measurements of yoghurts show that overlapping polymer molecules cause viscoelastic behaviour even at dilute concentrations, resulting in a sharp increase in viscosity with concentration \cite{benezech1994characterisation}. Yoghurts are shear thinning in addition to being viscoelastic. Shear thinning or thickening cannot be explained using linear constitutive equations, thus more complex models are needed to quantify their behaviour. Non-Newtonian effects manifest themselves particularly in the material properties that become dynamic quantities and in particular depend on the shear rate $\dot{\gamma}$\index{shear rate}. 

The Weissenberg number\index{Weissenberg number} is another dimensionless group that quantifies the ratio of elastic to viscous forces. For a fluid with a characteristic stress relaxation time $\lambda$ under shear, we write this ratio as
\begin{equation}
    \text{Wi} 
    \equiv \frac{\text{elastic forces} }{\text{viscous forces}} = \lambda \dot{\gamma}.
\end{equation}
Although seemingly similar to $\text{De}$, the Weissenberg number has a different interpretation because it captures the degree of anisotropy introduced by the deformation, rather than the effect of time-dependent forcing \cite{poole2012deborah}. The two numbers span a phase space interpolating between purely viscous and purely elastic deformations, with linear (typically at moderate $\text{De}$ and low $\text{Wi}$) and nonlinear viscoelasticity (at higher $\text{Wi}$) in between.

A first step into the non-linear territory is the {\it generalised Newtonian fluid} model\index{generalised Newtonian fluid}, in which the stress depends only on the instantaneous flow, but the viscosity in Eq.~\eqref{eq:NewtonViscosity} is replaced by a shear-dependent function $\mu(\dot{\gamma})$. Its form is usually derived empirically from the available data. Some common approximations include a {\it power law fluid}\index{power law fluid} with $\mu(\dot{\gamma}) = k (\dot{\gamma})^{n-1}$, where $k$ and $n$ are fitting parameters. If $n>1$, the fluid is shear-thickening (dilatant)\index{shear-thickening fluid}\index{dilatant fluid}, and if $n<1$, it is shear-thinning\index{shear-thinning fluid}. Most fruit and vegetable purees belong to the latter category and can be efficiently described by this model \cite{krokida2001rheological}.  Another set of examples are Carreau-Yasuda-Cross models\index{Carreau-Yasuda-Cross model} \cite{carreau1972rheological,yasuda1981shear,cross1965rheology}, which interpolate between the different zero and infinite shear rate viscosities ($\mu_0$ and $\mu_\infty$, respectively) by $\mu(\dot{\gamma}) = \mu_\infty + (\mu_0-\mu_\infty)[1+(\lambda \dot{\gamma})^{a}]^{(n-1)/a}$, with additional fitting parameters $\lambda,a,n$. Such models successfully describe, for example, the flow of skim milk concentrate \cite{karlsson2005relationship} or semi-solid Spanish dairy desserts called {\it natillas} \cite{tarrega2005rheological}.  An important category are yield fluids, which flow only above some critical stress $\sigma>\sigma_c$. Within these, the Bingham models\index{Bingham model} satisfy $\mu(\dot{\gamma}) = \mu_0 + \sigma_c/\dot{\gamma}$ \cite{bingham1922fluidity}. This type of behaviour is seen commonly, e.g., in tahini, curry and tomato pastes \cite{rao1992rheological}. The Herschel Bulkley\index{Herschel Bulkley model} models use $\mu(\dot{\gamma}) = k \dot{\sigma}^{n-1} + \sigma_c/\dot{\gamma}$ \cite{herschelbulkley1926}, and have proven useful in aptly describing the rheology of stirred yoghurts \cite{ramaswamy1991rheology}. 
 
The truly non-linear shear-dependent properties of kitchen matter can be seen in the context of mixing and whisking. The \textit{``Weissenberg effect''} \cite{freeman1948some}\index{Weissenberg effect} is an illustrative proxy of viscoelasticity. As depicted in Fig.~\ref{fig:nonnewtonian}d, it is seen when a spinning rod is inserted into an elastic fluid: Instead of the meniscus curving inwards, the solution is attracted towards the rod and rises up its surface. This is due to normal stresses in the fluid acting as hoop stresses and pushing the fluid towards the rod \cite{muller1961weissenberg}. When egg whites are whisked with a mixer \cite{walker1978amateur}, we see the solution rise up close to the mixer shaft, rather than move outwards in a parabolic shape characteristic for Newtonian liquids, as described in \S\ref{subsec:RotatingLiquid}. \citet{reiner1949weissenberg} observed this effect in sweetened condensed milk after thickening, when it becomes a highly thixotropic gel; they linked it with the uncoiling of globular protein molecules during heat denaturation of the albumin-globulin fraction. \citet{muller1961weissenberg} suggested that the extent to which the cake batter exhibits the Weissenberg effect depends on its egg content, and isolated the so-called thick white as the egg component that exhibits it most markedly. The rheology of an inside of an egg was also investigated by \citet{bertho2022egg}, who determined its shear-thinning properties and, motivated by the famous problem of distinguishing between raw and hard-boiled eggs, observed its residual rotation on a table to deduce its viscosity.

Importantly, the viscosity may also change as a result of a change in the chemical composition under external stimuli, such as heat. This complex landscape is particularly important in the kitchen environment, where we often work with thickening agents such as roux or Xanthan gum (also found in bubble gum), gently heat up egg yolks to make Hollandaise sauce, or milk for the b\'echamel. For demonstration purposes in the home laboratory, \citet{wiegand1963demonstrating} showed the transition from Newtonian to Weissenberg-like behaviour of a gelatine solution when cooling down from 32$^\circ\text{C}$ to 26$^\circ\text{C}$. 

Composite food products typically respond in a non-Newtonian manner to deformation. Dynamic quantities that characterise the rheology of food products are of paramount importance for food processing, in which appropriate length and time scales should be chosen for the expected result. For example, pasta products such as curly spaghetti can be thought of as a hydrated gel \cite{dangelo2022shaping}. A chef can gauge pasta by its texture, but a rheological study of pasta cooking by \citet{hwang2022swelling} showed that an objective measure for spaghetti can be derived from its time-dependent length, and the measurement can be adjusted for other types of pasta. In a rheological study of lutefisk (Norwegian dry cod, soaked in lye to re-hydrate), \citet{feneuil2022elastic} measured the elastic modulus of the material to identify the key element of the preparation process that determines the mouthfeel. Working on plant-based meat surrogates,  \citet{ghebremedhin2022meat} studied the structure and rheological properties of meat-, vegetarian-, and vegan sausages.
Another subjective textural property of food products is spreadability, which can be effectively measured using the vane method and can be related to the torque on a knife during application \cite{daubert1998quantitative}. 

Numerous articles quantify the rheological response of food products; \citet{nelson2018extending} measured the extensional yield-stress of Nutella and showed that it can be modelled as a classical yield-stress fluid. However, other food products can have a more complex rheological signature; \citet{martinetti2014critical} characterised a range of bubble and chewing gums under shear, finding that while in small-amplitude oscillatory and steady shear they behave like power-law critical gels in the linear regime, in start-up flows or large-amplitude oscillatory shear nonlinear viscoelastic effects were observed. They found in particular that extensional thickening, more pronounced in bubble than chewing gums, stabilises film blowing, which is important for the ability to blow large bubbles, desired by consumers. Steady-shear measurements of the properties of various commercially available salad dressings by \citet{elliott1977salad} showed that they can be described as modified Bingham bodies. On the other hand, the choice of a particular constitutive model depends on the physical situation – for example, the transient flow of mayonnaise in a coaxial viscometer is similar to that of polymer melts, and a similar mathematical description can be used \cite{campanella1987analysis}.  These are only a few examples of a wide current that aims to describe transient effects in the flow of food products \cite{kokini1981attempt}.

Since the beginnings of rheologic characterisation of materials, it has been tempting to answer the inverse question: Can we produce a material with the desired properties? Such questions have particular implications for cooking and food science. What properties do we need in terms of texture, ease of processing, mouthfeel? What effect do we want the consumer to see or taste when they interact with it? And finally, once these qualities are identified, how can we get there from the basic microstructure or processing?  The design of complex fluids has evolved to be a prominent research field \cite{ewoldt2022designing}, with many open questions that can potentially be answered when physicists and cooks team up and interact.

We also note here that the kitchen might play an important educational role in the teaching of rheology. In a recent paper, \citet{hossain2022yourself} proposed a do-it-yourself home rheology course, where students were encouraged to select complex fluids and produce the desired flow types to infer their rheological properties, such as yield stress, extensional viscosity, or shear viscosity. Inspired by the similarity between a sandwich cookie, composed of two wafers separated by a cream filling, and a parallel plate rheometer, \citet{owens2022oreology} investigated `oreology' and the deformation of cream by torsional rotation and the subsequent separation of the cookie into two pieces, as shown in Fig.~\ref{fig:nonnewtonian}e. They then presented a home-made rheometer (or `oreometer') that can be used for such measurements, thus providing an inspiring example with an accessible home experiment. 

Rheology is a vast field, discussed widely in classic textbooks, also in the context of the applicability and structure of different models \cite{macosko1994rheology, mccrum1988}. 
The rheological properties of food are also important for microbial motility in complex fluids \cite{spagnolie2022swimming} and medical conditions including dysphagia \cite{nita2013matching}, where fluid dynamics can help predict the ease of swallowing \cite{marconati2019review}.
For a more detailed description of food rheology, we recommend reading the following books and reviews  \cite{borwankar1992rheology, brummer2006rheology, fischer2011rheology, ahmed2016advances, bird1987dynamics}, as well as these works on soft matter \cite{piazza2011soft, ubbink2008food, vilgis2015soft, assenza2019soft, pedersen2019soft, mcleish2020soft}, and references therein.

\subsection{Mixing up a sauce}
\label{subsec:Blending}

\begin{figure}[t]
    \includegraphics[width=1\linewidth]{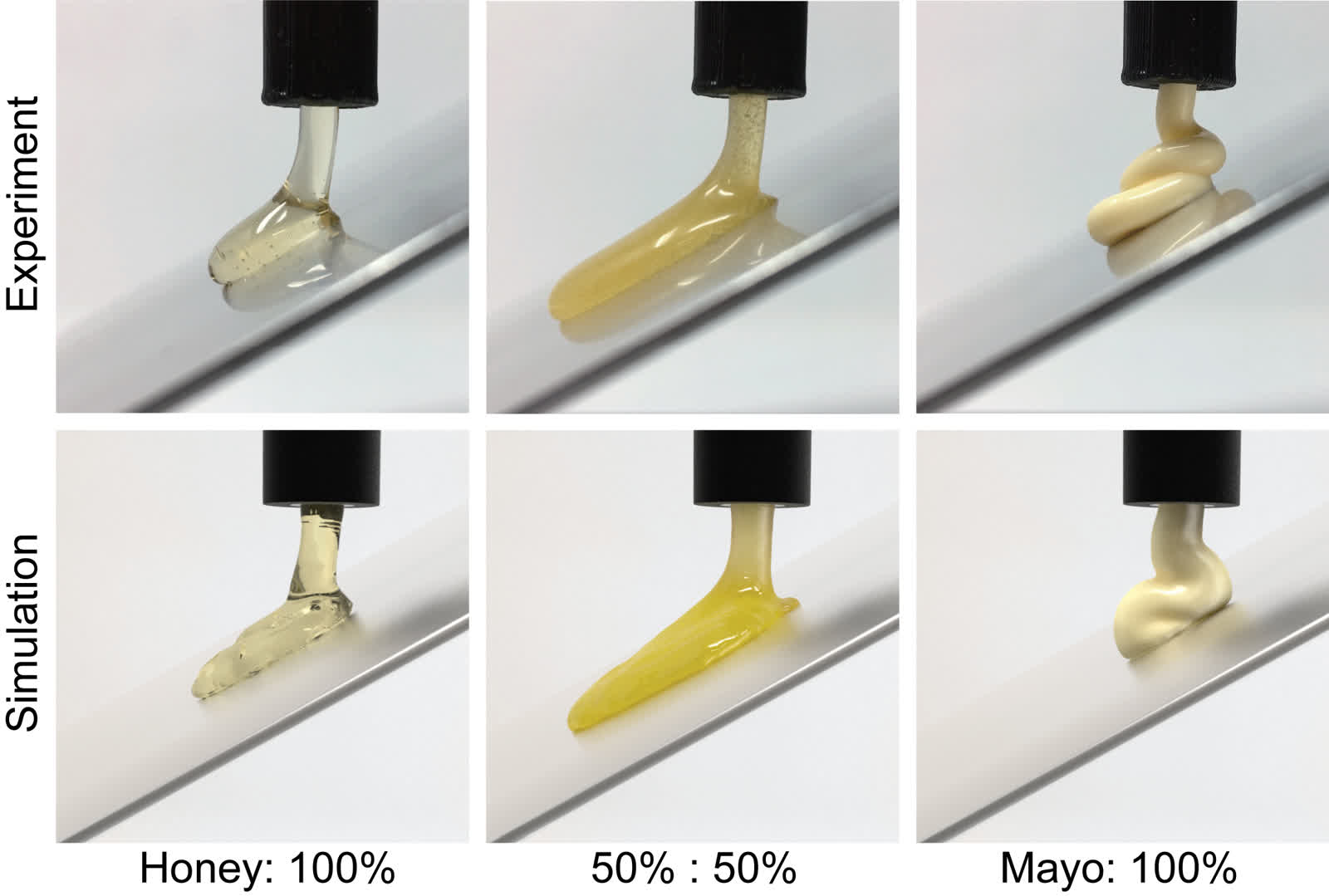}
    \caption{Blending sauces.
    Left: pure honey. Middle: 50:50 mixture. Right: pure mayonnaise.
    Top row: Rheometry experiments of the sauces mixtures flowing down an inclined plane. While honey and mayo move slowly, the mixture runs down fast.
    Bottom row: The same dynamics simulated with a shear-thinning mixing model and displayed using computer-generated imagery (CGI).
    Image courtesy of Yonghao Yue.
    }
    \label{fig:blendingSauces}
\end{figure}

When we make a sauce, rather counter-intuitive effects can emerge: the combination of two thin liquids can suddenly lead to a thick mixture, or vice versa.
Indeed, as discussed in \S\ref{subsec:WhiskyTasting}, we saw that an ethanol-water blend has a higher viscosity than both pure liquids.
In general, the viscosity $\mu_{12}$ of most binary mixtures is not a linear function of their relative composition \cite{bingham1914viscosity}.
Instead, a first approximation is given by the Arrhenius equation,
    \begin{equation}
        \label{eq:ArrheniusEquation}
        \ln \mu_{12}
        =x_{1} \ln \mu_{1}+x_{2} \ln \mu_{2}
    \end{equation}
where $x_i$ and $\mu_i$ are the mole fraction and viscosity of the $i$th component, respectively. 
This expression holds for an ideal binary mixture, where the volume of the components is conserved, i.e. the excess volume of mixing is zero.
Building on this work, a more accurate description was given \citet{grunberg1949mixture} and \citet{oswal1998studies}, which reads
    \begin{align}
        \label{eq:GrunbergNissan}
        \ln \mu_{12}
        &= 
        x_{1} \ln \mu_{1}+x_{2} \ln \mu_{2} + \epsilon x_1 x_2
        \nonumber \\
        &+ K_{1} x_{1} x_{2}\left(x_{1}-x_{2}\right)+K_{2} x_{1} x_{2}\left(x_{1}-x_{2}\right)^{2},
    \end{align}
where $\epsilon, K_1, K_2$ are empirical parameters that account for molecular interactions.
While there is no universal theory that accurately predicts the viscosity of a liquid blend, more extended models have been derived that are important for many industrial applications including food science \cite{zhmud2014viscosity, schikarski2017direct}.

In terms of mixing sauces, most ingredients each have different elasto-viscoplastic properties.
To describe the material properties of food mixtures,
\citet{nagasawa2019mixing} considered a wide range of theoretical models and derived a viscosity blending model for shear thinning fluids. 
Using rheometry experiments, they also tested these models for various sauces including honey, mustard, mayonnaise, ketchup, hot chilli sauce, condensed milk, chocolate syrup, sweet bean sauce, oyster sauce, Japanese pork cutlet sauce, and BBQ sauce.
When mixed together, unexpected behaviours can arise:
The top row of Fig.~\ref{fig:blendingSauces} shows experimentally that pure honey flows down an inclined slope slowly because of its high viscosity (left) and pure mayonnaise remains stagnant because of its yield stress (right) \cite{balmforth2014yielding}. 
However, a 50:50 mixture runs down the slope quickly, with a much lower viscosity $\mu_{12}$ than its constituents (middle).
In the bottom row, the authors accurately reproduced these surprising dynamics using numerical simulations combined with high-end computer-generated imagery (CGI) techniques.



\subsection{Suspensions}
\label{subsec:Suspensions}

\index{suspension}
Drinks and foods often take the form of a particle suspension \cite{moelants2014review}, with  examples ranging from unfiltered coffee and wine to Turkish pepper paste and Kimchi. The texture and mouthfeel of suspensions 
depend on their rheology [see \S\ref{subsec:FoodRheology}], which is sensitive to  the particle size, microstructure and concentration. \index{rheology} Dilute suspensions such as coffee and juices, are typically Newtonian fluids, while concentrated suspensions such as pastes and purees typically display non-Newtonian behavior due to both long-ranged hydrodynamic interactions between the particles [see \S\ref{subsec:Stokeslet}] and various short-ranged interactions including friction \cite{guazzelli2011physical, zenit2018hydrodynamic}. Indeed, these rheological properties have been studied to optimise food paste 3D printing \cite{zhu2019extrusion}.

The influence of internal structure on macroscopic properties of suspensions has been actively investigated since the birth of statistical physics. \citet{einsteinviscosity} established that the viscosity\index{Einstein viscosity} of a dilute suspension increases by adding solute according to the Einstein viscosity,
\begin{equation}
    \mu = \mu_0\left(1 + \frac{5}{2}\phi\right),
\end{equation}
where $\mu_0$ is the (dynamic) viscosity of the solvent, and $\phi$ the volumetric concentration of particles. This relationship which was later developed in the context of other transport coefficients, included the diffusion and sedimentation coefficients of suspended particles, and to account for higher volume fractions and different interactions between the particles \cite{guazzelli2011physical}. This is reviewed in the context of food suspensions by \citet{genovese2007rheology} and \citet{moelants2014review}.

When we grind coffee beans or otherwise create a suspension, the particles are not all of the same size, but instead they follow a size distribution. The width of this size distribution is called the dispersity, and it can be tuned to control the rheology of a suspension.  Notably, the Farris effect \cite{farris1968}\index{Farris effect} explains how the viscosity of a suspension decreases when the dispersity increases; that is, a broader distribution of particle sizes yields a lower viscosity as compared to a narrow distribution of particle sizes.
In food science, the Farris effect has been exploited to adjust the rheological properties of edible microgel suspensions such as cheese \cite{hahn2015adjusting}, and it  has been used to minimise the apparent viscosity of cooked cassava pastes \cite{ojijo2008minimization}. Conversely, by narrowing the particle dispersity in coffee and unfiltered wine, it should be possible to enhance the mouthfeel by the opposite mechanism, but we are not aware of any reports on this topic.   
The Kaye effect is a phenomenon that occurs when a complex fluid is poured onto a flat surface, where a jet suddenly spouts upwards \cite{kaye1963bouncing}. Many non-Newtonian liquids feature this effect, including shampoo, and recent experiments have explained it by using high-speed microscopy to show that the jet slips on a thin air layer \cite{versluis2006leaping, lee2013shampoo, king2019kaye}.


\subsection{Emulsions}
\label{subsec:Emulsions}


Food emulsions consist of oil drops dispersed in water (oil-in-water emulsion) or vice versa (water-in-oil emulsion) and are ubiquitous in gastronomy and food science \cite{dickinson2010food, berton2018formation}, with everyday products such as cream, yoghurt, mayonnaise, salad dressing, and sauces \cite{mcclements2015food}. 
Emulsions are easy to make, and even an amateur chef can make an emulsion in seconds by vividly shaking or stirring the oil and water phase, which efficiently breaks up the dispersed phase into droplets by a Rayleigh-Plateau instability [see \S\ref{subsec:RayleighPlateau}]. But, as anyone who have tried to make a Hollandaise sauce would painfully know, such colloidal systems are by design thermodynamically unstable and prone to phase separation, sometimes called a `broken sauce'. 




In the food industry, phase separation is an even bigger problem as it can severely degrade the food product and shorten the shelf life, but fortunately stabilizers such as emulsifiers, texture modifiers, ripening inhibitors and weighting agents can be added to keep the system in a metastable state (by creating a free energy barrier), efficiently extending the lifetime to hours, days, months, or even years \cite{friberg2003foodSjoblom}. Protein molecules are particularly good stabilizers \cite{ferrari2022egg}, as can be readily observed by adding a small amount of mustard to vinaigrette dressing, where the stability is due to a reduction in the interfacial tension (thereby reducing the capillary driving force for drop-drop coalescence), and by the formation of viscoelastic networks that act as physical barriers against coalescence \cite{mcclements2004protein, tcholakova2006coalescence}.

In addition to the stabilization, these surface proteins control the complex rheology of emulsions \cite{brummer2006rheology}, which is responsible for the appearance and our sensory perception of food products \cite{fischer2011rheology}. Salad dressing is a widely studied kitchen emulsion, which has been examined in the context of its complex rheological response \cite{barnes1994rheology} and processing \cite{franco1995rheology}, stability, and linear viscoelasticity \cite{franco1997linear}. 
Due to the enormous surface area of emulsion drops, the overall rheology and stability is controlled by interfacial properties, and in particular, by the surface coverage and structure of adsorbed protein layers \cite{fischer2011rheology}. Animal proteins such as whey and casein readily form viscoelastic networks with high surface coverage, leading to excellent emulsion stability, while plant-based proteins such as those from cereals and pulses are less efficient stabilizers, and this is mainly due to their poor solubility in the aqueous phase \cite{solubilityOfPlantProtein}. While heating can be used to increase the stabilizing abilities by denaturing the proteins  \cite{amagliani2017globular}, such treatment can degrade the taste as well as the texture and the nutritional value of plant based foods. As such, the ability to control the interfacial properties of plant-based emulsions  without denaturing the protein is an important goal, and an exciting new direction in food science for vegetarians and vegans \cite{liu2021comparison}.
Pickering emulsions \cite{pickering1907cxcvi,ramsden1904separation} are a special type of emulsion that are stabilised with solid nanoparticles that sit at the drop interface. This field has promising applications in drug delivery and structured nanomaterials \cite{chevalier2013emulsions,zanini2017universal}, but also in food science. For example, \citet{cuthill2021colloidal} used cocoa shells, which are a food grade industry co-product, to produce colloidal lignin-rich ready for use as Pickering-type stabilisers.

Emulsions are often used in food for their creamy texture, where tribo-rheology can be linked with mouthfeel \cite{mu2022creamy}. As such, they even appear in forms that you may not expect. Espresso crema is an emulsion of coffee oil in water that floats on the coffee like a foam [\S\ref{subsec:brewingCoffee}], while foie gras and p{\^{a}}t{\'{e}} are fatty liver-based emulsions \cite{via2021microscopic}.
The process of converting separate fluids into an emulsion is called homogenization \cite{haakansson2019emulsion} and is industrially realized with high-energy mechanical methods, such as blenders or ultrasonics \cite{kentish2008use}, where strong shear forces break up the dispersed phase into droplets. The degree of homogenization that translates to the distribution of sizes of constitutive droplets or suspended particles affects the mouthfeel. For example, ice cream is a frozen emulsion made with water, milk fat and air. Upon repeated heating and cooling when taking the box out from the fridge multiple times, the size of ice crystals within the mixture changes, giving a more gritty and crunchy texture while keeping the same chemical composition [see \S\ref{subsec:IceCream}]. Another example is rice flour batter, shown by \citet{ichikawa2020phenomenological} to change the bubble size distribution during whipping, with the potential aim to create new textures for rice-flour products. The manufacturing of an emulsion does not necessarily need mechanical processing. Another special type is spontaneous emulsification, as we will discuss in the next section.

\subsection{Ouzo effect}
\label{subsec:OuzoEffect}

Ouzo, raki, arak, pastis, and sambuca are popular aperitifs in Southern Europe. They are known for their anise aroma and the remarkable change in turbidity: 
Clear when pure, they turn milky-white when clear water or ice is added, which has been termed the `ouzo effect' \cite{vitale2003liquid}. The key to this puzzle lies in the chemical composition of the drink, being mostly a mixture of water, alcohol, and essential oils, of which anethole is a prominent part. Anethole (also known as anise camphor) is highly soluble in ethanol but not in water \cite{ashurst2012food}, thus an undiluted spirit has a completely clear appearance. Upon the addition of small amounts of water, however, the oils start separating and create an emulsion of fine oil droplets which act as light scattering centres, resulting in the final cloudiness. 

The ouzo effect, also called louching or the louche effect, can be regarded as spontaneous emulsification. Such emulsions are highly stable and require little mixing \cite{sitnikova2005spontaneously}. In these multi-component mixtures, the thermodynamic stability of the emulsion comes from the trapping between the binodal and spinodal curves in the phase diagram. The ouzo effect has been widely studied to elucidate its mechanisms \cite{vitale2003liquid}. However, the microscopic dynamics are still under active investigation.  Small-angle neutron scattering studies in Pastis \cite{grillo2003small} and Limoncello \cite{chiappisi2018looking} measured the size of the demixing oil droplets to be of the order of a micron, a bit larger than the wavelengths of visible light, giving rise to Mie scattering [see \S\ref{subsec:PoiseuilleFlow}].
\citet{sitnikova2005spontaneously} established the mechanism for oil droplets growth to be Ostwald ripening without coalescence and observed the ripening rate to be lower at higher ethanol concentration, with stable droplets reaching an average diameter of three microns. \citet{lu2017universal} tried to disentangle the effects of concentration gradients from the extrinsic mixing dynamics by following the nanodroplet formation in a confined planar geometry and observed universal branch structures of the nucleating droplets under the external diffusive field, analogous to the ramification of stream networks in the large scale \cite{Devauchelle2012ramification,Cohen2015path}, and the enhanced local mobility of colloids driven by the emerging concentration gradient. The ouzo effect can be triggered not only by the addition of water but also by the  evaporation of ethanol, e.g. in sessile ouzo droplets \cite{tan2016evaporation,diddens2017evaporating,tan2017self}, leading to an astoundingly rich drying dynamics involving multiple phase transitions.

The remarkable stability of the spontaneously formed emulsion gives hope for potential generation of surfactant-free microemulsions without resorting to mechanical stabilisation, for example high-shear stabilisation that is often used in fat-filled milk formulations \cite{osullivan2018use}. Thus, the ouzo effect has been used for the creation of a variety of pseudolatexes, silicone emulsions, and biodegradable polymeric capsules of nanometric size \cite{ganachaud2005nanoparticles}. Nanoprecipitation can also be used for drug delivery and the design of nanocarriers \cite{lepeltier2014nanoprecipitation}. Particles created using the ouzo effect are kinetically stabilised, and provide an alternative to thermodynamically stabilised micelles formed using surfactants \cite{almoustafa2017technical}.
Thus, this field offers many interesting directions of future research.

\subsection{Cheerios effect}
\label{subsec:CapillaryFloating}


In the previous sections we have seen complex fluid effects in the bulk, but a similar complexity can arise at interfaces.
Imagine sprinkling pepper on your soup.
Interestingly, small objects that are more dense than the fluid may still float at the air-fluid interface because of surface tension \cite{vella2015floating}. 
Moreover, floating objects tend to aggregate at the surface, brought together by capillary forces induced by the presence of a curved meniscus around floating objects. Aptly named the `cheerios effect' \cite{vella2005cheerios}\index{cheerios effect}\index{floating}\index{capillarity}, this is seen not only with corn flakes, but also bread crumbles \cite{singh2005fluid}, foams, and generally objects that are large enough to create menisci of considerable size. The mechanism of lateral capillary interaction due to interfacial deformation admits a universal theoretical description for particle sizes ranging from nanometers to centimeters \cite{Kralchevsky2000capillary}.  
Initially, the interaction of widely spaced particles may be regarded as a two-body problem \cite{Paunov1993lateral}, but eventually multiparticle rafts are formed \cite{lagarde2019capillary}. 
The dynamics of these aggregates are more complex, since they may undergo internal redistribution and destabilisation \cite{abkarian2013gravity}. 
An interesting example is an active assembly of dozens of fire ants on a water surface \cite{Mlot2011fire}. 
The presence of surface tension allows to sustain deformed surfaces which can support a load of an insect walking on water \cite{gao2004water, bush2006walking, childress2010walking}, or a biomimetic water-walking device \cite{hu2010water}. 
Similar behaviour, termed the inverted cheerios effect, is seen when water droplets sit on a soft, deformable substrate, and the induced deformation drives their assembly \cite{karpitschka2016liquid, pandey2017dynamical}. 

By a combination of capillary forces and externally controlled fields, e.g. electromagnetic field, both static and dynamic assembly can be achieved in capillary disks \cite{wang2017dynamic,koens2019near}. Capillary forces between spherical particles floating at a liquid-liquid interface have also been quantified to show a qualitatively similar behaviour \cite{vassileva2005capillary}. Same guiding principles are used in micro-scale for colloidal self-assembly, driven not by gravity but by an anisotropically curved interface \cite{ershov2013capillarity}. Finally, in active microrheology \cite{furst2017microrheology,squires2010fluid,mackintosh1999microrheology,mizuno2008active,zia2018active}, an external force field (usually magnetic or optical) is used to distort surface active or bulk probes in order to extract viscoelastic responses of complex materials, with direct applications in food science \cite{yang2017application}. 

A separate class of interfacial interaction involves the dynamic problem of stone skipping, known in Britain as `ducks and drakes', where the interfacial properties determine the optimal angle of attack for the most successful rebound and therefore maximal range \cite{lorenz2007spinning, clanet2004secrets, hewitt2011continual}. Moreover, elastic `stones' have been shown to demonstrate superior skipping ability by assuming hydrodynamically optimal shapes during the collision \cite{belden2016elastic}. 
Although seemingly unrelated, the dynamics of interfacial deformation by contact with a boundary might be of importance in the development of ergonomic kitchen utensils such as spatula and scrapers, optimal coatings for baking surfaces \cite{magens2017adhesion}, dealing with food adhesion in industrial processing \cite{frabetti_adhesion_2021} and may also inspire novel approaches to these procedures. 

\section{Hot Main Course: Thermal Effects}
\label{sec:PotsPans}

The word `cooking' refers to the preparation of food in general, but specifically to the operations involving temperature and heat such as boiling, frying, baking and poaching, to transform food products into a final dish.
Thus, many kitchen flows are subject to thermal effects that alter their taste, texture and mouthfeel. Below we discuss such culinary processes involving heat transfer, the Leidenfrost effect, temperature-driven flows, the physics of seared steaks, and even hot vapours, smoke and fire. 
Nothing beats a warm meal, but be careful not to get burned. 
In the words of William Shakespeare (c.1564-1616),
\begin{quote}
    \emph{Heat not a furnace for your foe so hot, that it do singe yourself.}
\end{quote}

\subsection{Feel the heat: Energy transfer}
\label{subsec:heatEquation}

Heat is transported in fluids in a way similar to momentum [\S\ref{subsec:NavierStokes}]. Fluid parcels are advected with the flow, and additionally exchange heat by conduction. Local variations in temperature can additionally induce density gradients, which can drive macroscale convective motion. 

The relevant quantity characterising the thermal properties of the fluid is the scalar temperature field, $T(\vec{r},t)$. The spread of temperature is described by the heat equation, which for a fluid of density $\rho$ and heat capacity at constant pressure $c_p$ can be written as
\begin{equation} \label{eq:heatEquation}
  \rho c_p  \frac{DT}{Dt} = k \nabla^2 T + h + Q,
\end{equation}
where $k$ is the thermal conductivity, governing the diffusive spread of temperature by thermal conduction, $h$ is a source term accounting for local heating (e.g. by chemical or nuclear reactions), and $Q$ is the viscous dissipation term, which can be neglected in most practical situations. In the absence of local heat sources, the heat equation becomes simply a Fourier's diffusion equation, with the thermal diffusivity $D_T = k/(\rho c_p)$. The dominant (heat) transport mechanism is determined by the (thermal) P\'eclet number,
    \begin{equation}
    \label{eq:PecletNumber}
    \text{Pe}_{(T)}
    \equiv \frac{\text{Diffusion time} }{\text{Convection time}}
    = \frac{L_0 U_0}{D_{(T)}},    
    \end{equation}
which besides thermal diffusion can equally be used to characterise molecular diffusive transport. 

The concept of heat diffusion  suffices to explain several kitchen processes. Baking a cake requires the heat to reach the inner parts of the dough but changing either the dimensions of the cake or the amount of batter used alters the baking time in a way that can indeed be predicted from the diffusion equation \cite{olszewski2006baking}. Heat flow considerations can guide the development of an optimal flipping schedule when frying burgers \cite{thieffault2022mathematics}. And the problem of perfectly boiling an egg can also be quantified in terms of the energy equation \cite{roura2000long} to aid many breakfast table discussions. 

\subsection{Levitating drops: Leidenfrost effect}
\label{subsec:Leidenfrost}

\begin{figure}[t]
    \centering
    \includegraphics[width=\linewidth]{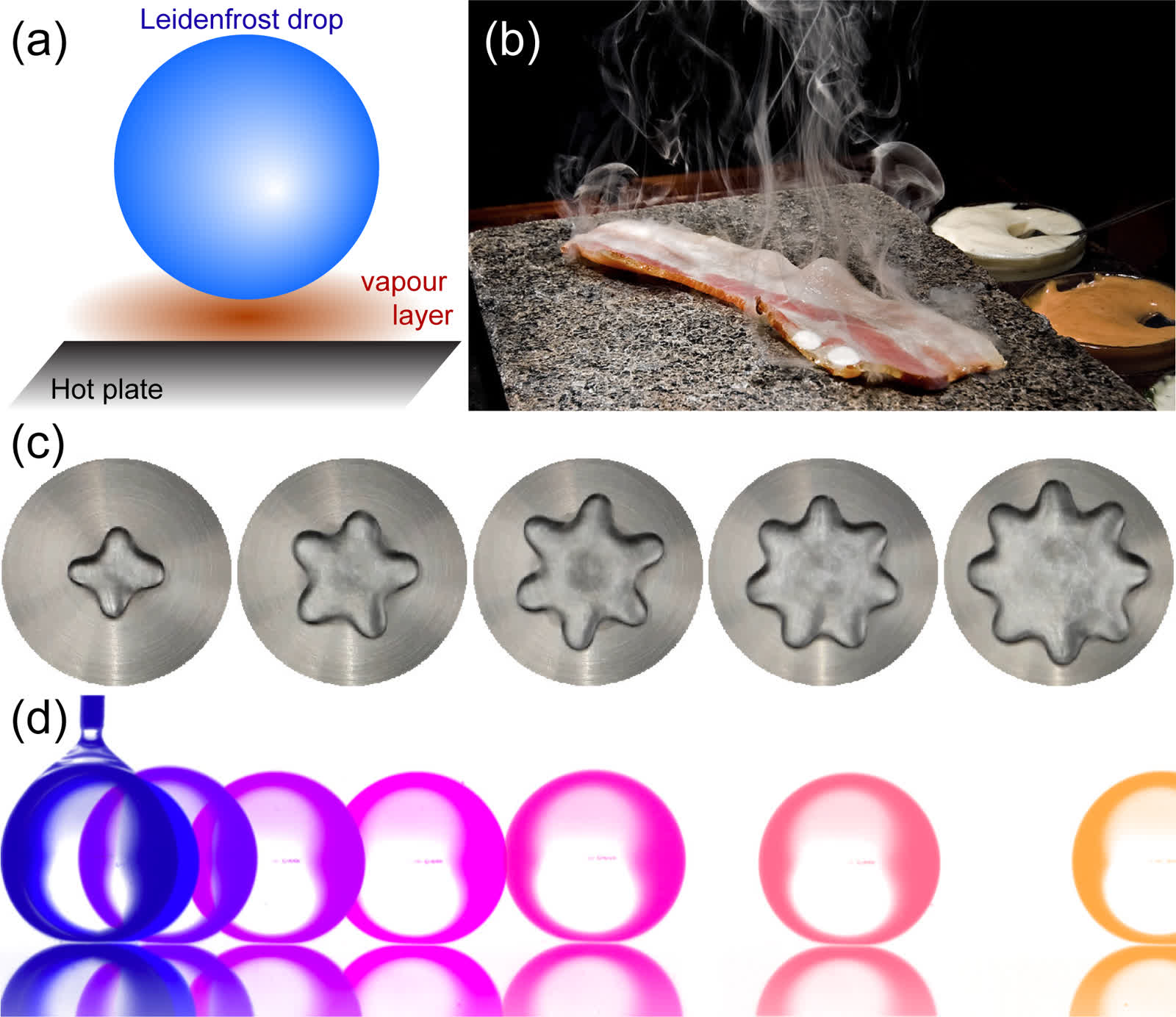}
        \caption{Leidenfrost effect.
        \textbf{(a)} Diagram of a droplet levitating on a cushion of evaporated vapour above a heated surface.
        \textbf{(b)} The Leidenfrost effect prevents meat from sticking to a hot plate. Artwork entitled `Bacon Prelude' by Pedro Moura Pinheiro, licensed under CC BY-NC-SA 2.0.
        \textbf{(c)} Star-shaped oscillations of Leidenfrost drops \cite{ma2017star} that make characteristic sounds. Top view. From \citet{singla2020sounds}. 
        \textbf{(d)} Time-lapse image of a self-propelled Leidenfrost drop on a reflective wafer heated at $\SI{300}{\celsius}$. Side view. From \citet{bouillant2018symmetry}. 
        }
    \label{fig:leidenfrost}
\end{figure}

The famous Maillard reaction \cite{brenner2020science} is what gives browned food its nutty, delicious flavor and is known for turning uncooked, raw meat into tender, delicious steaks or sausages \cite{ghebremedhin2022meat}. 
When grilling steaks, a simple way to assess whether the frying pan is sufficiently hot is to sprinkle a handful of water droplets onto it. When the surface temperature slightly exceeds the water boiling point, the droplets start vigorously evaporating, producing a sizzling sound. However, if the pan is left on full heat for a while and becomes considerably hotter, small droplets change their behaviour completely and start levitating above the hot surface without boiling [Fig.~\ref{fig:leidenfrost}a,b]. 
This levitation can help with preventing the meat from sticking \cite{herwig2014ach}.

This phenomenon was first known to be observed by a Dutch scientist H. Boerhaave in 1732, and later described in detail by a German doctor Johan Gottlob Leidenfrost in 1756. He provided a record of water poured onto a heated spoon that `{\it does not adhere to the spoon, as water is accustomed to do, when touching colder iron}' \cite{leidenfrost}. The Leidenfrost effect, as it was later termed, has been studied extensively in the scientific context \cite{curzon1978leidenfrost, thimbleby1989leidenfrost, quere2013leidenfrost}, and even became a plot device in Jules Verne's novel \textit{Michel Strogoff} in 1876. \index{Leidenfrost effect} \index{temperature}\index{droplet}\index{boiling}\index{heat transfer}

The explanation of this effect boils down to the analysis of heat transfer rate between a hot plate and a droplet. For intermediate excess temperatures above the boiling temperature (between 1$^\circ$C and ca.\,100$^\circ$C) the droplets undergo either nucleate boiling, with vapour bubbles forming inside, or transition boiling, when they sizzle explosively upon impact on the plate. However, above the Leidenfrost temperature, which for water on a metal plate is approximately 150-180$^\circ$C, the heat transfer dynamics change when a thin vapour layer is created between the droplet and the plate. This thin cushion both insulates the droplet and prevents it from touching the substrate which would cause nucleate boiling inside the droplet [Fig.~\ref{fig:leidenfrost}a]. Due to the competition between evaporation and film draining, the typical thickness of the insulating layer is about $\SI{100}{\micro\metre}$.
Because of this effect, the lifetime of droplets on a substrate can increase by an order of magnitude \cite{biance2003leidenfrost}. Further increase in the substrate temperature naturally decreases the lifetime but the decrease is slow.
The minimum temperature required for the Leidenfrost effect to occur on smooth surfaces was characterised recently by \citet{harvey2021minimum}.

The presence of a thin lubricating vapour layer, which is characterised by a low Reynolds number [\S\ref{subsec:LowReynolds}], prevents meat from sticking to a hot plate [Fig~\ref{fig:leidenfrost}b]. It makes water droplets highly mobile due to the diminished friction [Fig.~\ref{fig:leidenfrost}d]. As soon as a spontaneous instability causes a slight difference in the thickness of the vapour layer, a flow emerges which triggers self-propulsion by rolling motion \cite{bouillant2018leidenfrost, leon2021self}. The interaction with a structured substrate can also be used to induce directed motion, e.g. across ratcheted grooves \cite{linke2006leidenfrost, jia2017reversible, wurger2011leidenfrost}, and the motion may further be controlled with thermal gradients \cite{sobac2017self}.
A video featured on BBC Earth shows how the Leidenfrost effect can be used to make water run uphill\index{ratchet}\index{self-propulsion} \cite{bbcearth2013leidenfrostuphill}.
Besides self-propulsion, the energy injected by droplet heating can cause droplet vibration with star-shaped droplet modes \cite{ma2017star} that lead to distinct Leidenfrost sounds \cite{singla2020sounds} [Fig.~\ref{fig:leidenfrost}c].
 
The Leidenfrost phenomenon is seen all across the temperature scale and is controlled mainly by the temperature difference between the substrate and the droplet, and surface roughness.
Interestingly, the substrate need not be solid: a similar effect is observed with acetone droplets (nail polish remover) on a bath of hot water \cite{janssens2017behavior}.
More generally, the Leidenfrost state, in which an object hovers on a solid or on a liquid due to the presence of a vapour layer, can be seen in a variety of contexts. 
For example, it occurs when a block of sublimating solid carbon dioxide (dry ice) is placed on a plate at room temperature \cite{lagubeau2011leidenfrost}, which is also termed the \textit{inverse Leidenfrost effect} \cite{hall1969inverse}, or when room temperature ethanol droplet falls on a bath of liquid nitrogen \cite{gauthier2019self} and starts moving\index{acetone}\index{vapour layer}.
Furthermore, when two water droplets are placed on a hot plate, the vapour layer between them prevents their coalescence, which is called the triple Leidenfrost effect \cite{pacheco2021triple}.
Frequently demonstrated in popular lectures and science fairs, the Leidenfrost effect allows a person to quickly dip a wet finger in molten lead or blow out a mouthful of liquid nitrogen without injury \cite{walker2010boiling}.
\index{liquid nitrogen}

\subsection{Heating and Boiling: Rayleigh-B\'{e}nard convection}
\label{subsec:HeatingBoilingRBC}

\begin{figure}[t]
        \centering
        \includegraphics[width=\linewidth]{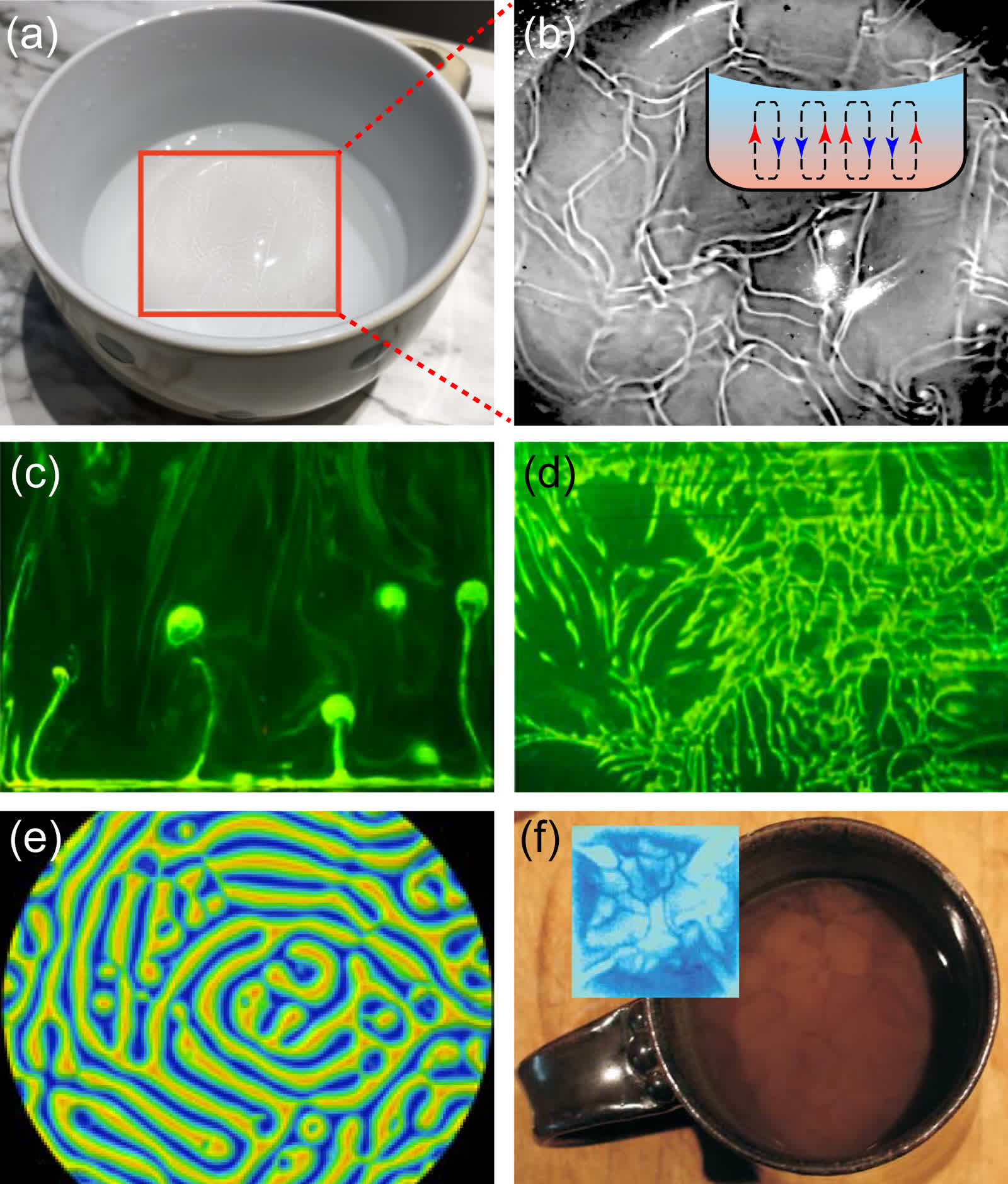}
            \caption{Rayleigh-B\'{e}nard convection.
            \textbf{(a)} Rising plumes in a pot of water heated from below, visible because the refractive index changes with temperature differences.
            \textbf{(b)} Contrast-enhanced magnification.
            \textbf{(c)} Side-view of mushroom-like plumes in a high-viscosity fluid. The green line at the bottom is the boundary layer. 
            \textbf{(d)} Top-view of dendritic line plumes.
            (c,d) From \citet{Prakash2017}. 
            \textbf{(e)} Temperature field in a simulation of Rayleigh-B\'{e}nard convection at $\text{Ra}=5000$ and $\text{Pr}=0.7$. From \citet{emran2015large}. 
            \textbf{(f)} Vortex structures in a coffee cup with milk at the bottom, which gets displaced by cold plumes that sink down from the evaporating interface. Inset: IR thermograph showing convection cells of colder (downwelling) and warmer (upwelling) regions. 
            From \citet{wettlaufer2011universe}. 
            }
        \label{fig:RBC}
    \end{figure}

Let's cook some pasta \cite{audoly2005fragmentation, heisser2018controlling, tao2021morphing, hwang2022swelling}.
We place a pot with water on the stove and start heating. 
Heat from the stove is transferred to the water, first through conduction, and then through natural convection\index{convection} \cite{batchelor1954heat}, which we see as characteristic structures called `plumes' near the bottom wall [Fig.~\ref{fig:RBC}a,b].
The fluid layer adjacent to the heated surface becomes unstable and starts to rise [Fig.~\ref{fig:RBC}c,d], since it is lighter than the bulk fluid [see also \S\ref{subsec:LayeredCocktails}]. 
This fundamental process has been widely studied in many different configurations. One of the most well-studied is the Rayleigh-B\'{e}nard Convection (RBC) system\index{Rayleigh-B\'{e}nard convection}, consisting of a fluid layer bound between two horizontal plates, heated from below and cooled from above, as reviewed by \citet{LohseRMP2009, LohseARFM2010, Kadanoff2001}. It occurs ubiquitously in natural contexts, including astrophysics \cite{moore_1973,Cattaneo_2003}, geophysics \cite{mckenzie1974, Prakash2017}) and in engineering applications such as metallurgy, chemical, and nuclear engineering \cite{Brent1988, zhong2010streaks}.    

The key non-dimensional parameters governing natural convection is the Rayleigh number\index{Rayleigh number}, 
\begin{equation}
    \label{RayleighNumber}
    \text{Ra} 
    \equiv \frac{\text{heat diffusion time}}{\text{heat convection time}}
    = \frac{g \beta_{T} \Delta{T} H^{3}}{\nu D_T},
    \end{equation} 
and the Prandtl number\index{Prandtl number},
    \begin{equation}
    \label{PrandtlNumber}
    \text{Pr} 
    \equiv \frac{\text{momentum diffusivity}}{\text{thermal diffusivity}}
    = \frac{\nu}{D_T},
    \end{equation} 
where $g$ is gravity, $\beta_{T}$ is the coefficient of thermal expansion, $\Delta{T}$ is the temperature difference between the walls, $H$ is the height of the fluid layer, $\nu$ is the kinematic viscosity of the fluid, and $D_T$ is the thermal diffusivity of the fluid. 
Note that the Prandtl number only depends on the inherent properties of the liquid. Most oils have $\text{Pr} \gg1$, which means that heat diffuses very slowly in oils.
Then, depending on the magnitude of $\text{Ra}$, we can identify different regimes of natural convection. The heat transfer through the fluid is solely through conduction until a critical $\text{Ra}_\text{cr} \sim 1708$ \cite{krishnamurti_1970_a,krishnamurti_1970_b}, 
above which the convection consists of steady `laminar' rolls [Fig.~\ref{fig:RBC}e]. 
At $\text{Ra} \sim 10^4$, these convection rolls become unsteady, and beyond $\text{Ra} \sim 10^5$ the convection is characterized as turbulent natural convection. 
A clear example of this is seen when pouring milk in coffee [Fig.~\ref{fig:RBC}f].
Another important manifestation of thermally driven flows is found in deep-fat fryers, where plumes transport oxygen from the air interface into the hot oil, leading to reactive hydroperoxides and toxic compounds \cite{touffet2021coupling}.

Now we come back to our water pot. As we continue to supply heat, the temperature will eventually reach the boiling point of water. At this point, the boiling process begins with vapour bubbles forming in a superheated layer adjacent to the heated surface \cite{Dhir1998}. In this two-phase system, the vapour bubbles enhance the convective heat transfer in the standard Rayleigh-B\'{e}nard Convection system \cite{Lakkaraju2013}. This boiling process is so complicated that we currently only have an empirical understanding of it \cite{Dhir1998}, and theoretical progress has been lacking. However, cutting-edge numerical simulations seem to be a promising direction to model these processes accurately \cite{Dhir2005}. We put the pasta into boiling water and allow it to cook for approximately 10 minutes (or until it extends to a desired length, as discussed by \citet{hwang2022swelling} and in \S\ref{subsec:FoodRheology}. Once the pasta is cooked, we drain the excess water and serve with a favorite sauce [see \S\ref{subsec:Blending}].

\subsection{Layered latte: Double-diffusive convection}
\label{subsec:DoubleDiffusiveConvection}

\begin{figure}[t]
    \includegraphics[width=\linewidth]{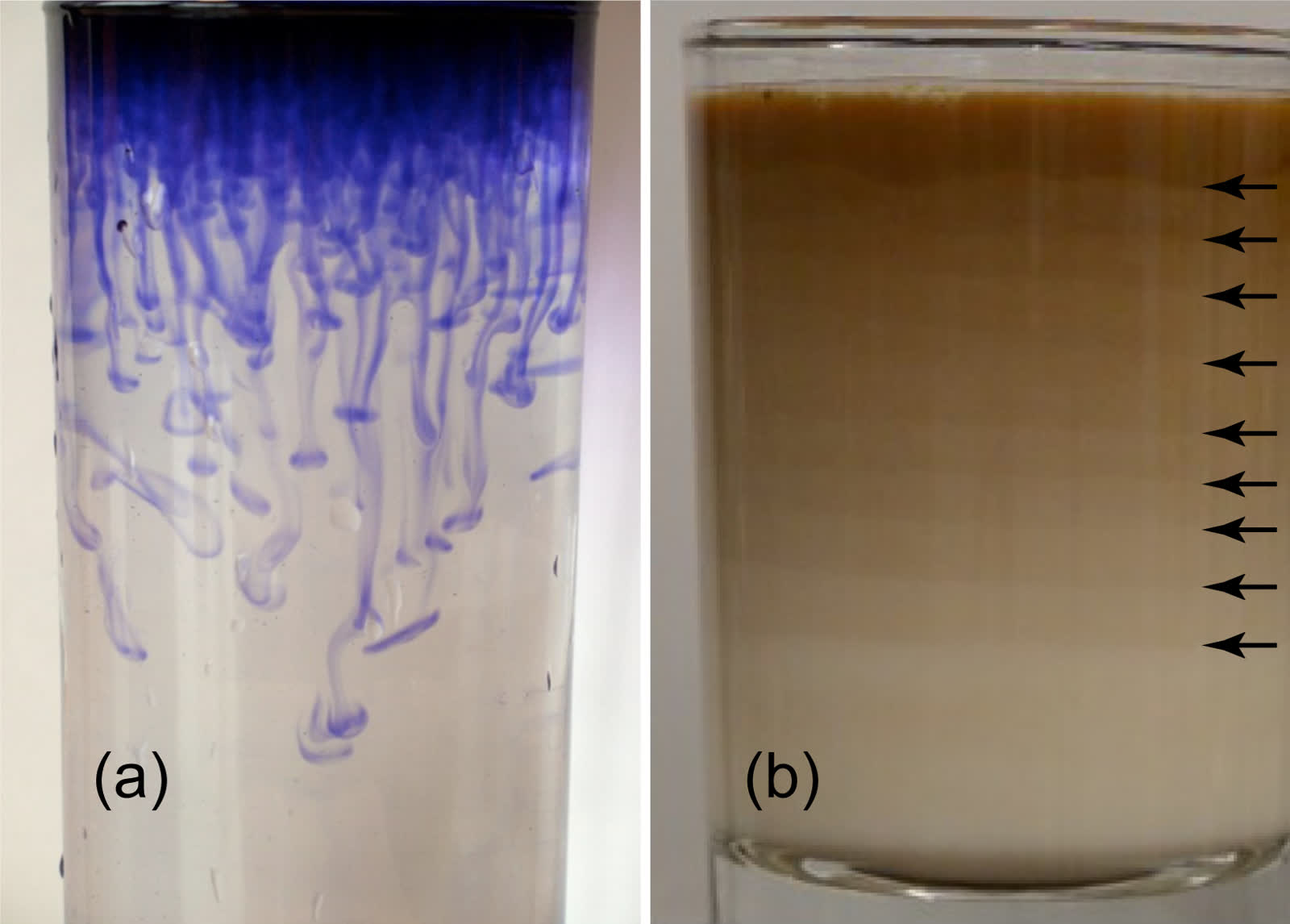}
    \caption{Double-diffusive convection phenomena.
    \textbf{(a)} A cocktail with blue salt fingers, produced by warmer salty water resting on colder fresh water of a higher density. Image courtesy of Matteo Cantiello (Flatiron Institute). 
    \textbf{(b)} Layered caff\`e latte. Adapted from \citet{xue2017laboratory}. Black arrows are added to highlight the layer boundaries. 
    }
    \label{fig:DDC}
\end{figure}

In \S\ref{subsec:HeatingBoilingRBC} we discussed Rayleigh-B\'enard convection, which occurs because the fluid density depends on temperature. 
Often the density also depends on a second scalar, like salt or sugar concentration.
Importantly, their molecular diffusivity, $D_S$, is smaller than the thermal diffusivity, $D_T$, so that heat is transported faster that mass. This explains why your tea cools down before the sugar diffuses up, but it can also lead to double-diffusive convection (DDC)\index{double-diffusive convection} that can cause a range of unexpected phenomena, as reviewed initially by \citet{huppert1981double}, and more recently by \citet{radko2013double} and \citet{garaud2018double}. 
Besides the Rayleigh and the Prandtl numbers [Eqs.~\eqref{RayleighNumber}, \eqref{PrandtlNumber}], DDC also depends on the Schmidt number\index{Schmidt number}, which is often used to characterize convective flows involving simultaneous momentum and mass diffusion processes:    \begin{equation}
    \label{eq:SchmidtNumber}
    \text{Sc} 
    \equiv \frac{\text{momentum diffusivity}}{\text{molecular diffusivity}}
    = \frac{\nu}{D_S}.
    \end{equation} 
One surprising DDC phenomenon that is readily observed in the kitchen \cite{heavers2009sugar} is called salt fingering\index{salt fingering}, which happens when warmer saltier water rests on colder fresher water of a higher density \cite{yang2016convection}.
Then, in the words of \citet{stern1960salt}, `\textit{[the] `gravitationally stable' stratification \dots is actually unstable}'.
If a parcel of warm salty water is perturbed to move down a bit, it loses its heat quicker than its salinity, so it will keep sinking further.
Hence, salt fingers (vertical convection cells) spontaneously start growing downwards, accelerated by thermal diffusion, which gives rise to strong mixing.
\citet{kerr2002salt} jokes that one might have a Martini cocktail ``fingered, not stirred'', which could in fact be quite a spectacle with coloured layers [Fig.~\ref{fig:DDC}a].
This mixing effect is likely important for nutrient transport in the ocean and climate change \cite{johnson2009ocean, fernandez2015importance}.
In astronomy, thermohaline mixing can occur in evolved low-mass stars \cite{cantiello2010thermohaline}.
DDC can equally feature in porous media \cite{griffiths1981layered}, which might be relevant for heat and mass transfer in porous materials under microwave heating \cite{dinvcov2004heat}.

Another striking example of DDC is the formation of distinct layers in a caff\`e latte \cite{xue2017laboratory} [Fig.~\ref{fig:DDC}b]. 
To make one, warm a tall glass with \SI{150}{\milli\litre} of milk up to \SI{50}{\celsius}, and pour \SI{30}{\milli\litre} of espresso at \SI{50}{\celsius} into it.
The milk is denser than espresso, so the dynamics will follow an inverted fountain effect [\S\ref{subsec:LayeredCocktails}] leading to stratification with a vertical density gradient. 
Subsequently, a horizontal temperature gradient is established because the glass slowly cools down from the sides.
This double gradient leads to stacked convection rolls separated by sharp interfaces, as seen in creaming emulsions \cite{mueth1996origin}.
The coffee pouring (injection) velocity sets the initial density gradient, and thus the Rayleigh number, which much exceed a critical value for layers to form \cite{xue2017laboratory}.
This was investigated further with direct numerical simulations by \citet{chong2020cafe}, who also discussed the mechanism how the layers merge over time: they found that as the circulation weakens, hot fluid accumulates by the hot sidewall, where buoyancy forces, larger for hotter fluid, eventually break the layer interface. The two layers merge, and a new circulation pattern is established within the thicker new layer. Besides caf\'e au lait, similar stacked layers are observed in the ocean, meters thick and kilometers wide, called thermohaline staircases\index{thermohaline staircase} \cite{tait1971thermohaline, yang2020multiple}.

\subsection{Tenderloin: Moisture migration}
\label{subsec:MoistureMigration}

So far, we have described various fluid mechanical phenomena related to drinks or liquids. When cooking solid foods, it is also important to understand moisture migration to achieve a tender result \cite{hwang2022swelling}.
We will consider the example of meat cooking (e.g. tenderloin), where two important physical aspects include time-dependent protein denaturation and cooking loss (water loss). To reach the desired meat textures, the meat must be cooked at well-defined temperatures to ensure selective protein denaturation \cite{zielbauer2016physical}. Since our focus here is on fluid dynamics, we will discuss water loss during the heat treatments, which also depends on the temperature \cite{zielbauer2016physical}. This cooking loss has been described using the Flory-Rehner theory of rubber elasticity \cite{vilgis1988flory, van2007moisture}. This theory models the transport of liquid moisture due to denaturation and shrinkage when the protein is heated. The moisture transport is due to a shrinking protein matrix, similar to a `self-squeezing sponge'. It is assumed that the poroelastic theory applies here. Then, this theory describes moisture transport by Darcy's law [see \S\ref{subsec:brewingCoffee}], where the fluid flow rate is linear with pressure gradient. The pressure is due to the elasticity of the solid matrix of the porous material, and here it is referred to as the `swelling' pressure, $p_\text{swell}$. According to Flory-Rehner theory, the swelling pressure can be decomposed into two components, 
\begin{equation}
    \label{eq:F-R1}
    p_\text{swell} = p_\text{mix} + p_\text{el},
\end{equation}
where $p_\text{mix}$ is the mixing or osmotic pressure, and $p_\text{el}$ is the pressure due to elastic deformation of the cross-linked polymer gel \cite{vilgis2000polymer}. In equilibrium, the network pressure opposes the osmotic pressure, and the swelling pressure is zero. The temperature rise during cooking causes an imbalance between the osmotic pressure and network pressure, leading to the expelling of excess fluid from the meat. Hence, the swelling pressure in meat is proportional to the difference between moisture content and water holding capacity. The gradient of this swelling pressure will drive the liquid moisture flow, and is used in Darcy's Law [Eq.~\eqref{eq:Darcy}]. This model has been found to agree well with experimental data, proving that the Flory-Rehner theory provides a sound physical basis for the moisture migration in cooking meat \cite{van2007moisture}.

\subsection{Flames, vapours, fire and smoke}
\label{subsec:flamesandfire}

\begin{figure}[t]
    \includegraphics[width=\linewidth]{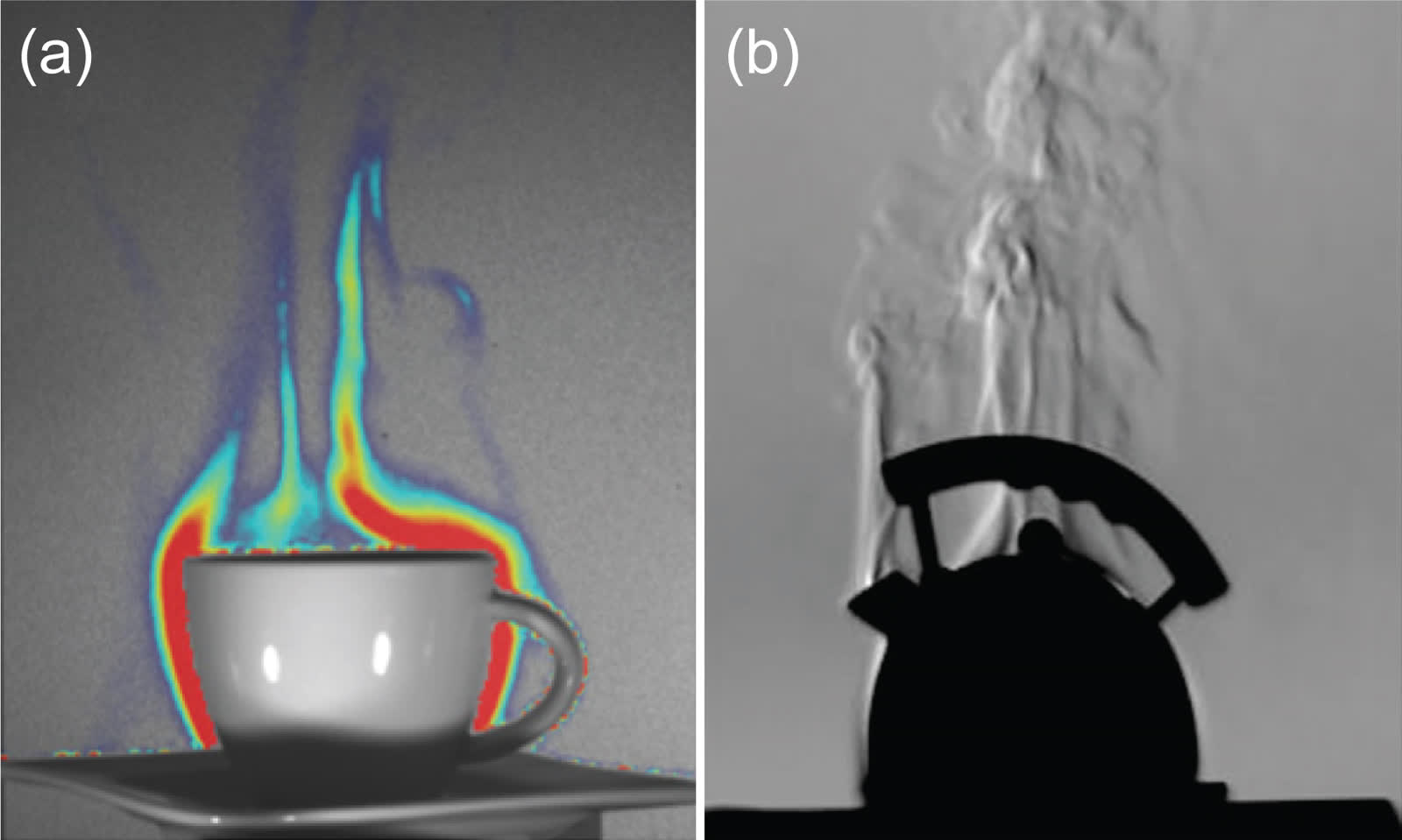}
    \caption{Buoyancy-driven plumes. 
    \textbf{(a)} Flows developing over an espresso cup visualised with schlieren imaging. Colors indicate fluid displacement, where red is the highest, and blue the lowest. From \citet{cai2021flow}. 
    \textbf{(b)} Plume around a hot tea kettle captured with schlieren imaging. From \citet{settles2017review}. 
    }
    \label{fig:espressocup}
\end{figure}

Next, we consider the flows generated around the hot cookware, utensils, or hot beverages. There are various heat transfer processes in the kitchen that can give rise to vapours, fumes, fires, and smoke. We of course enjoy the smell of a good dish that is cooking, and unexpected smoke is oftentimes an indicator that something is burning. Hence, in addition to increasing the overall room/kitchen temperature, the vapors/smoke also play an important role in giving us positive or negative feedback on how the cooking is going. Hot utensils transfer heat into the surrounding air setting up buoyancy-driven convection or natural convective flows around them. Figure~\ref{fig:espressocup} shows the beautiful flow of vapours and convection around a hot espresso cup \cite{cai2021flow}, and around a hot tea kettle, visualised with Schlieren imaging as reviewed by \citet{settles2001schlieren, settles2017review}. Such convective flows are present around all heated objects, and one can imagine how different geometries of vessels and cookware can give rise to complicated flows around them.

The kitchen is our safe place to prepare food, but we must remember that it is also a place where several safety hazards exist. At some point in our lives, most of us have forgotten to turn off the kitchen stove and suffered the consequences. When food is overheated, it starts to burn and eventually the temperature gets so high that the carbon content gets converted to soot that give rise to smoke and fumes. Smoke decreases the overall air quality and inhaling it can adversely affect our health \cite{comstock1981respiratory}, especially if the burning becomes intense and the heating is continued. While the dispersion of smoke as a general pollutant in the atmosphere has been studied extensively, smoke in the kitchen has also been the subject of several studies \cite{rogge1991sources, comstock1981respiratory}. A critical issue related to this is ventilation, i.e. how well a kitchen is designed to get rid of harmful smoke and fumes. Most modern kitchens feature a ventilation or exhaust `hood' right above the stove. Both Experimental Fluid Dynamics (EFD) and Computational Fluid Dynamics (CFD) techniques such have been utilized to study the flow through kitchen hoods in order to maximise their performance \cite{Chen20182dpiv}. Fires in the kitchen are the most dangerous safety hazard \cite{gao2014analytical}, and can result in destruction of property and loss of life. Given the importance of minimizing safety hazards in the kitchen, fire engineers rely on fluid dynamics modeling to develop safe kitchen ventilation \cite{yeoh2009computational, norton2013computational, chen2020air}.

\subsection{Melting and freezing}

We tend to naturally associate cooking with heat. Boiling, melting, freezing, solidifying, dissolving and crystallizing food products need the external heat stimulus to transform. Thermally driven reconfiguration can happen on a molecular level but more often it is enough to consider phase transitions due to the heating or cooling of a substance. 

Melting is another fundamental process of phase transition. From thawing to cooking, solid substances are transformed into soft matter or liquid products. The sole process of melting and flows induced therein are still of fundamental interest for physicists. The mechanism of heat transfer relies, in the simplest case described in \S\ref{subsec:heatEquation}, on the Fourier law, where the heat flux $q$ across a surface is locally proportional to the temperature gradient $\nabla T$. The temperature field thus satisfies the advection-diffusion equation \eqref{eq:heatEquation}, which can then be solved numerically or analytically in specific geometric configurations. The geometry itself inspires questions on how the process of thawing can be exploited to create desired forms, i.e. seen in constantly evolving ice sculptures or in ordinary ice cubes melting in a cocktail. Although the initial shapes of frozen structures may be arbitrary, macroscopic objects such as melting ice cubes and growing stalactites can approach non-intuitive geometric ideals. For the dissolution of non-crystalline objects, a paraboloidal shape was shown to be the geometric attractor \cite{nakouzi2015dissolving}. Even the simple melting of icicles is governed by a combination heat transfer from the air, the latent heat of condensation of water vapour, and the net radiative heat transfer from the environment to the ice \cite{neufeld_goldstein_worster_2010}, which emphasizes the complexity of phase change processes. In kitchen flows, an additional factor is the microstructure of food products, which renders their flow non-Newtonian, and their response to temperature variations non-linear. Even though the flow of molten chocolate in a fountain is aptly described by a power-law fluid \cite{townsend2015fluid}, the process of melting involves different crystalline phases and is thus highly complex. Subsequent freezing typically leads to a change in structure and appearance, with different physical properties and even taste.  The quality of chocolate products, in particular their gloss, their texture and their melting behaviour, depends primarily on two processing steps: the precrystallisation of the chocolate mass, and the eventual cooling process \cite{Mehrle2007SolidificationAC}.  These factors must be considered in confectionery manufacturing and, fundamentally, in the modelling of crystallization and melting kinetics of cocoa butter in chocolate \cite{reverend2009modelling, bhattacharyya2022effect}.

Butter (plant or animal-based) itself, being a vital product for cooking, responds strongly to external temperature, transiting from completely solid and brittle when taken out from the refrigerator, to pleasantly spreadable at intermediate temperatures, to liquid. This empirical feeling can be related to its viscoelastic characteristics \cite{Hayashi1994,landfeld2000viscosity}, which additionally depend on the substance and on the method of production \cite{shukla1994physicochemical}. Rheology and texture [see  \S\ref{subsec:FoodRheology}] are often the basic characteristics when heating or baking cheese \cite{lucey2003perspectives}, such as in the melting and browning of mozzarella in the oven \cite{rudan1998model}. The temperature can be also coupled to the nonlinear viscoelastic properties, for instance when considering starch gelatinisation while making gravies and thick sauces \cite{ratnayake2008starch} or when freezing and thawing a gelatin-filtered consomm\'e \cite{lahne2010gelatin}.

\subsection{Non-stick coatings} 
\label{subsec:nonStickCoatings}

Many non-stick pans or cookware\index{non-stick surfaces} are designed to be hydrophobic (water repellent) and oleophobic (oil repellent), which together is called amphiphobic \cite{williams2017waterproof, tehrani2019waterproof}.
Waterproof fabrics often use chemical coatings such as polyurethane (PU), polyvinyl chloride (PVC), or fluoropolymers.
However, there are many concerns for these materials regarding toxicity and other environmentally damaging effects \cite{sajid2017ptfe}, and to limit their destructive impact, the European Union has announced a ban on such chemicals by 2030 \cite{EUstrategyPFAS}. 
As such, interest has risen for purely physical coatings that make use of the `lotus effect'\index{lotus effect} \cite{marmur2004lotus, barthlott2017plant}.
Like leaves of the lotus plant (\textit{Nelumbo} genus), micron-sized structures can be designed that give rise to superhydrophobicity \cite{feng2002super, lafuma2003superhydrophobic, dupuis2005modeling}.
A droplet that impacts these surfaces can bounce off without wetting them \cite{richard2002contact, reyssat2007impalement}. 
Moreover, decorating sub-millimetric posts with nanotextures can lead to `pancake bouncing'\index{pancake bouncing}, where the contact time of droplets with the surface is significantly reduced \cite{liu2014pancake, liu2015symmetry}.
This is important for anti-icing \cite{mishchenko2010design} and self-cleaning surfaces \cite{blossey2003self}, with possible applications for airplanes and cars.
Remarkably, superamphiphobic coatings can also be made with candle soot \cite{deng2012candle}.

\section{Honey Dessert: Viscous Flows}
\label{sec:HoneyMicro}

\begin{figure}[t]
    \includegraphics[width=1\linewidth]{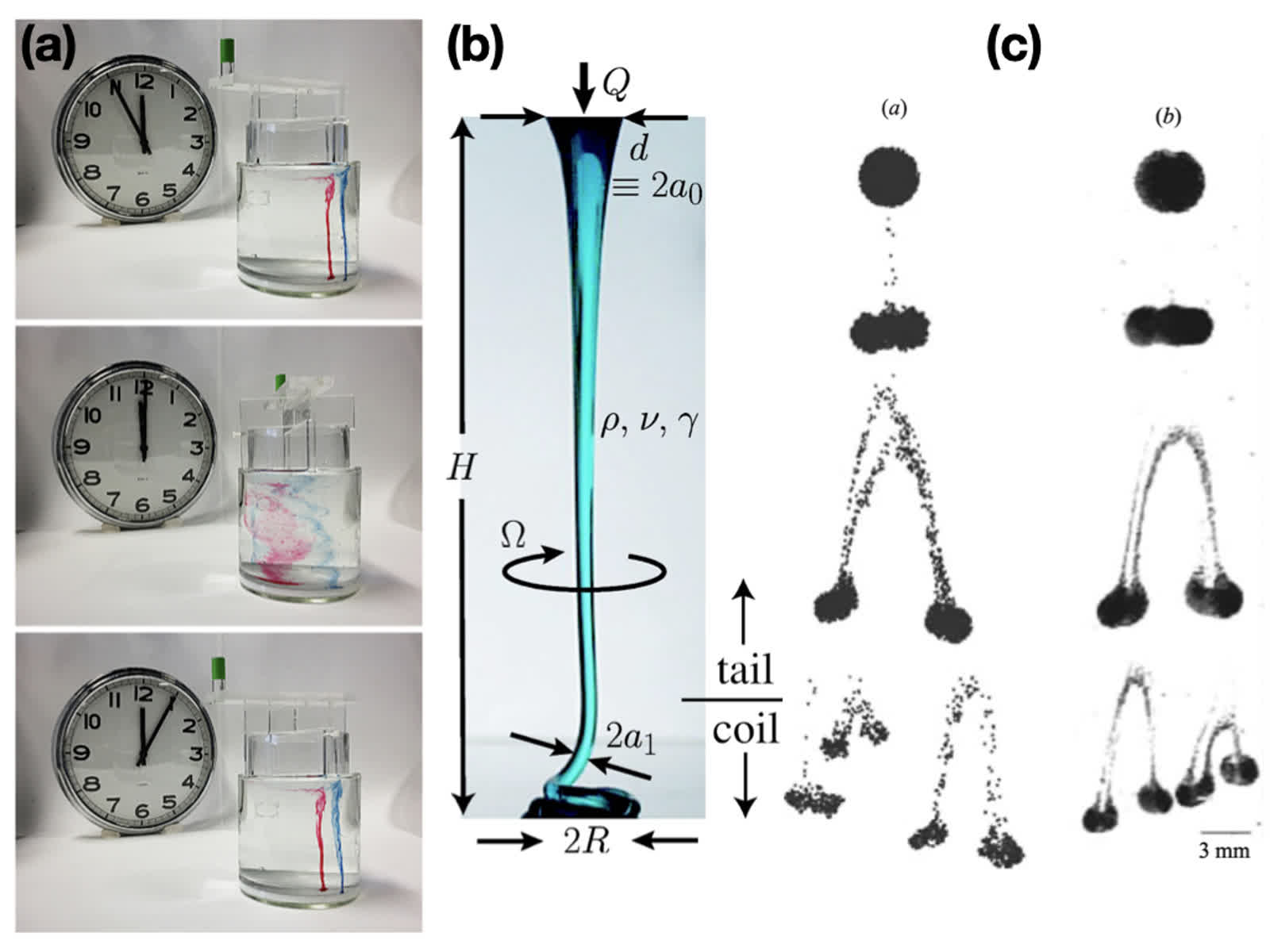}
        \caption{Examples of Stokes flows relevant to kitchen setting. 
        \textbf{(a)} Experimental demonstration of kinematic reversibility in an annular cylindrical space filled with silicone oil. Upon rotating the inner cylinder slowly a number of times and then reversing the forcing, the dyed fluid blobs remain unchanged, thus illustrating the difficulties in mixing viscous fluids. From Wikimedia Commons, licensed under CC BY 2.0. 
        \textbf{(b)} Coiling of a liquid rope made of a viscous fluid (corn syrup) demonstrates the complexities of free-surface gravity flows and also of pouring honey on the pancakes. From \citet{ribe2006stability}. 
        \textbf{(c)} Snapshots of the falling cloud: (left) in point-particle Stokesian dynamics simulation with 3000 particles and (right) in sedimentation experiments using \SI{70}{\micro\metre} glass beads in silicon oil. From \citet{metzger_nicolas_guazzelli_2007}. 
        }
    \label{fig:viscousflows}
\end{figure}

Viscosity shapes our notion of `thick' substances such as coconut oil or honey \cite{wray2020reduced}. 
This notion is hard to grasp, because when a liquid is cooled by just a few degrees, its viscosity can increase by a factor of a million.
Quoting the Nobel laureate Edward M. Purcell~\cite{purcell1977life},
\begin{quote}
    \emph{The viscosity of a fluid is a very tough nut to crack}
\end{quote}
Besides the nature of viscosity itself, the motion of thick liquids, often called `creeping' flow, can be equally puzzling.
A famous example of this is G.I.~Taylor's kinematic reversibility experiment, as depicted in Fig.~\ref{fig:viscousflows}a:
A mixing device with a cylindrical stirrer is filled with a viscous fluid containing dye streaks. 
When the stirrer is rotated in one direction, the dye streaks seem to mix with the fluid. 
However, the streaks are recovered when the stirrer is rotated back.
This experiment, shown explicitly in this video \cite{taylor1967video}, demonstrates that a time-reversible driving force will lead to time-reversible particle trajectories in Stokes flow. 
Another example is pouring honey or golden syrup on a pancake [Fig.~\ref{fig:viscousflows}b],
showing unexpected coiling dynamics because of the rich interplay between gravity, viscosity, and surface tension \cite{ribe2006stability}. 
Finally, as seen in Fig.~\ref{fig:viscousflows}c, the complex shapes of sedimenting clouds of food colouring added to a cocktail can be aptly described using many-body hydrodynamic interactions between microparticles \cite{zenit2018hydrodynamic, metzger_nicolas_guazzelli_2007}. 
Indeed, the theory of microhydrodynamics underlies the behaviour of most composite food products, such as emulsions, suspensions, and particle-laden fluid substances. 
Below, we will discuss how these viscous flows manifest themselves in a myriad of culinary aspects, and how they improve our understanding of science at a more fundamental level.

\subsection{Flows at low Reynolds number}
\label{subsec:LowReynolds}
\label{subsec:Stokeslet}

As discussed in Sec. \ref{subsec:NavierStokes}, flows dominated by viscosity have a very small Reynolds number, $\text{Re}=\rho U_0 L_0/\mu \ll 1$. This condition is satisfied either in highly viscous fluids, at very low velocities, or at small length scales.
For example, when humans swim in water at a velocity of $U_0 \approx \SI{1}{\metre\per\second}$ with a dynamic viscosity $\mu \approx \SI{0.9}{\milli\pascal \second}$ and density $\rho \approx \SI{e3}{\kilo\gram\per\metre\cubed}$, we find $\text{Re}\sim10^6$. This means that the fluid inertia is much more important than viscosity, and the flow is likely turbulent.  
For microbes ($L_0 \approx \SI{1}{\micro\metre}$) swimming in the same medium at typically $U_0 \approx \SI{10}{\micro\metre\per\second}$, however, we have $\text{Re}\sim10^{-5}$, so the viscosity is overwhelmingly dominant \cite{purcell1977life}. 
For such flows, the Navier-Stokes equations (Eqs.~\eqref{eq:NavierStokes}) reduce to the Stokes equations,\index{Stokes equations} 
    \begin{equation}
    \label{eq:StokesEqn}
    0 = - \vec{\nabla} p + \mu \nabla^2 \vec{u} + \vec{f}, 
    \quad \quad 
    \vec{\nabla} \cdot \vec{u} = 0.
    \end{equation}
These equations have several important properties. 
Firstly, they are linear, which makes them much easier to solve than the Navier-Stokes equations. This also means that, for given set of boundary conditions, there is only one unique solution \cite{kim2013microhydrodynamics}.
Secondly, the Stokes equations do not depend explicitly on time.
Viscosity-dominated flows will therefore respond essentially instantaneously to changes in the applied force, pressure, or the boundary conditions.
In other words, disturbances to the flow field spread much faster than the flow itself. 
Thirdly, Eqs.~\eqref{eq:StokesEqn} are kinematically reversible, i.e. they are invariant under the simultaneous reversion of the direction of forces and the direction of time. 
This means that if the forces driving the flow are reversed, the fluid particles retrace their trajectories in time, as seen in Fig.~\ref{fig:viscousflows}a, where a droplet of dye remains undeformed upon shearing it reversibly between two cylinders. 
As a consequence, it is notoriously difficult to mix fluids at low Reynolds number [see \S\ref{subsec:ChaoticAdvection}].
Apart from beautiful demonstrations of mixing and demixing under reversed forcing in famous video experiments by G.I. \citet{taylor1967video}, this issue bears great significance for microfluidic flows as $L_0 \approx \SI{1}{\micro\metre}$), which are becoming increasingly relevant for food science [see \S\ref{subsec:Emulsions} and \S\ref{subsec:microfluidicsInFoodScience}]. 


The fundamental solution to the Stokes equations, the Green's function, is called the Stokeslet. 
It is the flow $\vec{u}_{S}(\vec{x},t)$ at position $\vec{x}$ and time $t$ due to a point force, with the distribution $\vec{f}(\vec{x},\vec{y},t) = \delta(\vec{x} - \vec{y}) \vec{F}(t)$. The force has a time-dependent strength $\vec{F}(t)$ exerted on the liquid, and is located at position $\vec{y}(t)$, as expressed by the Dirac delta function. For an unbounded fluid, the boundary condition is $\vec{u}=0$ as $|\vec{x}| \to \infty$.
Physically, one could think of this point force as a small particle being dragged through the liquid, such as a sedimenting coffee grain [see \S\ref{subsec:Sedimentation}].
The flow generated by this point force is given by
    \begin{align}
    \vec{u}_{S}(\vec{x},t) 
    \label{eq:flowStokeslet}
    &= \mathcal{J}(\vec{x} - \vec{y}(t)) \cdot \vec{F}(t), 
    \end{align}
where the Oseen tensor $\mathcal{J}_{ij}(\vec{r})$ has Cartesian components
    \begin{align}
    \label{eq:oseenTensor}
    \mathcal{J}_{ij}(\vec{r}) &= 
    \frac{1}{8\pi\mu} \left( \frac{\delta_{ij}}{r} + \frac{r_i r_j}{r^3}\right),
    \end{align}
with indices $i,j \in \{1,2,3\}$, the relative distance is $\vec{r} = \vec{x} - \vec{y}$, and $r = |\boldsymbol{r}|$. 
There are different ways to derive this tensor, as summarised by \citet{lisicki2013four}.

This Stokeslet solution is powerful, analogous to Coulomb's law for electric point charges. 
Like the Stokes equations, it also reveals important properties of viscous flows.
First, Eq.~\eqref{eq:flowStokeslet} shows that the applied force is directly proportional to the flow velocity.
As opposed to Newtonian mechanics, where the forces are proportional to acceleration, this reflects Aristotelian mechanics \cite{van2016aristotelian}, where there is no motion in the absence of forces.
Inertia vanishes at low Reynolds number, that is, so the dynamics are overdamped.
Second, hydrodynamic interactions are very long ranged.
The Stokeslet flow decays as $1/r$ with distance, as opposed to gravitation or electrostatics that both follow inverse-square laws.
This has interesting consequences in the kitchen.
For example, in a particle suspension such as sedimenting coffee grains, the motion of one grain will produce a flow that moves other grains, which in turn generate flows that affect the first grain again [see \S\ref{subsec:Suspensions}].


\index{Stokes law}
The fundamental solution can also be used to derive the widely celebrated Stokes law, 
    \begin{align}
    \label{eq:StokesLaw}
        \vec{F} = 6 \pi \mu a \vec{U},
    \end{align}
which describes the viscous drag force that a sphere of radius $a$ moving with velocity $\vec{U}$ exerts on a viscous fluid \cite{stokes1851effect, batchelor2000introduction, yeomans2014introduction,guazzelli2011physical}. The significance of Stokes' law can hardly be overemphasized.
\citet{dusenbery2009living} writes that it is directly connected to at least three Nobel Prizes. 
Can you name them? 

\subsection{Coffee grounds in free fall: Sedimentation}
\label{subsec:Sedimentation}

In the words of \citet{pendergrast2010uncommon}, \textit{`Possibly the cradle of mankind, the ancient land of Abyssinia, now called Ethiopia, is the birthplace of coffee'}.
Thus, before discussing coffee brewing further in \S\ref{subsec:brewingCoffee}, we analyse here the oldest method of preparation: Sedimentation of coffee grounds under gravity. 
To optimise our daily cup, we may want to assess the size distribution of the coffee particles after grinding.
This can be achieved using the Stokes law we just discussed in \S\ref{subsec:LowReynolds}. 
The gravitational force, $\vec{F}_{g} = m\vec{g}$, pulling on a sphere of radius $a$ is proportional to the mass differential with the surrounding fluid, $m=\left(\rho_p-\rho \right) \frac{4}{3}\pi a^3$. The same grain is slowed by a viscous drag force, given by Eq.~\eqref{eq:StokesLaw}. When the drag force balances the gravitational force, the terminal velocity is 
\begin{equation}
    U_\infty = \frac{2}{9}\frac{a^2\left(\rho_p - \rho\right)g}{\mu}.
    \label{eq:terminalVelocity}
\end{equation}
To measure $U_\infty$ in the kitchen, we can use a mobile phone to videotape individual grains sedimenting. Eq.~\eqref{eq:terminalVelocity} can be rearranged to solve for the particle size $a$. 
Conversely, it is possible to solve for $\mu$ if $a$ is known, which is the principle of falling sphere viscometry \cite{sutterby1973falling}. 
Not least, \citet{millikan1913elementary} used a form of Eq.~\eqref{eq:terminalVelocity} to find the elementary electrical charge.
Note that the above equation was developed under several assumptions, and in the paragraphs to follow we assess the validity of these.

The first assumption is that the grain sediments at low Reynolds number [see \S\ref{subsec:NavierStokes}], given by $\text{Re} = aU_\infty\left(\rho_p-\rho \right)/\mu$. Using Eq.~\eqref{eq:terminalVelocity}, we find that a typical grain of radius $a=\SI{1}{\milli\metre}$ sediments at $\SI{0.1}{\milli\metre \per \second}$, yielding a Reynolds number just below order unity. We should therefore be cautious when using Stokes' law in this case \cite{arnold1911limitations}. In order to obtain the range of Reynolds numbers over which Eq.~\eqref{eq:terminalVelocity} gives a good approximation, one can turn to experiments. Numerous reports have been published on this topic \cite{ruby1933settling, flemmer1986drag}, and it is generally agreed that Eq.~\eqref{eq:terminalVelocity} gives good estimates up to $\text{Re}\approx 1$ \cite{guyon2001physical}. 

The second assumption is that the grain is spherical and smooth, as shape irregularities and surface heterogeneities increase the surface area, which normally increases the drag force. And since a coffee grain is neither smooth nor spherical, one would think that it falls slower than a Stokes sphere having the same density and volume. However, in the low Re limit, it turns out that the drag force is relatively insensitive to shape \cite{rubey1933settling}, and Eq.~\eqref{eq:terminalVelocity} is thus likely to approximate $U_\infty$ well even for irregular coffee grains. 
The boundary conditions will change of course if the particle is liquid instead of solid, as discussed in \S\ref{subsec:immiscibledropsdynamics}.


The sedimenting coffee grain generates a flow field that decays slowly with distance from its center [see \S\ref{subsec:Stokeslet}]. Close to the particle, viscous forces dominate, while in the far field the inertial terms dominate. It is the absence of walls that facilitates this shift from viscous to inertial dominance, as walls provide friction to the flow. By accounting for inertial effects, considering an unbounded fluid, \citet{oseen1910stokes} developed the following improved formula for the drag coefficient, $C_{D} = F/p_{dyn}A$,
    \begin{equation}
    C_{D} = \frac{24}{\text{Re}}\left( 1+\frac{3}{16} \text{Re} \right),
    \label{eq:OseenDrag}
    \end{equation}
where $p_{dyn} = 1/2\rho U_\infty^2$ is the dynamic pressure and $A$ the cross-sectional area of the sedimenting particle. The first term in Eq.~\eqref{eq:OseenDrag} is the Stokes drag close to the particle, while the second term stems from inertial effects far away from the particle. Oseen's formula agrees fairly well with experiments up to $\text{Re} \approx 10$ \cite{dey2019terminal}.

Beyond the Oseen regime, inertial effects in the flow surrounding the particle can no longer be neglected. This causes the nearby fluid streamlines to divert from the particle, causing the flow to separate. Then the pressure drop across the particle is reduced, which leads to a drag reduction. A simple experiment using coffee grains released in air (instead of water) can be performed to observe this behavior. By accounting for inertial effects around a sedimenting sphere, \citet{stewartson1956xxxii} found the drag coefficient to be approximately $1.06$, which may be compared to the value of $7.2$ using Eq.~\eqref{eq:OseenDrag} for $Re=10$. For more details, we refer the interested reader to the excellent review paper on Stokes' law, and its legacy, by \citet{dey2019terminal}. 

\emph{Wall effects:} We now return to the assessment of the validity of Stokes' law for sedimenting coffee grains in bounded water. In developing Eq.~\eqref{eq:terminalVelocity}, we assumed that the grain falls without influences of walls, and this is our third assumption. However, building on pionnering works of \citet{Lorentz1907} and \citet{Faxen1923},  \citet{oNeill1964slow} showed that hydrodynamic contributions due to walls can slow down a sedimenting grain (or a rising bubble) by as much as $5\%$ when the distance to the walls is ten times its size. The degree of retardation increases linearly as the grain gets closer to the vessel walls. Using matched asymptotic expansions, \citet{goldman1967slow} obtained a solution for the drag force acting on a sedimenting sphere moving parallel to a wall,
    \begin{equation}
        \frac{F_{\parallel\text{W}}}{F_{S}}=1-\frac{9}{16}\frac{a}{h} + \frac{1}{8}\left(\frac{a}{h}\right)^3 - \frac{45}{256}\left(\frac{a}{h}\right)^4 - \frac{1}{16}\left(\frac{a}{h}\right)^5, 
        \label{eq:modifiedStokesEqn}
    \end{equation}
which is normalized by the Stokes drag force [Eq.~\eqref{eq:StokesLaw}]. Thorough experiments have been performed to validate the above result, showing good agreement \cite{brown2003sphereWallEffects}.

For spheres very close to a wall, lubrication effects become important [\S\ref{subsec:lubrication}], yielding a logarithmic relationship between the force and the distance to the wall \cite{cichocki1998image}. This strong logarithmic dependence can be readily observed when making French press coffee: grains close to the vessel container sediment much slower than grains out in the bulk. Recently, \citet{rad2020flat} revisited this problem in non-Newtonian liquids, which has important biophysical implications. 

\emph{Collective effects:}
Our final assessment of Eq.~\eqref{eq:terminalVelocity} concerns the possible influence of multiple particles correlated by long-ranged hydrodynamic interactions [see \S\ref{subsec:Stokeslet}], and as Eq.~\eqref{eq:StokesLaw} does not include such effects, our fourth and final assumption is that these can be neglected. However, when two spheres sediment side-by-side, they fall slower than in the absence of the other particle, while if they are separated by a vertical line going through their centers, the opposite is true \cite{guazzelli2011physical, zenit2018hydrodynamic}. In a suspension containing several grains, the influence of other particles always leads to a decrease in sedimentation velocity. This observation is known as hindered settling\index{hindered settling} \cite{richardson1997sedimentation} and is mainly due to an upward flow generated by each particle as it sediments. In a dilute suspension, the hindered settling velocity depends on the particle concentration as $U \approx U_\infty\left(1-6.55\phi\right)$, as shown by \citet{batchelor1972sedimentation}. Other effects such as Brownian motion \cite{lin2019brownian} and shape \cite{ruby1933settling} can also affect the sedimentation velocity and introduce memory effects \cite{szymczak2004memory}. 
Finally, if particles sediment towards a surface covered with moving actuators, a self-cleaning effect can occur where hydrodynamic fluctuations repel the particles, leading to a non-Boltzmannian sedimentation profile \cite{guzman2021active}.
An important topic for future research is how such non-equilibrium distributions can evolve dynamically in active and living systems \cite{gompper20202020}.

\subsection{Pot stuck to stove top: Stefan adhesion and lubrication theory}
\label{subsec:viscousAdhesion}
\label{subsec:lubrication}

If a pot of pasta overboils, a common subsequent problem is that the pan is ``stuck'' to the surface.
This effect does not require any glue or the formation of molecular bonds. Instead, it stems from the viscous liquid film that is sandwiched between the objects, which requires a large force to be displaced.\index{Stefan adhesion}
\citet{stefan1874versuche} first described this ``apparent adhesion'', and later \citet{reynolds1886lubricationtheory} quantified it with a detailed treatise on lubrication theory. 

It is important to notice a separation of length scales:
The thickness $h$ of the fluid layer is much smaller than the radius $R$ of the pot.
Thus, we can define a small parameter, $\varepsilon=h/R \ll1$.
In a simplified 2D system, where the directions $z$ and $x$ are perpendicular and parallel to the substrate, we can expand the Stokes equations to leading order in this parameter, giving
    \begin{equation}
        \label{eq:lubricationEquations}
        \frac{\partial p}{\partial z} =0, \qquad  \frac{\partial p}{\partial x} = \mu \frac{\partial^2 u_x}{\partial z^2}.
    \end{equation}
The first expression tells us that the pressure changes little with height in the thin gap above the substrate, and the second expression says that the pressure change along the substrate is related to the flow variation across the gap. 
The general three-dimensional case of these expressions are referred to as the Reynolds equations\index{Reynolds equation}, as reviewed by \citet{oron1997long, batchelor2000introduction, szeri2010fluid}.
These equations may be solved analytically or numerically for a range of different applications.

For a cylindrical pot stuck to the surface, the Reynolds equations can be used to show that the force required to lift the pot, the Stefan adhesion force, is given by
\begin{equation}
    \label{eq:StefanAdhesionForce}
    F = \frac{3 \pi \mu R^4}{2h^3} \frac{dh}{dt}.
\end{equation}
Because it is hard to squeeze a viscous liquid through a narrow gap, this force can be very large for thin films.
For example, we estimate that for a pan of radius $R\sim\SI{12}{\centi\metre}$, film thickness $h\sim\SI{10}{\micro\metre}$, and separation speed $\frac{dh}{dt}=\SI{1}{\milli\metre \per \second}$, the force is $F \sim \SI{e6}{\newton}$, which is orders of magnitude larger than the weight of the pan filled with water.
Conversely, it is very difficult to bring two surfaces into close contact. 
This is further described by squeeze flow theory, with many generalisations for viscoelastic fluids \cite{engmann2005squeeze}. 

Indeed, lubrication theory is used ubiquitously whenever one system dimension is significantly smaller than the others.
In tribology, it is crucial for reducing wear and friction between bearings \cite{khonsari2017applied}.
In biology, tree frogs can climb vertical walls using liquid film flows, which has inspired new tire technology \cite{barnes1999tree}, and biomimetic materials that adhere under wet conditions \cite{barnes2007biomimetic, meng2019tree}. 
Thin films significantly alter the motion of trapped microorganisms \cite{mathijssen2016hydro}, governing their surface accumulation. 
Note that many biological fluids are viscoelastic [see \S\ref{subsec:FoodRheology}], which changes the Reynolds equations, but the basic concepts still hold.
Using ferrofluids, adhesion can also be made switchable for use smart adaptable materials \cite{wang2018multifunctional}.
Recent developments on the nanoscale physics can improve lubricant design \cite{xu2018squeezing}, and enhance painting and coating flows \cite{ruschak1985coating}.
Additionally, lubrication theory has been used to model air hockey \cite{weidman2015steady} and, in the spirit of kitchen experiments, Reynolds also used it to determine the viscosity of olive oil \cite{reynolds1886lubricationtheory}. 
Pan lubrication is also an important step in industrial baking \cite{cruz}. 
Finally, the lubrication theory can also be used to model hand washing, as explained in \S\Ref{subsec:dryingHands}.


\subsection{Making perfect cr\^epes: Viscous gravity currents} 
\label{subsec:ViscousGravityCurrents}

\textit{``Open the front door of a centrally-heated house and a gravity current\index{gravity current} of cold air immediately flows in.''}
These opening words by \citet{huppert1982propagation} describe many processes in the kitchen: 
Opening the fridge or the oven door, but also the spreading of oil in a frying pan.
To understand how long this spreading takes, we need to determine the evolution of the height profile, $h(\vec{r}, t)$, which strongly depends on the dynamic viscosity and density of the oil, $\mu$ and $\rho$, gravity, $g$, and the pouring rate, $Q$. 

The presence of inertia makes predictions of the radial spreading velocity difficult. 
If we instead pour the oil really slowly, at low Reynolds number, assuming that both the oil layer thickness $h$ and the `jet' radius $R_j$ are small compared to the current $R$, then we can use a lubrication approximation [see \S\ref{subsec:lubrication}] to obtain a simplified version of the radial force balance:
\begin{equation}
\frac{\partial h}{\partial t}-\frac{1}{3}\frac{\rho g}{\mu}\frac{1}{r}\frac{\partial}{\partial r}\left(rh^3\frac{\partial h}{\partial r} \right)=0.
\label{eq:lubricationEqn}
\end{equation}
By adding a mass conservation equation to Eq.~\eqref{eq:lubricationEqn}, and by introducing a similarity variable, Huppert showed that the radial extent of the evolving puddle, $R$, is given exactly by:
\begin{equation}
    R = 0.715\left(\frac{\rho g Q^3}{3 \mu}\right)^{1/8}t^{1/2}.
    \label{eq:HuppertViscousGravityCurrent}
\end{equation}

The dominant balance of forces in Huppert's analysis is between the hydrostatic pressure head that drives the fluid, and viscous stresses that slow the fluid down. 
Huppert discarded effects of surface tension, $\gamma$, and Eq.~\eqref{eq:HuppertViscousGravityCurrent} requires that the Bond number be large, $\text{Bo}=\rho R^2/\gamma \gg 1$ [see \S\ref{subsec:HangingDrop}]. 
Many geophysical flows are characterized by large Bond numbers, and the scalings by Huppert have seen widespread use for predicting spreading rates of saltwater currents into freshwater, lava flows, and many other geophysical gravity currents \cite{huppert2006gravity, craster2009dynamics, meiburg2010turbidity}. The success of these scalings in $\emph{miscible}$ fluids might seem surprising. However, geophysical flows are often characterized by high P\'eclet numbers [Eq.~\eqref{eq:PecletNumber}], such that the transport of momentum outpaces the transport of mass [see \S\ref{subsec:heatEquation}]. On the fast time scale of the flow, diffusion does not have sufficient time to blur the interface separating miscible liquids, resulting in the liquids displaying immiscible behavior. Recently, Eq.~\eqref{eq:HuppertViscousGravityCurrent} was found to accurately describe the spreading rate of miscible sessile drops of corn syrup and glycerol in water \cite{walls2018shape}. Other important contributions to the study of gravity currents include the spreading of French vinaigrette \cite{benabdelhalim2022phase}, the spreading of a saltwater current under a bath of freshwater \cite{didden1982viscous}, spreading hot plumes in  cold environments \cite{britter1979spread}, and oil spreading on the sea \cite{hoult1972oil}. These and other works are described in detail in a number of review papers \cite{simpson1982gravity,huppert2006gravity} and in the book by \citet{ungarish2009introduction}.


\emph{Making the perfect cr\^epe:}\label{subsec:CrepeMaking} 
When pouring pancake batter into a frying pan, it is tempting to speed up the spreading by holding the pan inclined. 
However, viscous currents down a slope can be unstable, as shown in another famous paper by \citet{huppert1982flow}.
An initially uniform propagation front can break up into long fingers, leading to undesired stripes instead of a uniform cr\^epe.
%
Another problem arises because the pan is heated, so the spreading problem becomes more complicated as the viscosity is a non-linear function of time. Initially, a temperature rise is associated with a viscosity \emph{decrease}, but later in the spreading process, the batter starts to solidify, leading to a viscosity \emph{increase}. Since the current continually loses momentum as it spreads, the solidification can arrest the flow long before it extends the entire pan.
Motivated by the aim of making a perfectly flat and uniform cr\^epe, \citet{boujo2019pancake} recently approached the spreading problem with both time-dependent viscosity and gravity. Armed with numerical tools, they identified different swirling modes and measured the resulting pancake shape. Interestingly, a swirling mode that is naturally adopted in pancake making, namely draining all the batter in one place and then rotating the batter around the perimeter of the pan in one big swirling motion, appeared to optimise the pancake shape. To learn more about the rheology of pancake making and other cooking processes, see \S\ref{subsec:FoodRheology}.

\subsection{Microbial fluid mechanics}
\label{subsec:MicroSwimmers}

\begin{figure}[t]
    \includegraphics[width=\linewidth]{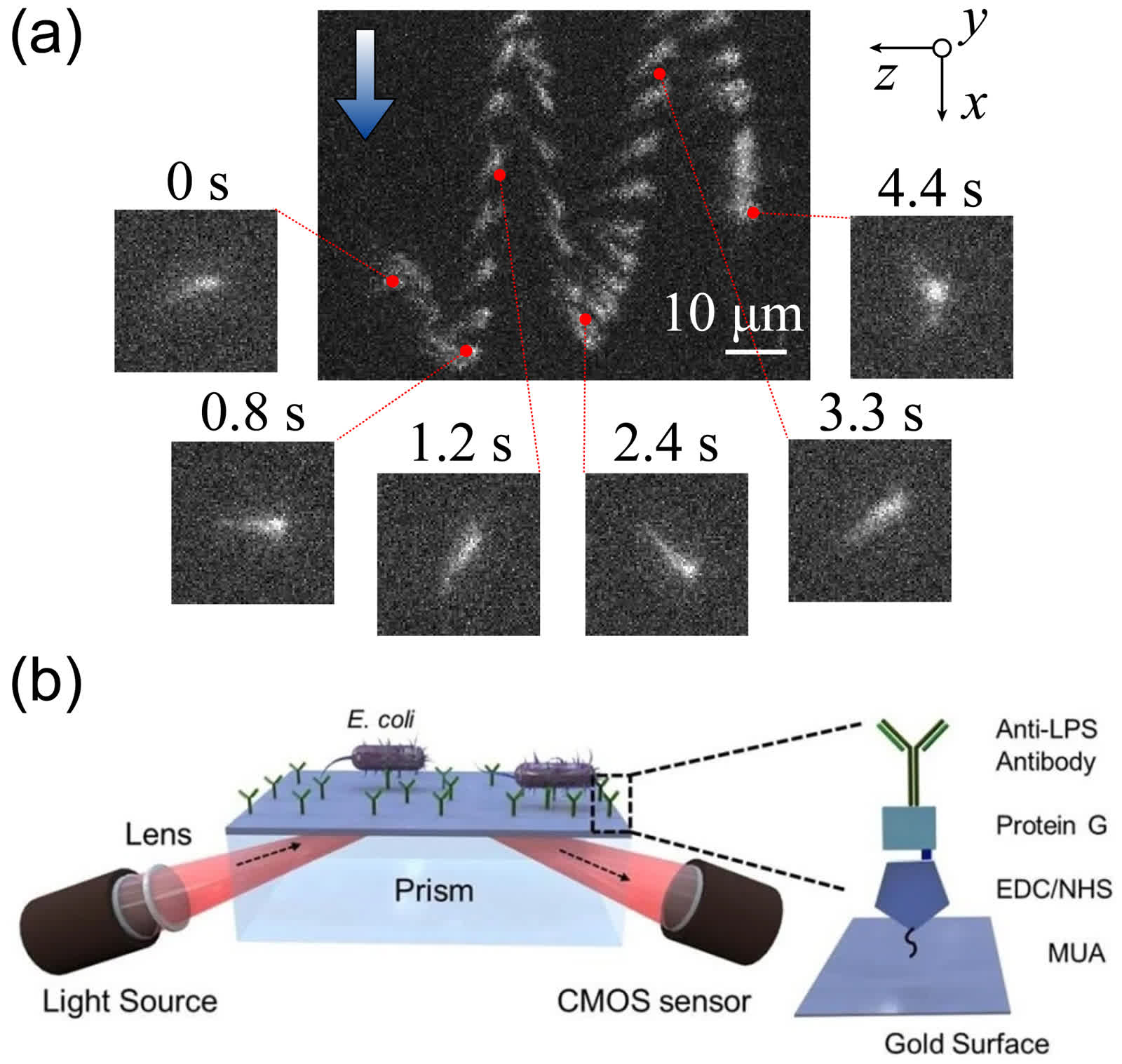}
    \caption{Microbial dynamics and food safety.
    \textbf{(a)} Time lapse of an \textit{E. coli} bacterium performing oscillatory rheotaxis, with fluorescently labelled flagella to reveal its reorientation with respect to the flow (blue arrow). From \citet{mathijssen2018oscillatory}.
    \textbf{(b)} Working principle of an optofluidic pathogen detector. When \textit{E. coli} bind to Y-shaped monoclonal antibodies, it induces a shift in refractive index which can be detected by an optical sensor. From \citet{tokel2015portable}. 
    }
    \label{fig:microbial}
\end{figure}

Microbes play an important role all across Eastern and Western cuisines \cite{tamang2020fermented}.
On the one hand, their presence can improve texture and taste of food, such as sourdough bread \cite{arendt2007impact}, yoghurt, and many fermented products \cite{csanlier2019health}. 
On the other hand, their growth and development can cause illness \cite{chitlapilly2020impact}. 
It is thus crucially important to understand the mechanisms of their growth, interactions, and locomotion to control their influence on food products and the living systems they inhabit \cite{doyle2019food}. 

Because of their size, the hydrodynamics of microorganisms is governed by the Stokes equations [see \S\ref{subsec:LowReynolds}].
As such, they experience the remarkable and profound consequences of living at low Reynolds number \cite{purcell1977life}. 
Unlike us, cells cannot use inertia to coast forward after each swimming stroke; viscous dissipation slows them down almost instantly.
Furthermore, microswimmers must obey the ``scallop theorem'', stating that a time-reversible swimming gait cannot lead to a net displacement \cite{elgeti2015physics, lauga_2020}. 
That is, a scallop would not be able to swim in viscous fluids; while closing and opening its shell, it would just move back and forth.
Thus, microorganisms must employ more sophisticated mechanisms in order to propel themselves. 

Most bacteria swim using helical flagella driven by a rotary motor, while eukaryotic swimmers often use whip-like beating organelles called cilia \cite{brennen1977fluid, lauga2009hydrodynamics, gilpin2020multiscale}.
Both these flagella and cilia can be modelled as slender filaments that exert viscous drag forces on the liquid.
To understand this better, we can consider a generalisation of the Stokes law [Eq.~\eqref{eq:StokesLaw}] for a cylinder of length $\ell$ and radius $a$, with $\ell \gg a$.
When it moves through the fluid sideways, perpendicular to its major axis, it will exert a viscous drag force
    \begin{align}
    \label{eq:StokesLawCylinderPar}
        F_\perp \approx \frac{4 \pi \mu \ell U}{\text{ln}(\ell/a)},
    \end{align}
as predicted by slender-body theory\index{slender-body theory} \cite{gray1955propulsion, batchelor1970slender, cox1970slender,johnson1980improved}.
Note, this expression depends only weakly on the filament radius, which is on the order of $a\approx \SI{10}{\nano\metre}$.
Thus, a typical cilium of length $\ell \approx \SI{10}{\micro\metre}$ and velocity $U \approx \SI{100}{\micro\metre \per \second}$ can exert a force $F_\perp \approx \SI{1}{\pico\newton}$. 
While this force is relatively large for a small cell, it is of little use if the same force is exerted forwards and backwards during periodic beating cycles. 
However, because of the structure of the Oseen tensor\index{Oseen tensor}, $\mathcal{J}_{ij}$ given by Eq.~\eqref{eq:oseenTensor}, the cylinder will exert a smaller drag force on the fluid when it moves parallel to its axis,
    \begin{align}
    \label{eq:StokesLawCylinderPerp}
        F_\parallel \approx F_\perp / 2.
    \end{align}
Crucially, both cilia and flagella rely on this fundamental principle.
A cilium can exert more force on the liquid during its perpendicular ``power stroke'' and less force during its parallel ``recovery stroke''.
Similarly, a bacterial flagellum can force liquid backwards because each rotating helix element moves with velocity components parallel and perpendicular to the filament \cite{lauga2016bacterial}.
This result [Eq.~\eqref{eq:StokesLawCylinderPerp}] lies at the heart of numerous transport processes for viscous fluids.
Indeed, numerous synthetic swimmers and microrobots are based on this hydrodynamic anisotropy \cite{bechinger2016active}.
Moreover, humans also use cilia to remove pathogens from our lungs \cite{ramirez2019multi} and to transport liquid in our brain ventricles \cite{faubel2016cilia}.

Microbes cannot rely on external forces to swim: 
They are force free, such that their propulsion and drag forces balance.
Therefore, instead of a Stokes monopole [Eq.~\eqref{eq:flowStokeslet}], the flows they generate are described by Stokes dipoles and higher multipoles \cite{mathijssen2015tracer}.
These flows can be used to model their long-ranged interactions with each other and with their environment \cite{wensink2012meso, elgeti2013emergence, lauga_2020}.
Cell motility takes on additional complexity because they often reside in viscoelastic fluids \cite{sznitman2015locomotion, spagnolie2022swimming}.

Because the hydrodynamics of microorganisms has been studied so extensive recently, it is now possible to connect it with food science in terms of infectious disease transmission \cite{mittal2020flow}, bacterial contamination by upstream swimming \cite{mathijssen2015upstream}, and the microbiology of bacterial coexistence \cite{gude2020bacterial}.
For instance, Fig.~\ref{fig:microbial}a shows how \textit{E. coli} bacteria can reorient with respect to externally imposed rinsing flows, a phenomenon called rheotaxis \cite{mathijssen2018oscillatory}.
Moreover, recent engineering improvements have lead to sensors capable of detecting harmful microbes. 
For example, a porous silk microneedle array can be used to sense the presence of {\it E. coli} in fish fillets \cite{kim2021microneedle}. 
In general, a wide spectrum of potential pathogens stimulates future research on food safety \cite{ali2020application}, as we discuss next.

\subsection{Microfluidics for improved food safety}
\label{subsec:microfluidicsInFoodScience}


Foodborne pathogens such as \emph{E. coli} and \emph{Salmonella} bacteria cause approximately 420,000 deaths and 600 million illnesses yearly \cite{world2015estimates}. Such pathogens often originate from bacterial biofilms in food processing plants \cite{vidakovic2018dynamic, mathijssen2018nutrient} and poor hygiene [\S\ref{subsec:dryingHands}, \S\ref{subsec:SoapFilmDynamics}]. Therefore, on-site rapid detection is necessary to prevent these harmful bacteria from entering our grocery stores and ending up in foods. The traditional way of detecting pathogens is by cultivating them in Petri dishes, but slow bacterial growth limits the usefulness of such `babysitting' in food plants. This has led to alternatives such as nucleic-acid based methods (including polymerase chain reactions (PCR) \cite{kant2018microfluidic}, which is the primary method to test patients for COVID-19), but extensive training requirements, expensive equipment and labor intensive steps prevent a robust and efficient implementation in food production facilities.  

Fortunately, microfluidic-based biosensors can detect or ``sense'' pathogens on much faster timescales thanks to small volumes and flow-mediated transport, with high-speed imaging enabling real-time monitoring \cite{skurtys2008applications, mairhofer2009microfluidic, he2020application}. Biosensors work by measuring an electrical or optical signal induced by a chemical reaction as a target molecule (pathogen) binds to a bioreceptor molecule \cite{karunakaran2015introduction, van2010recent}. The bioreceptors are designed to exactly match the surface elements of the pathogen (these elements are called antigens) like a lock and key fit. Due to their excellent specificity, monoclonal antibodies (mAbs) are the most widely used bioreceptor molecules. Figure \ref{fig:microbial}b shows the working principle a biosensor functionalized with Y-shaped mAbs molecules for detecting \textit{E. coli} \cite{tokel2015portable}.

The speed and accuracy of surface-based biosensors depend on a number of factors where the most important are the pathogen concentration, the number of available binding sites on the sensor,  the transport of pathogens from the bulk fluid to the sensor via flow and diffusion, and finally, the kinetics of the binding reaction. Without flow, the sensing time is usually limited by the time it takes for a pathogen or a virus to diffuse to the sensor, which can take several hours in a microchannel due to the low diffusivity of such large particles. However, by leveraging microfluidic flow, solute transport can be sped up by several orders of magnitude, leading to reaction-limited kinetics. We recommend the pedagogical reviews on transport and reaction kinetics in surface based biosensors by \citet{gervais2006mass}, \citet{squires2008making} and \citet{sathish2021toward}.

 

\subsection{Ice creams}
\label{subsec:IceCream}

Although first mentions of this classic dessert as 'flavoured snow or ice' date back to the Persian and Roman Empires, evidence suggests that dairy iced products originate from 12th century China \cite{marshall2003ice}. From a chemical point of view, ice cream is an emulsion [see \S\ref{subsec:Emulsions}] made with water, ice, milk fat and protein, sugar and air. The ingredients are mixed together and turned into foam upon the addition of air bubbles. The colloidal emulsion is then frozen to preserve the metastable mixture. The details of the process have been extensively studied within the food science community \cite{goff2013ice,arbuckle2013ice}, but also from the physics and general science viewpoint \cite{clarke2003physics,clarke2015science}. Special attention has been paid to the colloidal character of the emulsion \cite{goff1997colloidal}. 

An appealing texture and rheology are crucial aspects of ice cream quality. Due to its popularity, ice cream was apparently the first food product to have its extensional viscosity measured, as early as in 1934 \cite{leighton1934apparent}. Since then, rheological properties of ice cream have been examined in detail, and various dynamical models have been proposed to account for their behaviour \cite{martin2008rheology}. The breadth of related topics has even inspired an interdisciplinary undergraduate course taught within the physics programme \cite{trout2019science}.

The production of ice cream involves a flow undergoing a structural and phase transition by a combination of mechanical processing and freezing. The dynamics of ice crystallisation therein are not fully understood. Because an ice cream mix is opaque, \emph{in situ} crystallisation has not been observed and its mechanism is debated \cite{cook2010mechanisms}. In a typical ice cream freezer, ice is formed on externally cooled walls by surface nucleation and growth, and is then scraped off to the bulk fluid, where secondary nucleation and ripening take place \cite{hartel1996ice}. The number and size of ice crystals formed is also heavily dependent on the mixture composition, and also affects the melting rate and hardness of the final food product \cite{muse2004ice}. Studying the characteristics of ice cream production leads to the development of novel methods for rapid freezing and thawing of foods \cite{li2002novel}. The structure of ice cream can be practically controlled in most common conditions. However, environments with large temperature fluctuations remain problematic, so developing a detailed description of the underlying processes remains an interesting pathway for future research.

Interestingly, as opposed to frozen ice cream, one sometimes sees `hot ice cream' that seems to melt upon cooling. This effect can be achieved using methyl cellulose, a thickener that acts in high temperatures to produce a product with surprising, `opposite' melting properties \cite{hoticecream}. Perhaps not the obvious choice on a hot summer day, but definitely worth a taste!


\section{Coffee: Granular Matter \& Porous Media}
\label{sec:CoffeeSugar}

The Hungarian probability theorist Alfr\'ed R\'enyi (1921-1970) once said \cite{suzuki2002history},
    \begin{quote}
        \emph{A mathematician is a device for turning coffee into theorems.}
    \end{quote}
Coffee is arguably one of the most popular beverages world-wide, and many of us look forward to our cup of joy several times a day! It is not surprising that we are fascinated with the fluid dynamics of coffee -- the focus of this section. We start with the flow of coffee beans and other granular materials, including avalanches, hoppers, and the Brazil nut effect. We then consider brewing coffee using different methods in the context of porous media flows and percolation theory, and we finish with the illustrious coffee ring effect. In the words of \citet{wettlaufer2011universe}, there is a `\textit{universe in a cup of coffee}'.

\subsection{Granular flows and avalanches}
\label{subsec:AvalancheDynamics}

Granular materials\index{granular matter} \cite{gennes1999granular} are found everywhere in the kitchen. Take, for example, flour, rice, nuts, coffee beans, sugar or salt.
Indeed, the food industry processes billions of kilograms of granular material every year \cite{gray2018particle}.
They are composed of discrete, solid particles (grains) over a wide range of sizes. 
Therefore, the grain diameter $D$ is often denoted on the Krumbein phi scale, $\varphi = - \log_2(D/D_0)$, with $D_0 = \SI{1}{\milli\metre}$ \cite{krumbein1934size}\index{Krumbein phi scale}. 
For example, $\varphi=-6$ and $6$ correspond to oranges and powdered sugar, respectively.

Granular matter can have many surprising properties \cite{herminghaus2005dynamics, mehta2012granular}. 
One such example is the Janssen effect\index{Janssen effect}: The hydrostatic pressure at the bottom of a cylindrical container does not grow linearly with the filling height of grains, unlike in a liquid [\S\ref{subsec:Eureka}]. Instead, the pressure saturates exponentially to a value much less than the weight of the grains,
because they are partially supported by the vertical silo walls due to friction forces \cite{aguirre2010pressure}. 
Another example is that granular matter can behave like solids, liquids, or even gases, depending on the amount of kinetic energy per grain \cite{lun1984kinetic, jaeger1996granular, forterre2008flows}. 
For this reason, granular flows are challenging to analyze and predict.

\begin{figure}[t]
    \includegraphics[width=\linewidth]{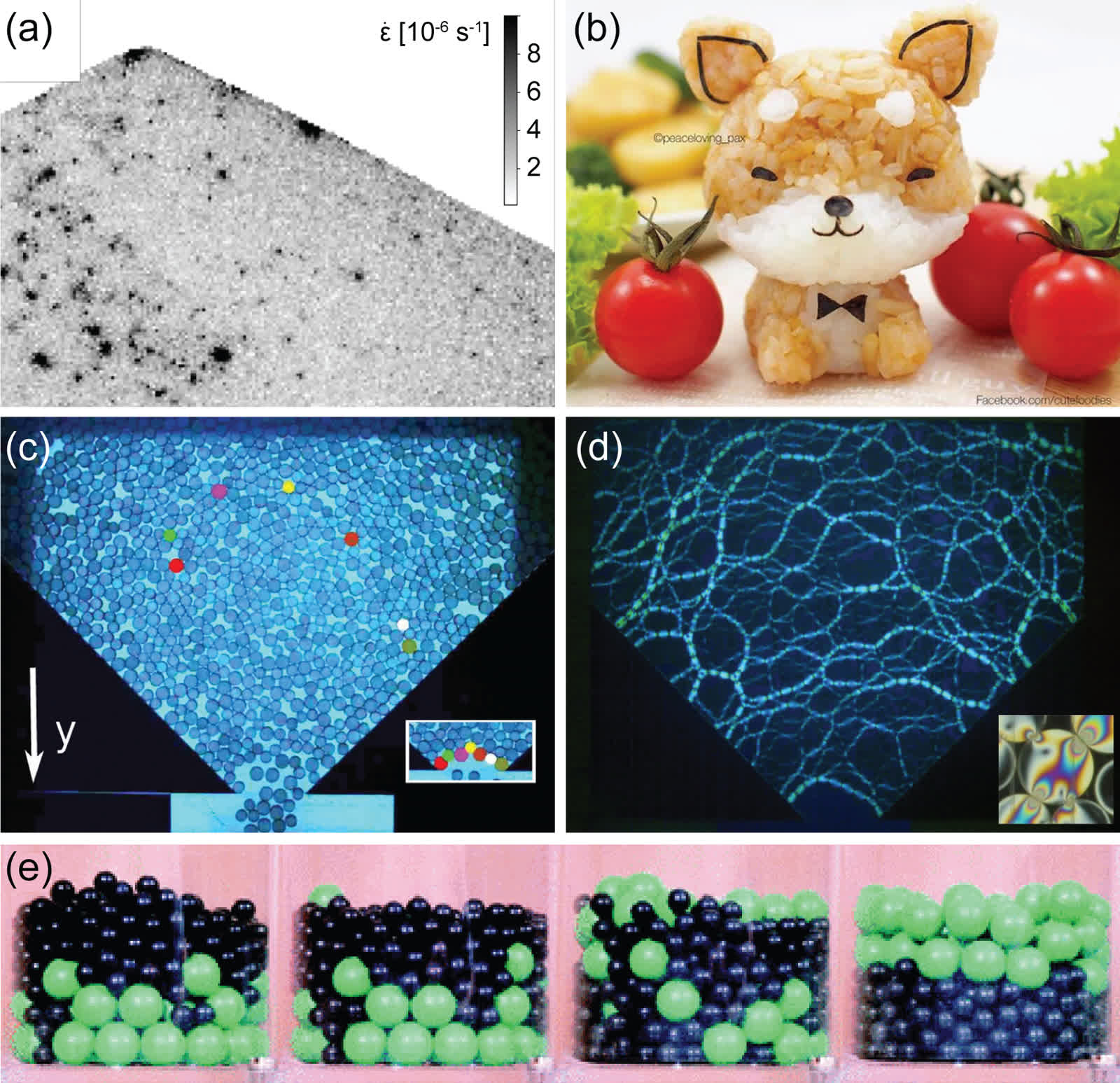}
    \caption{Granular matter.
    \textbf{(a)} In a pile of grains, the motion is quantified by the rate of strain $\dot{\varepsilon}$ using spatially-resolved diffusing wave spectroscopy (DWS). From \citet{deshpande2020perpetual}.
    \textbf{(b)} Rice sculpture by doctor and food artist \citet{peacelovingpax}. 
    \textbf{(c)} The velocity field of grains flowing in a 2D hopper is measured by direct particle tracking. The seven particles that are marked in colour form a jamming arch, shown in the inset. 
    \textbf{(d)} Force chains in a jammed state, visualised by photoelastic particles that show intensity fringes under stress, highlighted in the inset.
    From \citet{tang2011granular}. 
    \textbf{(e)} The Brazil nut effect, with \SI{8}{\milli\metre} black glass beads and \SI{15}{\milli\metre} green polypropylene beads. Evolution in time from left to right. From \citet{breu2003reversing}.
    }
    \label{fig:Granular}
\end{figure}

A notorious example thereof is avalanche dynamics \cite{nagel1992instabilities, paczuski1996avalanche, frette1996avalanche, hunt2010booming}.
A pile of grains [Fig.~\ref{fig:Granular}a] is held together by `chains' of frictional and compressive forces \cite{mueth1998force}, but the pile will suddenly collapse if the slope exceeds a maximum angle, $\theta_m$. This sliding will only stop after the slope has reduced below the critical angle of repose, $\theta_r$, of which typical values range between $45^\circ$ for wheat flour to $25^\circ$ for whole grains \cite{al2018review}.
The angle of repose is important across food industry, from silo roof design to conveyor belt transport \cite{song2003conveyor} and the geology of hillslopes \cite{deshpande2020perpetual} that limits farming.
It also sets a fundamental rule for food plating, which in turn affects the perception of taste \cite{zellner2011neatness}, and the design of food sculptures [Fig.~\ref{fig:Granular}b].
In this artwork the grains are sticky, but the weakest links are still susceptible to avalanche dynamics.
From a fundamental point of view, avalanche dynamics exhibits self-organized critical (SOC) behaviour for rice grains with a large aspect ratio, as shown by \citet{frette1996avalanche}.
Moreover, \citet{einav2018tracking} used a column of puffed rice to investigate the crumbling of a brittle porous medium by fluid flow. These `ricequake' experiments may give insight how to prevent the collapse of rockfill dams, sinkholes, and ice shelves.
Needless to say, avalanches can be extremely dangerous, also in food science such as entrapment in grain storage facilities \cite{issa2017contributing}, where a silo of grains can clog [Fig.~\ref{fig:Granular}c] because of force chains forming between grains that prevent them from relative motion [Fig.~\ref{fig:Granular}d]. However, a disruption of this delicate balance might lead to large-scale avalanche flows. Thankfully there are helpful rescue strategies in case of grain entrapment \cite{roberts2011summary}.

Sometimes it is desirable to make granular matter flow faster. 
This can often be achieved with granulation\index{granulation} \cite{cuq2013agglomeration}, where a powder or small grains are made to clump together.
Perhaps counter-intuitively, these aggregates can flow more easily. 
This has many important examples in the kitchen, such as granulated sugar: 
Compared to powdered sugar, granulated sugar has a smaller angle of repose and considerably better flow characteristics \cite{teunou1999characterisation} because the larger grains are less cohesive and more easily fluidised \cite{krantz2009characterization}. 
Moreover, it is much easier to compactify granulated materials than powders, which is of vital importance for making tablets in the pharmaceutical and food industry \cite{kristensen1987granulation}. 
Food for thought when you next sweeten your coffee.

\subsection{Hoppers: Grains flowing through an orifice}
\label{subsec:GrainsOrifice}
An important quantity across food science is how quickly a granular material can pass through an orifice \cite{beverloo1961flow}.
Chefs experience this daily when dispensing spices or grains from a hopper.
The flow rate $Q$ is given empirically by the Beverloo law\index{Beverloo law}, and depends on the friction coefficient, the hopper angle, the grain bulk density, the neck width, and the grain diameter.
    \begin{equation}
    \label{eq:BeverlooLaw}
        Q = C(\mu, \theta) \rho_b \sqrt{g} (D-kd)^{3/2},
    \end{equation}
where the constant $C$ depends on the friction coefficient $\mu$ and the hopper angle $\theta$, where $\rho_b$ is the grain bulk density, $D$ is the neck width, $d$ is the grain diameter, and the parameter $k\approx \mathcal{O}(1)$.

The Beverloo law can be explained partially by dimensional analysis, scaling arguments and hourglass theory \cite{nedderman2005statics}, but its foundations are still under active investigation, as discussed recently by \citet{alonso2021beverloo}.
This problem is hard to solve because the grains behave both like a liquid and a solid near the opening \cite{zuriguel2014invited}.
This is the result of jamming \cite{liu1998jamming, liu2010jamming}\index{jamming}, where the effective viscosity increases dramatically above a critical particle density, so the flowing grains suddenly form a rigid arch or vault \cite{tang2011granular}, as shown in Fig.~\ref{fig:Granular}c,d.
This clogging becomes exponentially unlikely as the opening size is increased, so all hoppers have a non-zero probability to clog \cite{thomas2015fraction}.
Clogging is particularly common in grain hoppers \cite{to2001jamming} and microfluidic devices \cite{dressaire2017clogging}, but also in sheep herds and pedestrian crowds moving through a bottleneck \cite{zuriguel2014clogging}.
Jamming is equally essential for food processing \cite{barker2013granular}, for biological tissue development \cite{lawson2021jamming} and for food structuring \cite{van2012soft}.

The non-linear nature of jamming can lead to surprising consequences.
For example, the observed flow rate does not depend on the filling height of the hopper [Eq.~\eqref{eq:BeverlooLaw}].
This was assumed to be due to the Janssen effect [\S\ref{subsec:AvalancheDynamics}], but \citet{aguirre2010pressure} showed that the grain flow rate remains constant even if the pressure at the orifice decreases during discharge, and that different flow rates can be achieved with the same pressure.
More counterintuitively, inserting an obstacle just above the outlet of a silo can in fact help with clogging reduction \cite{zuriguel2011silo}.
In fact, it has been shown that `the sands of time run faster near the end', which is caused by a self-generated pumping of fluid through the packing \cite{koivisto2017sands} -- you may want to take this into account for your kitchen timer.
By varying the softness of the grains, \citet{tao2021clogging} showed that clogging occurs more often for stiffer particles, and that clogging arches are larger for particles with larger frictional interactions.
Hence, understanding jamming dynamics can help with the design of clog-free particle separation devices \cite{mossige2018separation}.

\subsection{Brazil nut effect}
\label{subsec:BrazilNutEffect}
When you repeatedly shake a box of cereal mix, the larger (and heavier) grains often rise to the top [Fig.~\ref{fig:Granular}e]. 
This phenomenon was called the `Brazil nut effect' after the seminal work by \citet{rosato1987brazil}, also known as granular segregation \cite{ottino2000mixing, kudrolli2004size, gray2018particle}.
One explanation is buoyancy, but the effect can happen even when the larger grains are denser than the smaller ones.
A second contributing factor is called percolation (not to be confused with brewing coffee, \S\ref{subsec:brewingCoffee}). 
Here, smaller grains fall into the gaps below the larger grains during shaking.
However, a container shape can be designed where the larger grains move downwards \cite{knight1993vibration}.
This was explained by a third effect called `granular convection'\index{granular convection}, where the shaking leads to a flow pattern of grains moving upward along the walls and downward in the middle of the container \cite{knight1993vibration}. 
Soon after, granular convection was imaged directly using Magnetic Resonance Imaging (MRI) \cite{ehrichs1995granular, knight1996experimental} and studied using extensive computer simulations \cite{gallas1996molecular}.
However, the scientific trail was mixed up again by the discovery of the `reverse Brazil nut effect' by \citet{shinbrot1998reverse}, verified experimentally by \citet{breu2003reversing}. Here larger but lighter particles can sink in a shaken bed of smaller grains, which cannot be explained by granular convection alone, nor by percolation nor buoyancy. 
Furthermore, \citet{mobius2001size} showed that particle segregation depends on the interstitial air between the grains, and depends non-monotonically on the density. 
\citet{huerta2004vibration} then found that there are two distinct regimes: At low vibration frequencies, inertia and convection drive segregation, where inertia (cf. convection) dominates when the relative density is greater (cf. less) than one. At high frequencies, segregation occurs due to buoyancy (or sinkage) because the granular bed is fluidised and convection is suppressed.

Interestingly, granular convection can occur even in very densely packed shaken containers, on the brink of jamming, where unexpected dynamic structures can arise under geometrical restrictions \cite{rietz2008brink}.
\citet{murdoch2013granular} studied granular convection in microgravity during parabolic flights, revealing that gravity tunes the frictional particle-particle and particle-wall interactions, which have been proposed to drive secondary flow structures.
Recently, \citet{ortana2020self} discovered that a self-induced {R}ayleigh-{T}aylor instability [see \S\ref{subsubsec:RayleighTaylorInstability}] can occur in segregating granular flows, where particles continuously mix and separate when flowing down inclines.

Granular separation is of vital importance in the food industry.
It might be convenient if grains need to be sorted, but the effect is often undesirable when we require an even grain mixture.
This is especially problematic when the product must be delivered within a narrow particle-size distribution or with specific compositions of active ingredients \cite{gray2018particle}. 
However, most food storage facilities (heaps or silos) and processing units (chutes or rotating tumblers) are prone to grain segregation \cite{baxter1998stratification}. 
Researchers are learning how to prevent separation, but this is hard because it strongly depends on details of the flow kinematics.
One technique is to add small amounts of liquid to make the grains more cohesive \cite{li2003controlling}, but at the cost of reduced food longevity due to rot.
Another strategy is to use modulation of the feeding flow rate onto heaps \cite{xiao2017controlling}.
The Brazil nut effect might also be suppressed using a system with cyclical shearing, where grains remain mixed or segregate slowly \cite{harrington2013suppression}. 
There are also new developments in machine vision systems for food grain quality evaluation \cite{vithu2016machine}.
Granular flows are often hard to image with opaque particles, but powerful techniques to measure the 3D dynamics include MRI \cite{ehrichs1995granular}, Positron Emission Particle Tracking (PEPT) \cite{windows2020positron} or X-ray Computed Tomography (CT) \cite{gajjar2021size}. 

Finally, coming back to our kitchen, the Brazil nut effect may occur too when stir-frying. This is why chefs often toss the ingredients into the air repeatedly, fluidising the grains, to mix them evenly and to avoid burning them at very high temperatures, for e.g. \citet{ko2020physics} recently described the physics of tossing fried rice.
There is also an interesting connection between the Brazil nut effect and popcorn \cite{virot2015popcorn, hoseney1983mechanism, da1993makes}, where the grains that have not popped conveniently sink to the bottom.
Granular physics hence plays a key role in several culinary contexts, both in the food industry and in our kitchens.

\subsection{Brewing coffee: Porous media flows} 
\label{subsec:brewingCoffee}

\begin{figure}[t]
    \includegraphics[width=\linewidth]{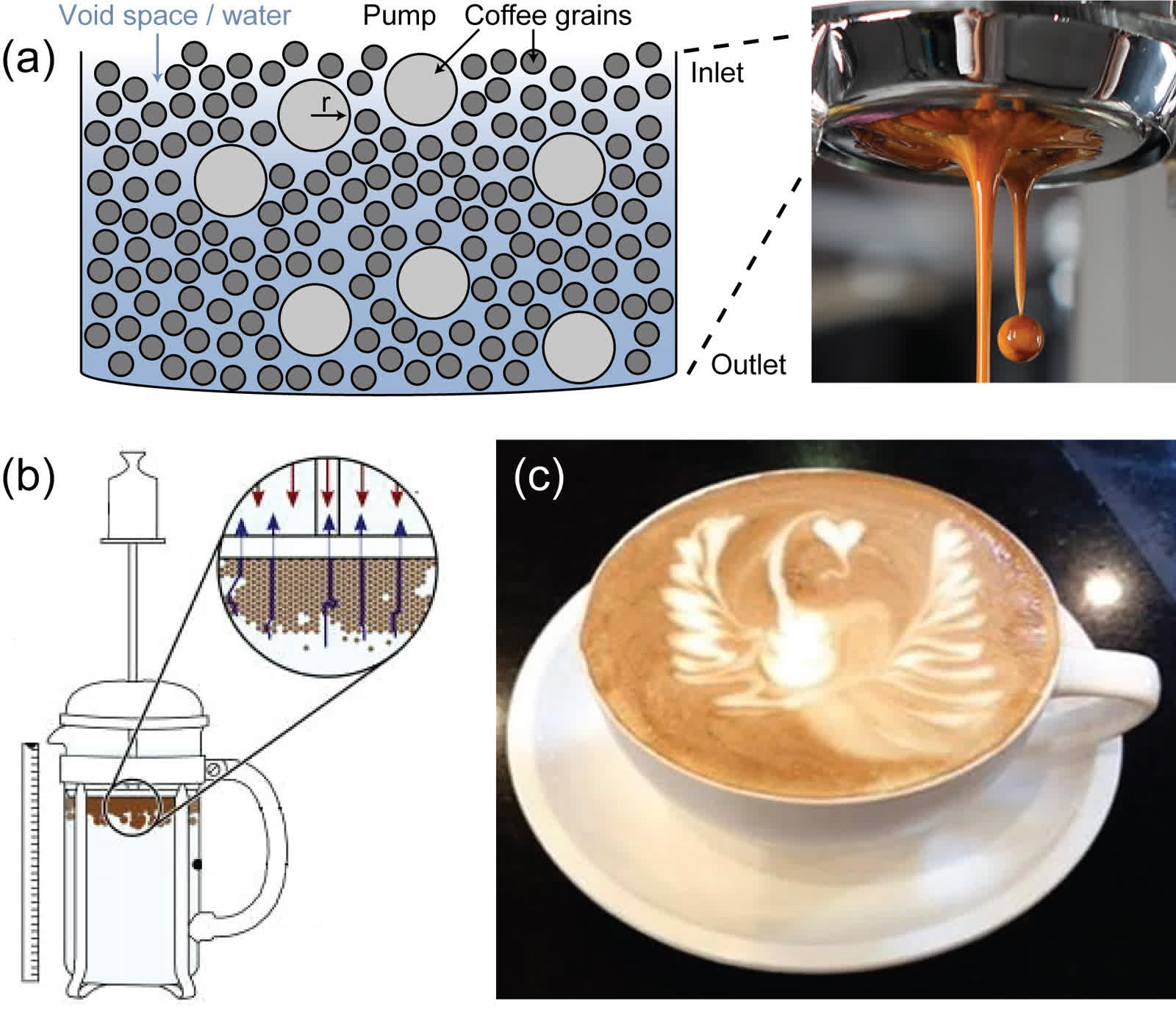}
    \caption{Coffee brewing. 
    \textbf{(a)} Schematic of percolation in an espresso machine basket. A pressure differential pushes water down through the pore spaces. Image courtesy of Christopher H. Hendon. 
    Inset: Espresso drops by photographer theferdi, licensed under CC BY-NC-SA 2.0.
    \textbf{(b)} Diagram of a French press. The water moves up around the coffee grounds (blue arrows) by applying a constant gravitational force on the plunger (red arrows). From \citet{wadsworth2021force}.
    \textbf{(c)} A cappuccino with latte art in the shape of a phoenix. From \citet{hsu2021does}.
    }
    \label{fig:Coffee}
\end{figure}

Henry Darcy (1803-1858) was a French hydrologist who studied the drinking water supply system of Dijon, a city also known for its mustard. 
In an appendix of his famous publication \cite{darcy1856fontaines}, he describes experiments on water flowing through a bed of sand.
From the results he obtained Darcy's law\index{Darcy's law}, an empirical expression for the average velocity $\vec{q}$ of the liquid moving through the bed.
It was refined by \citet{muskat1938flow}, and can be written as
\begin{equation}
    \vec{q} = - \mathcal{K} (\vec{\nabla} p - \rho \vec{g}) / \mu,
    \label{eq:Darcy}
\end{equation}
where $\mathcal{K}$ is a tensor that describes the permeability of the material in different directions, sometimes replaced by a constant $k$ for isotropic materials. 
Note that the local velocity in the pores is $\vec{u} = \vec{q}/\phi$, where $\phi \in [0,1]$ is called the porosity.
Darcy's law can indeed be derived theoretically \cite{whitaker1986flow}, and it is integral to many industrial processes in food science, such as sand filtration and antimicrobial water treatment \cite{yang2020development, hills2001microstructural}. 
Moreover, it can be applied to a much broader class of porous media flows \cite{philip1970flow, bear2013dynamics, blunt2001flow}, which are found everywhere in gastronomy: Examples include percolation in a salad spinner, drizzle cake, squeezing a sponge, and not least, making coffee \cite{thurston2013coffee, egger2016home}.

For espresso brewing \cite[see e.g.][]{giacomini2020water}, one can use the first term in Darcy's law for pressure-driven flow [Fig.~\ref{fig:Coffee}a], while drip coffee \cite[see e.g.][]{angeloni2019characterization} can be described using the second term for gravity-driven flow in Eq.~\eqref{eq:Darcy}.
In both cases, the permeability $\mathcal{K}$ can be changed by tamping the grounds or adjusting the grain size, in order to tune the flow rate and thus the extraction time \cite{corrochano2015new}.
Another particularly well-studied method of making coffee is using the Italian `moka' pot [see Fig.~\ref{fig:TantalusBowl}c].
Its sophisticated design was patented by Alfonso Bialetti (1888-1970) \cite{binder2020moka}. 
\citet{gianino2007experimental} used a moka pot to measure the flow rate through a bed of grains and applied Darcy's law to measure its permeability. Later, \citet{navarini2009experimental} improved this method by accounting for the decreasing permeability as aromatic substances dissolve into water. In a third study, \citet{king2008physics} investigated the effect of packing and coffee grain temperature on the permeability. 
Percolation in a moka pot was visualised beautifully by looking through the metal using neutron imaging \cite{youtube2012neutron}.

Most studies on coffee extraction showcase some direct applications of Darcy's law under idealized conditions, but they do not attempt a description beyond the Darcy regime \cite{Fasano2000}. 
For instance, they ignore the initial stage of percolation, when the water invades the (initially dry) coffee grain matrix. 
This process can be described with percolation theory \cite{stauffer2018introduction}, where the pores between the coffee particles can be considered as a random network of microscopic flow channels \cite{blunt2001flow}.
If the coffee is coarsely ground or tamped lightly, the many open microchannels (bonds) allow the water to find a connecting path through the coffee.
Then this process is first characterised by capillary wetting \cite{singh2019capillary} [see \S\ref{subsec:wettingCapillaryAction}] and considerable swelling of coffee grains \cite{mo2022modeling}, after which Darcy's law becomes applicable. 
For intermediate tamping, as the permeability $\mathcal{K}$ decreases, we approach the percolation threshold.
Then the extraction time will significantly increase, which can result in over-extraction \cite{severini2017much}.
Finally, if the coffee is tamped too strongly or too finely ground, the water cannot find a path between the grounds.
Then, either there is nothing to drink, or the pressure builds up until we see hydraulic fracturing \cite{adler2013fractured}. 
In that case, large flow channels suddenly crack open that bypass the microchannels \cite{berre2019flow}. Channelization might be induced by flow-mediated rearrangement of the porous coffee bed \cite{mahadevan2012flow,derr2020flow}. 
This `fracking' can give the coffee a bad taste because it leads to uneven extraction \cite{moroney2019analysing}.
Symptoms that your espresso is fractured are liquid spraying through the bottom of the basket, irregular flow, and a cracked coffee cake.
Percolation theory has many other applications, including predictions for forest fires, disease spreading, and communication in biology \cite{sahini2014applications, mathijssen2019collective}. With these efforts to perfect the taste of resulting beverage, there are still aspects of coffee brewing that are understood only qualitatively and described empirically. Microscopic details of swelling and extraction remain to be explored in order to fully understand the complex physics of reactive transport in espresso machine brewing.

Coffee can be made in many ways, each involving different fluid mechanics. Here we only mention a few preparation methods, with some recent results:
To make an espresso, it is important to note that coffee beans are often prone to variations in quality. Even if the theory is perfect, the ingredients are not. 
To overcome this, \citet{cameron2020systematically} offer advice to systematically improve espresso brewing by proposing a set of guidelines towards a uniform extraction yield. Interestingly, they also found that the smallest grains do not give the highest yield as they tend to clump together and form aggregates [see granulation in \S\ref{subsec:AvalancheDynamics}]. 
For drip brew coffee, it is common to use a precise temperature-controlled kettle, but \citet{batali2020brew} surprisingly found that the brew temperature may not have quite so much impact on the sensory profile at fixed brew strength and extraction.
Considering the French press, \citet{wadsworth2021force} recently determined the force required to operate the plunger [Fig.~\ref{fig:Coffee}b]. They recommend using a maximum force of \SI{32}{\newton} to complete the pressing action in \SI{50}{\second}, using \SI{54}{\gram} of coffee grounds for \SI{1}{\litre} of boiling water.
Looking at cold brew coffee, \citet{cordoba2019effect} recently evaluated the extraction time and flavour characteristics, \citet{rao2018acidity} investigated its acidity and antioxidant activity, and \citet{ziefuss2022ultrafast} showed that cold-brewing can be achieved rapidly using picosecond-pulsed laser extraction.
Finally, Greek and Turkish coffee rely on the sedimentation of fine particles, which we discuss in \S\ref{subsec:Sedimentation}, and Café de Olla is a Mexican coffee drink which is made with aromatic spices and sugar. 

After having made the perfect cup of coffee, it can be decorated with latte art [Fig.~\ref{fig:Coffee}c].
Indeed, \citet{hsu2021does} showed that coffee tastes sweeter with latte art, which they related to brainwave activity using electroencephalography (EEG).
The fluid mechanics of pouring steamed milk foam into the denser coffee can be described as an inverted fountain [see \S\ref{subsec:LayeredCocktails}], which depends on the jet diameter and the pouring height via the Froude number [Eq.~\eqref{eq:TurnerFountain}]. 
Large fountains lead to more mixing and brown foam, while gentle pouring gives white foam. After much practice, these colours can be combined in rapid succession to make exquisite patterns, including a heart and the phoenix \cite{latteartguide}. 
Some people prefer their coffee without milk, but with a thin layer of espresso crema \cite{illy2011neglected}. Undesirably, this coffee foam can agglomerate along the perimeter. This effect can often be suppressed by heating the cup beforehand since it is caused by B\'enard-Maragoni convection, which we discuss in the next section.

\subsection{Coffee ring effect}
\label{subsec:CoffeeRingEffect}

\begin{figure}[t]
    \centering
     \includegraphics[width=1\linewidth]{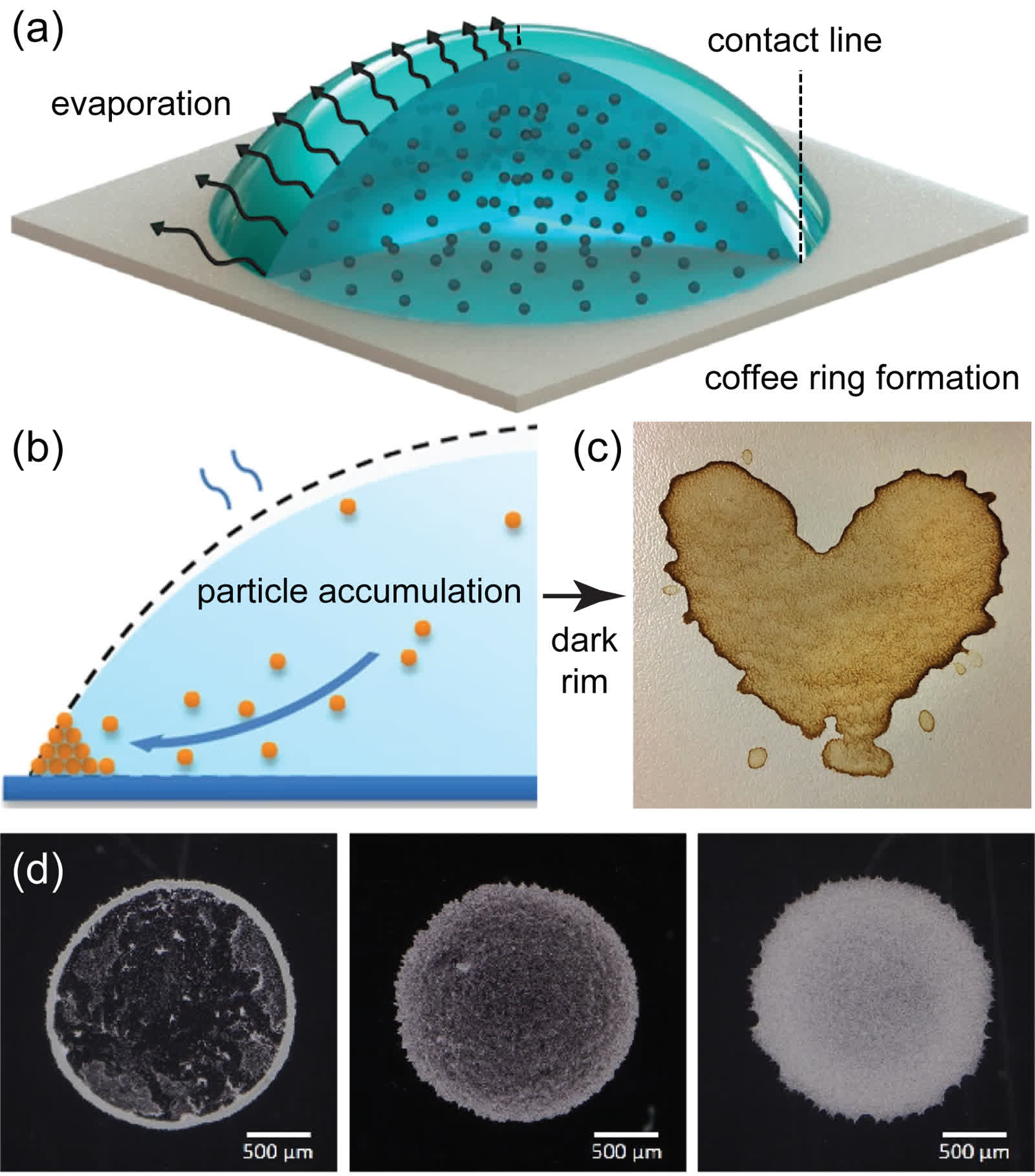}
         \caption{Coffee ring effect. 
         \textbf{(a)} Schematic of an evaporating droplet containing a suspension of microparticles. Stronger evaporation near the contact line drives an internal flow to the outer edge. From \citet{jafari2016alternative}. 
         \textbf{(b)} A dark rim is formed by particles that accumulate at the pinned contact line. From \citet{li2016rate}.
         \textbf{(c)} Example of a heart-shaped coffee ring. Image by Robert Couse-Baker, licensed under CC BY 2.0.
         \textbf{(d)} Suppressing the coffee ring effect by adding cellulose nanofibers (CNF) to a drop of 0.1 wt\% colloidal particles. From left to right: CNF concentration of 0, 0.01, 0.1 wt\%. From \citet{ooi2017suppressing}.
         }
     \label{fig:coffeering}
 \end{figure}

When the last sip of coffee or wine is left overnight, it dries out creating a stain with a brighter interior and much darker borders, where most residues are deposited\index{coffee-ring effect}\index{coffee}\index{wine}\index{droplet} [Fig.~\ref{fig:coffeering}].
The coffee ring effect, as it has been termed, can be observed in almost any kitchen mixture containing small particles. As explained by \citet{deegan1997capillary}, the coffee ring effect results from the drying dynamics of the droplet, combined with the pinning of its contact line with the substrate \cite{marin2011order, mampallil2018coffee}.\index{evaporation}\index{contact line}\index{sediment}
As the solvent evaporates, the outflow of matter decreases the thickness of the droplet at every point. If the contact line was not pinned, the droplet would shrink. This additional constraint, together with the surface tension requirement to keep the contact angle fixed [see \S\ref{subsec:wettingCapillaryAction}], induces an outward flow from the interior to replenish the evaporated liquid at the borders. This flow transports the sediment, which is then deposited at the outer ring, leaving the lower-concentration interior.

Instead of a solid particle suspension, the coffee ring effect also emerges if the dissolved component is another liquid. In that case, the edge of the puddle remaining after a volatile solvent evaporates forms either ``fingers'' or spherical ``pearls'', or some combination of the two \cite{mouat2020tuning}. Understanding the underpinning dynamics of coffee ring formations remain an active research topic. An example of the many excellent publications in recent years is the study by \citet{moore2021nascent} on the effects of diffusion of solute from the pinned contact line to the bulk of the drop on the pattern formation. 

Interestingly, by adding some alcohol to the drop to make it more volatile, the coffee ring effect can be suppressed by consequential Marangoni flows from the contact line to the drop's interior \cite{hu2006marangoni}. These flows are induced by a surface tension gradient along the surface of the drop [see \S\ref{subsec:Tears}], which is in turn caused by nonuniform cooling as the droplet evaporates, because the surface tension increases as the temperature decreases. This is referred to as B\'enard-Marangoni convection\index{B\'enard-Marangoni convection} [also see \S\ref{subsec:HeatingBoilingRBC}]. The deposition pattern then depends on the strength of the relative magnitude of thermocapillary stresses to viscous ones, as expressed by the dimensionless Marangoni number, Ma, as defined in Eq.~\eqref{eq:MarangoniNumber} with $\Delta \gamma = (\partial \gamma/\partial T) \Delta T$. For large Ma, the particles end up in the center, for intermediate Ma, they are deposited evenly along the substrate, while for small Ma, the coffee ring effect is fully recovered. Therefore, by tuning Ma, for example through tuning the alcohol percentage in the coffee, it is in principle possible to control the deposition pattern.
Moreover, the coffee-ring effect can also be suppressed by shape-dependent capillary interactions \cite{yunker2011suppression, yunker2013effects} or the addition of cellulose nanofibers [Fig.~\ref{fig:coffeering}d].

The applications of the coffee-ring effect and the drying of thin colloidal films in general go far beyond kitchen experiments \cite{routh2013drying, mampallil2018coffee}.
For example, it is the basis of Controlled Evaporative Self‐Assembly (CESA), which is used to create functional surfaces with controllable features \cite{han2012learning}.
The coffee-ring effect can also be used for controlled ink-jetting of a conductive pattern of silver nanoparticles \cite{zhang2013controlled}, and particle deposition on surfaces could be controlled further using light-directed patterning by evaporative optical {M}arangoni assembly \cite{varanakkottu2016light}.
Moreover, the effect can be used for nanochromatography to separate particles such as proteins, micro-organisms, and mammalian cells with a separation resolution on the order of \SI{100}{\nano\metre} \cite{wong2011nanochromatography}.
Interestingly, the growth dynamics of coffee rings are altered when active particles such as motile bacteria move around in the evaporating droplet \cite{nellimoottil2007evaporation, sempels2013auto, hennes2017active, andac2019active}, and bacterial suspensions can feature active de-pinning dynamics. 
Hence, in the spirit of frugal science\index{frugal science}, it might be possible to exploit the coffee ring effect to detect antimicrobial resistance \cite{kang2020simple}.


\section{Tempest in a Teacup: Non-linear Flows, Turbulence and Mixing}
\label{sec:Turbulence}

What is turbulence? This intriguing question has fascinated fluid mechanicians throughout history. 
Leonardo da Vinci already eluded to two important properties of turbulence \cite{marusic2021leonardo}: The generation of motion at large scales, and the destruction due to viscosity at the smallest scales. The many scales of motion in turbulence is arguably its main signature; for example, a volcanic plume spans over several kilometers, with eddies all the way down to the Kolmogorov microscales \cite{kolmogorov1941local}\index{Kolmogorov length scale}. As the turbulent structures break up, energy is transferred from large whirls to smaller ones. The poem by \citet{richardson2007weather} beautifully describes this energy cascade:
    \begin{quote}
        \emph{Big whirls have little whirls} \newline
        \emph{That heed on their velocity,} \newline
        \emph{And little whirls have littler whirls} \newline
        \emph{And so on to viscosity.}
    \end{quote}
The consequences of turbulence are numerous in our everyday lives. It gives us the characteristic sound of a kettle whistle, it helps with mixing milk into our tea or coffee, and it gives us some frictional losses when biking home from the restaurant. In this section, we will catch a whiff of turbulence in the kitchen.

\subsection{Tea leaf paradox: Secondary flows}
\label{subsec:teaLeafParadox}
\label{subsec:secondary}

\begin{figure}[t]
    \includegraphics[width=\linewidth]{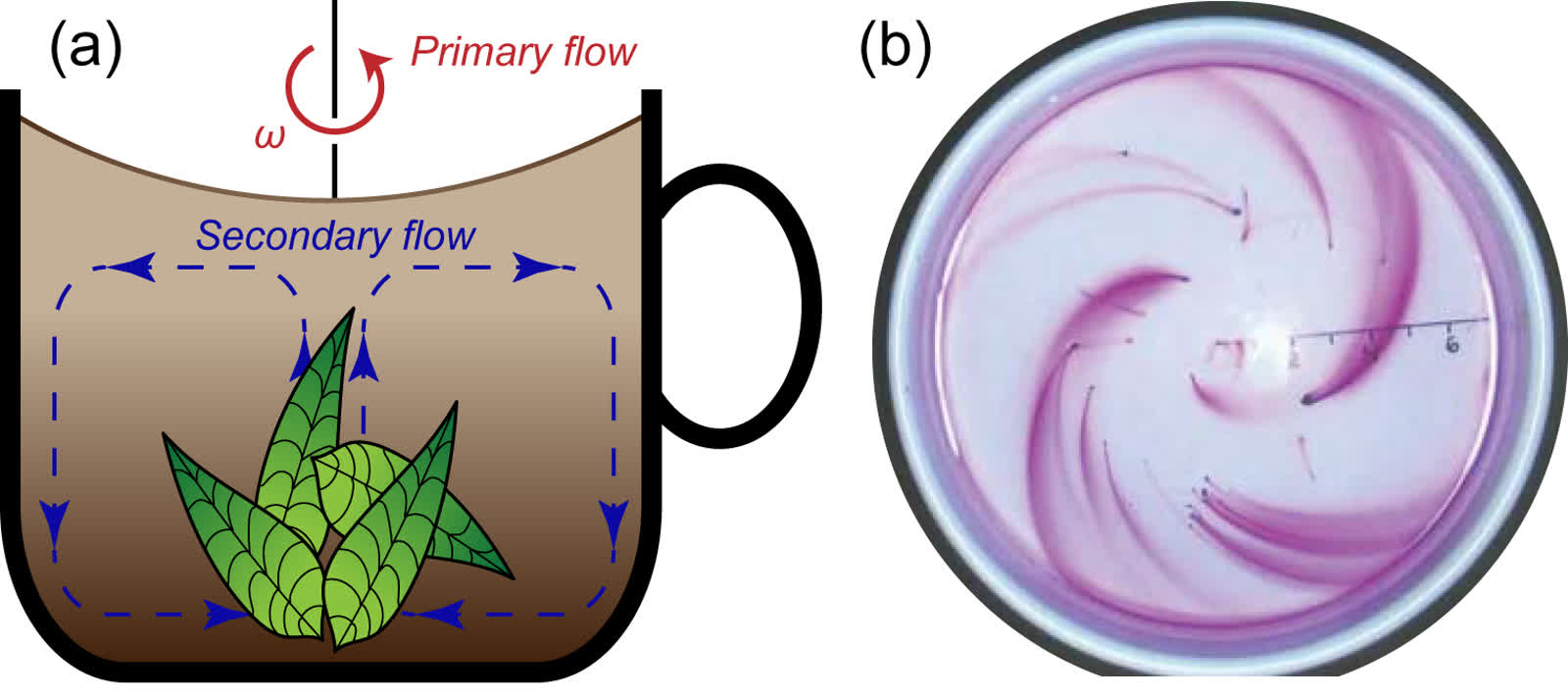}
    \caption{Tea leaf effect due to secondary flows.
    \textbf{(a)} After stirring a cup of tea, rotating liquid (primary flow) slowly comes to a halt because of friction with the walls (spin-down). This friction also induces toroidal recirculation (secondary flow) directed outwards at the top and inwards at the bottom, which causes the leaves to collect in the center. 
    \textbf{(b)} Spiral dye streaks due to secondary flows in a cake pan. Instead of slowing down, the liquid is rotated increasingly faster (spin-up), so the secondary flow is reversed, such that the dye spirals outwards at the bottom of the cake pan. From \citet{heavers2010ekman}.
    }
    \label{fig:TealeafEffect}
\end{figure}

Before diving into chaotic realms, we consider the surprising effects that the non-linearity of the Navier-Stokes equations [Eq.~\eqref{eq:NavierStokes}] can bear in laminar flow.
One such surprises is the ``tea leaf paradox''\index{tea leaf paradox}.
Biological tissues tend to be denser than water \cite{aoyagi1992measurement}, thus soaked tea leaves will sink to the bottom of a cup.
When the water is stirred around in circles, the leaves are expected to move towards the edge of the cup because of centrifugal action.
The opposite happens, however: The leaves always migrate to the center of the cup, as seen in Fig.~\ref{fig:TealeafEffect}a.
\citet{thomson1892xvii} first recognised that the solution of this paradox stems from `friction on the bottom'. 
Later, \citet{einstein1926cause} gave a detailed description of the tea leaf experiment itself, in order to explain the erosion of riverbanks. A detailed theoretical treatment was provided later by \citet{greenspan1963time}.

The paradox is resolved with fluid mechanics, as follows:
As the liquid rotates in the cup, the first approximation of the fluid flow is just a solid-body rotation.
Specifically, we have $\vec{u} = \vec{\Omega} \times \vec{r}$ with a uniform angular velocity $\vec{\Omega}$. 
On the fluid acts a centrifugal force, $\vec{F}_c \propto \vec{\Omega} \times (\vec{\Omega} \times \vec{r})$. If $\vec{\Omega}$ is constant in space, then this force does not modify the flow.
However, frictional drag with the cup walls slows the fluid down in the boundary layer. 
In particular, near the bottom surface the angular velocity $\vec{\Omega}(\vec{r})$ and thus the centrifugal force will be less than near the top water-air interface.
Consequently, an in-up-out-down recirculation emerges [Fig.~\ref{fig:TealeafEffect}a] as the liquid slowly stops spinning.
Interestingly, this recirculation can also be reversed [Fig.~\ref{fig:TealeafEffect}b], when the liquid is rotated increasingly faster (spin-up\index{spin-up}) from rest \cite{greenspan1963time, baker1968demonstrations}.

The tea leaf effect has many applications. 
In the kitchen, it can be applied conveniently when poaching eggs \cite{moore1989swirling, heavers2010ekman}:
Before cracking the egg into the pot, the hot water can be gyred to keep the egg whites together at the center of the pot. 
But be quick, because the flow ceases after a time 
    \begin{equation}
    \label{eq:EkmanTime}
    \tau_{E} \sim \sqrt{\frac{H^2}{\nu \Omega}},
    \end{equation}
called the Ekman time\index{Ekman time} \cite{ekman1906beitrage, greenspan1963time}, in terms of the kinematic viscosity $\nu$ and $H$ is the height of the liquid layer, the depth. 
The same technique is also used to separate out trub during beer brewing \cite{bamforth2009beer}, and to separate blood cells from plasma in microfluidics \cite{arifin2007microfluidic}\index{red blood cells}\index{microfluidics}.
Also in geophysical flows, the same in-up-out-down circulation is seen in tornadoes \cite{rotunno2013fluid}.
In modern additive manufacturing (AM) technologies, the Ekman time sets a limit to how quickly objects can be 3D printed \cite{kelly2019volumetric}.


When discussing the tea leaf paradox, we saw that rotating a liquid in a cup gives rise to (1) a horizontal rigid body motion, and (2) a vertical flow structure due to frictional drag with the surfaces [Fig.~\ref{fig:TealeafEffect}a].
Indeed, it is a powerful concept to understand such fluid flows in terms of a ``primary flow''\index{primary flow}, guessed from basic physical principles, and a ``secondary flow''\index{secondary flow}, a correction due to high-order effects such as obstacles in the main flow.

Another classical example of secondary flows can be used when we inactivate microorganisms during the transport of fruit juices in pipes.
This can be achieved by trapping the microbes in vortices, and exposing them to UV-C light \cite{muller2011uv}.
These so-called Dean vortices naturally emergence in a curved pipe \cite{dean1927note}.
The primary structure can be taken as a straight Poiseuille flow [see \S\ref{subsec:PoiseuilleFlow}], and the secondary structure consists of vortices that can be explained using a perturbation method based on Poiseuille flow accounting for centrifugal forces \cite{germano1989dean, boshier2014extended}.
To tune the vertices, we must quantify the relative strength of the secondary flow compared to the primary. 
This is determined by the balance between inertial and centrifugal forces with respect to viscous forces, which is given by the Dean number,\index{Dean number}
    \begin{equation} 
    \label{DeanNumber}
    \text{De} 
    = \text{Re} \sqrt{R_p/R_c},
    \end{equation}
where $\text{Re}$ is the Reynolds number [Eq.~\eqref{eq:ReynoldsNumber}], $R_p$ is the radius of the pipe, and $R_c$ its radius of curvature. 
For small Dean numbers the current is unidirectional, mostly primary flow.
Dean vortices emerge for intermediate values, and for large $\text{De}$ the flow turns turbulent \cite{kalpakli2012dean}.
This knowledge can also be used in applications to separate particles by size using inertia \cite{di2009inertial}.
These vortices also emerge naturally in straight channels when two stratified fluid layers such as air and water flow through them \cite{vollestad2020experimental}, which is associated with large pressure losses in (food) industrial pipelines. 

In the beer brewing industry, secondary flows are widely use in hot trub sediment removal \cite{jakubowski2015secondary}, where the suspension is injected tangentially to a cylindrical tank, called the whirlpool, where it gradually spins down, while the sediment migrates towards the center, following the so-called Ekman spirals, known from meteorological flow considerations \cite{bodewadt1940drehstromung}. A variant of this process involves a whilpool kettle, in which a heating rod is placed at the axis, which distorts the flow and leads to the formation of a ring of deposit instead of the central cone \cite{jakubowski2019cfd}.

Similar calculations can be performed to study secondary flows in many other applications, including kitchen sink vortices \cite{andersen2003anatomy}, in turbomachinery compressors and turbines \cite{langston2001secondary}, and oceanic and atmospheric currents with Ekman layers \cite{ekman1906beitrage, eliassen1982vilhelm, garratt1994atmospheric}. 
Ekman layers are associated with transport of biomaterials in the ocean through so-called Ekman transport processes. The secondary flow pattern of Ekman transport can lead to upwelling and downwelling of algae and nutrients that promote or growth of phytoplankton populations \cite{miller2012biological}. A thorough understanding of the underpinning mechanisms is crucial to mitigate the devastating implications of harmful algal blooms that impact fish production and aquaculture. Secondary flows can also contribute to bridge scour \cite{wang2017review}, by the removal of sediment such as sand and gravel from around bridge abutments or piers, leading to one of the major causes of bridge failure around the world.

Thus, understanding secondary flows is useful in many scenarios \cite{bradshaw1987turbulent}.
Of course, not all currents can be decomposed in a simple primary and secondary flow structure.
Care must be taken with such superpositions, since the Navier-Stokes equations are non-linear.
However, secondary flows can give quick insights and more advanced perturbation methods can often be followed \cite{vandyke1975perturbation}.

\subsection{Tea kettles: Turbulent jets}
\label{subsec:turbulentJets}

When the water in the tea kettle boils, a turbulent jet of steam emerges from the spout with a conical profile [Fig.~\ref{fig:Tea}a]. 
To describe the dynamics of a turbulent jet, it is useful to decompose the velocities into an average and a fluctuating component. This averaging procedure is named after its inventor, Osborne Reynolds, and is written as  $u_i$ = $\bar{u}_i$ + $u_i^\prime$ with $i$=$\{ x,y\}$ for the velocity components in two dimensions \cite{white2006viscous}. By Reynolds averaging we arrive at the famous equation for the conservation of momentum in turbulent flow,
    \begin{equation}
        \frac{\partial \bar{u}_i}{\partial t} + \bar{u}_j\frac{\partial \bar{u}_i}{\partial x_j} + \overline{u_j^\prime\frac{\partial u_i^\prime}{\partial x_j}}= -\frac{1}{\rho}\frac{\partial \bar{p}}{\partial x_i} + \nu\frac{\partial^2\bar{u}_i}{\partial x_j^2},
        \label{eq:Reynoldsequation}
    \end{equation}
using Einstein notation.
The third term is an apparent stress due to turbulent fluctuations, and the remaining ones are the averaged transport terms in the Navier-Stokes equation [see \S\ref{subsec:NavierStokes}], where the last (viscous stress) term can be neglected in inertia-dominated flows. 
Interestingly, the flux of momentum remains constant beyond a certain distance from the spout \cite{guyon2001physical}.
To maintain its momentum, the jet must continually entrain ambient air. 
This is why blowing on a finger burnt by a hot kettle has a cooling effect.

Moreover, fluid jets tend to follow convex surfaces, rather than being scattered off, which is called the Coanda effect \cite{barros2016bluff}.
This is sometimes demonstrated by extinguishing a candle by blowing around a tin can.
Similarly, when a steam kettle jet curves around another pot, it can pose an unexpected safety hazard.
An interesting application is robotic food processing using Coanda grippers \cite{lien2008novel}.
The Coanda effect should not be confused with the teapot effect \cite{lopez2009coandaBackOfSpoon, duez2010wetting, scheichl2021developed}\index{teapot effect}, where a liquid follows a curved surface like a teapot spout [Fig.~\ref{fig:Tea}b], because this flow is dominated by surface tension and wetting [\S\ref{subsec:wettingCapillaryAction}].
While the teapot effect can cause a mess when pouring too slowly, it can be advantageous for coating or making complex shapes \cite{jambon2019liquid}, perhaps in novel culinary decorations.  


Other examples of turbulent jets include the contrails produced by aircrafts. According to one study \cite{burkhardt2011global}, this warms the planet even more than the carbon emitted by the jet engines, but fortunately it seems these effects can be mitigated by avoiding certain altitudes \cite{teoh2020mitigating}. In any event, when flying it is best to  steer clear off plumes, such as those emanating from smoke stacks or volcanoes. As part of a safety assessment, we can use the predictions by \citet{taylor1946dynamics} for the shape and final height of such plumes. 
Taylor's theory is valid across many length scales: It could equally be used to estimate the shape of a plume rising from a cup of coffee [Fig.~\ref{fig:espressocup}].

\subsection{Sound generation by kitchen flows}
\label{subsec:SoundGeneration}


\begin{figure}[t]
    \includegraphics[width=\linewidth]{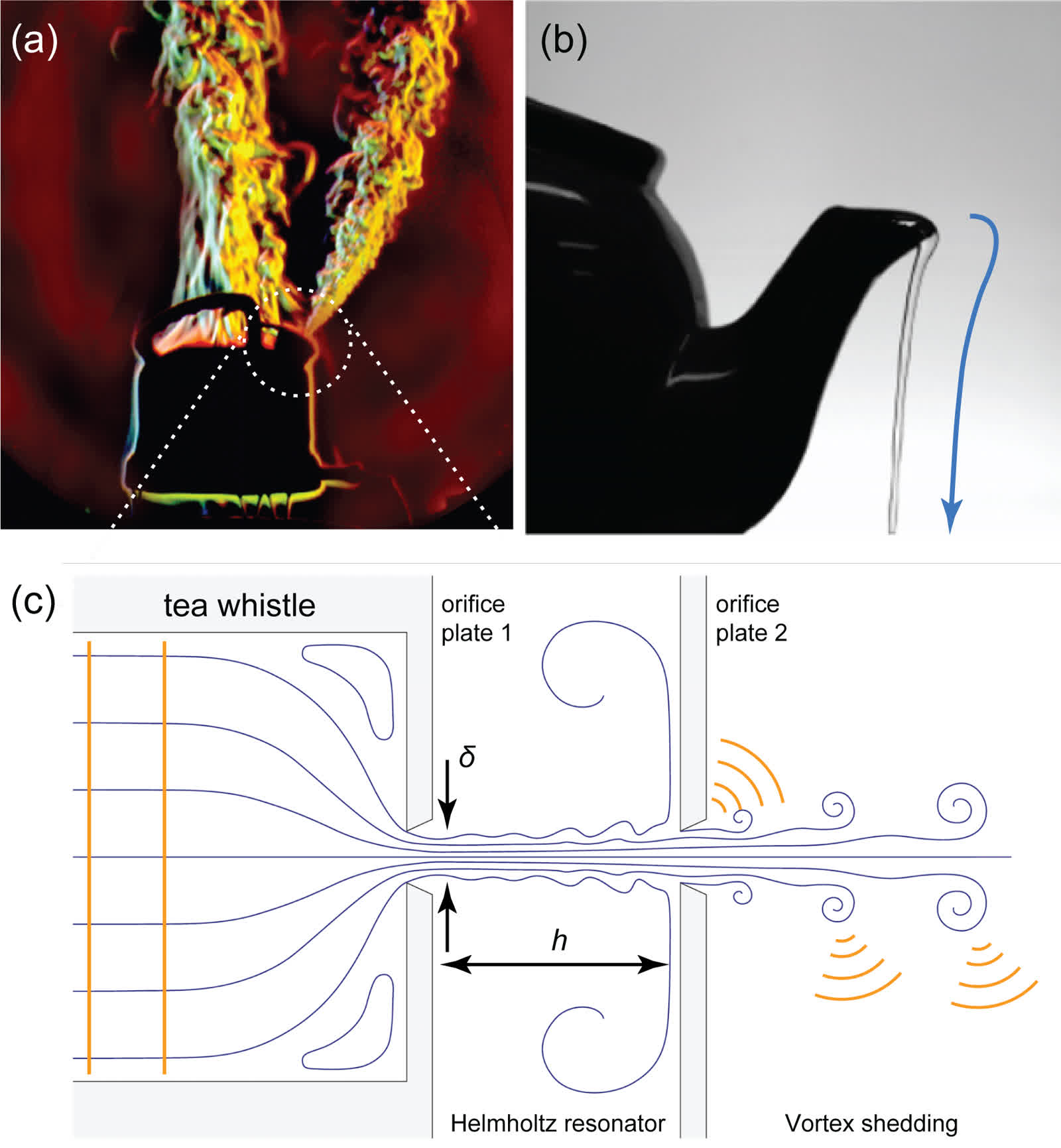}
    \caption{Tea time. 
    \textbf{(a)} A turbulent jet emanating from a tea kettle. Image courtesy of Gary S. Settles.
    \textbf{(b)} The teapot effect, showing a liquid stream following the curved surface of the spout (blue arrow). Pouring any slower will make the liquid stick to the pot entirely. From \citet{scheichl2021developed}.
    \textbf{(c)} Diagram of a tea kettle whistle. The steam passes through two orifice plates from left to right. From \citet{henrywood2013aeroacoustics}.
    }
    \label{fig:Tea}
\end{figure}

The tea kettle we just discussed can make a pleasant whistle  \cite{chanaud1970aerodynamic, henrywood2013aeroacoustics}, which propagates by pressure waves at the speed on sound \cite{rayleigh1896sound1}, approximately $\SI{343}{\metre\per\second}$ in air. 
Sir Isaac Newton (1642-1726) was the first known to measure the speed of sound, as reported in his book on classical mechanics \emph{Principia} \cite{newton1726philosophiae}. Since then, many creative attempts to measure this quantity have been reported, including the accurate experiments by the Reverend William Derham (1657-1735) involving a telescope and gunshots \cite{murdin2009full}.
According to \citet{rayleigh1896sound2}, the parameters determining the sound generation of a whistle are a characteristic length scale, $L_0$, the frequency, $f_0$, the fluid (steam) viscosity, $\nu$, and the steam jet velocity, $U_0$. They form two dimensionless groups, namely the Strouhal number, 
    \begin{equation}
        \label{StrouhalNumber}
        \text{St} = \frac{\text{Time scale of background flow}}{\text{Time scale of oscillating flow}} = \frac{f_0 L_0}{U_0}, 
    \end{equation}
and the Reynolds number [Eq.~\eqref{eq:ReynoldsNumber}].
\citet{henrywood2013aeroacoustics} found that for a typical tea kettle with two orifice plates [Fig.~\ref{fig:Tea}c], the frequency giving rise to the sound generation is sensitive to the jet diameter, $\delta$, and not the plate separation distance, $h$, in the diagram. 
So $L_0$ is to be replaced with $\delta$. 
Furthermore, the same authors found that the whistle's behaviour is divided into two regions:
For $\text{Re}_\delta \lesssim 2000$, the whistle operates like a Helmholtz resonator, with an approximately constant frequency (pitch). 
However, above a critical Reynolds number, $\text{Re}_\delta \gtrsim 2000$, the whistle's tone is determined by vortex shedding \cite{bearman1984vortex}, with a frequency that increases with $U_0$ at an approximately constant $\text{St}_\delta \approx 0.2$. This increases the whistle's pitch by the end and produces the characteristic shrieking sound.
Vortex shedding also occurs for cables or tall buildings in strong wind, which can be destructive if the aerodynamic driving frequency resonates with the structural eigenmodes \cite{irwin2010vortices}. 
To prevent damage from happening, newer buildings are designed to have several eigenfrequencies to effectively dissipate the energy, or to have roughness elements, as perfected by the glass sponge \textit{Euplectella aspergillum} \cite{fernandes2021mechanical}. 
The sound of the tea kettle whistle might inspire you to whistle for yourself while stirring your tea [\S\ref{subsec:teaLeafParadox}]. The physiology of mouth whistling was discussed by \citet{wilson1971experiments, azola2018physiology, shadle1983experiments}.

Plink. plink. plink. Another kitchen sound is the maddening noise of a leaky tap \cite{schmidt1997simple, leighton2012acoustic, speirs2018water}, which can also be annoyingly irregular because of its choatic behaviour \cite{schmidt1997simple, innocenzo2002experimental}.
The paradox of how a single drop impacting on a liquid surface can be so loud, compared to a more energetic continuous stream, is still not fully understood.
\citet{franz1959splashes} already discussed that the droplet can entrain air bubbles, which oscillate to make sound at a frequency $f=(1/2\pi a)\sqrt{3\gamma p_0/\rho}$ given by \citet{minnaert1933musical}, where $a$ is the bubble radius, $\gamma$ is the ratio of specific heats of air, and $p_0$ is the pressure outside the bubble.
However, not every drop makes a sound.
\citet{longuet1990analytic} and \citet{oguz1990bubble} developed the first detailed analytical models to explain this in terms of the Froude number [Eq.~\eqref{eq:FroudeNumber}] and the Weber number [Eq.~\eqref{eq:WeberNumber}] given the droplet radius and impact velocity.
The sound volume is set by the wave amplitude, but how does sound generated underwater cross the water-air interface?
\citet{prosperetti1993impact} reviewed the underwater noise of rain, and \citet{leighton2012can} discusses whether goldfish can hear their owners talking.
Looking at the dripping tap, \citet{phillips2018sound} tested previous theories by comparing sound recordings with direct high-speed camera imaging. They write that the airborne sound field is not simply the underwater field propagating through the water-air interface, but that the oscillating bubble induces oscillations of the water surface itself, which could explain the surprisingly strong airborne sound. 
Plink.

It is impossible to cook without making noise, often to the extend of breaking the sound barrier.
Indeed, we already mentioned the supersonic `pop' made by cracking open a Champagne bottle [\S\ref{subsec:BubblyDrinks}]. 
Similarly, supersonic shock waves can be generated by snapping a tea towel \cite{bernstein1958dynamics, lee1993does}.
Dropping an object in a filled kitchen sink can also create a supersonic air jet \cite{gekle2010supersonic}.
By investigating the popping sound of a bursting soap bubble, \citet{bussonniere2020acoustic} found a way to acoustically measure the forces that drive fast capillary flows.
\citet{kiyama2022morphology} investigated the morphology and sound generation of water droplets in heated oil baths such as deep-fat fryers.
Not least, numerous situations in food science involve hydrodynamic cavitation \cite{albanese2017beer, albanese2017gluten}, 
which can produce flashes of light called sonoluminescence \cite{jarman1964light, patek2004deadly} with internal temperatures reaching thousands of degrees Kelvin \cite{mcnamara1999sonoluminescence}.

The ``hot chocolate effect'' \cite{crawford1982hot} occurs when heating a cup of cold milk in the microwave, mixing in cocoa powder, and tapping the bottom with a spoon:
The sound pitch initially descends by nearly three octaves, comparable to the vocal range of an operatic soprano, after which the pitch gradually rises again. 
This happens because air is less soluble in hot liquids, so it becomes supersaturated with heating, and adding a fine powder provides nucleation sources for fine bubbles.
Air is more compressible than water, which lowers the speed of sound, and thus the pitch.
The same musical scales are heard when opening a fresh beer \cite{crawford1990hot}, which is supersaturated with CO\textsubscript{2} [\S\ref{subsec:BubblyDrinks}].
The hot chocolate effect was visualised directly by \citet{travnivcek2012visualization}.

\subsection{Making macarons: Chaotic advection}
\label{subsec:ChaoticAdvection}
\index{Chaotic Advection}
\index{Mixing at low Re}

A droplet of milk with a typical diffusivity of $D \sim \SI{e-9}{\metre^2\per\second}$ takes a long time to mix in a cup of coffee in the absence of fluid motion, typically $\tau=L_0^2/D$, so days, and much slower than the diffusion of heat [\S\ref{subsec:DoubleDiffusiveConvection}]. 
However, stirring reduces the mixing time dramatically, down to seconds, as turbulent eddies stretch the drop into thin filaments so diffusion can act efficiently \cite{dimotakis2005turbulent}. 
Moreover, hydrodynamic instabilities [\S\ref{subsec:LayeredCocktails}] can lead to turbulence, which enhances the mixing rate by maximizing the exposed surface area and the concentration gradient between adjacent fluids \cite{chandrasekhar2013hydrodynamic}.
However, turbulence does not occur in viscous fluids at low Reynolds numbers [\S\ref{subsec:NavierStokes}]. 
Instead, mixing can be achieved by combining diffusion with chaotic advection \cite{aref1984stirring}.  
This is beautifully demonstrated by the baker's map \cite{fox1997construction}. 
Imagine a piece of dough that is stretched out and folded on top of itself. Repeating these steps creates many exponentially thin layers, or laminae. A minimal amount of diffusion then mixes the layers together.
This is called `chaotic mixing' \cite{ottino1989kinematics, arnold1998topological}.

\begin{figure}
    \centering
    \includegraphics[width=1\linewidth]{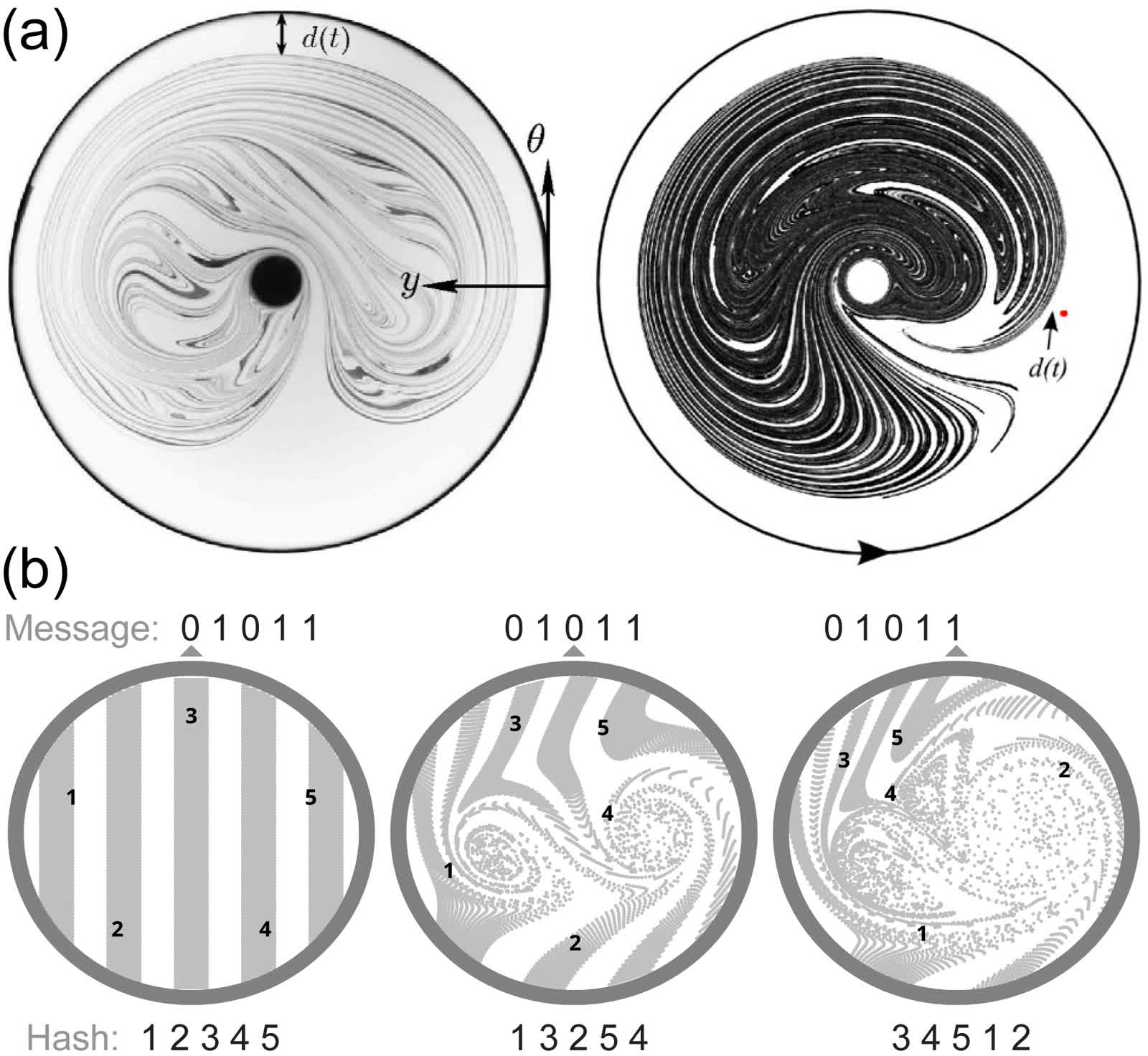}
    \caption{Chaotic stirring protocols can be used for a broad range of applications, from making macarons to cryptography.
    \textbf{(a)} Experiment (left) and simulation (right) of the figure-of-eight stirring protocol used to mix a blob of dye in sugar syrup. From \citet{thiffeault2011moving}. 
    \textbf{(b)} Chaotic advection used to create a digital message hashing function. Image courtesy of William Gilpin.
    }
    \label{fig:ChaoticAdvection}
\end{figure}

A culinary example of chaotic advection is making macarons, where highly viscous batter must be mixed gently to maintain its foam structure \cite{ozer2020influence}. The choice of stirring protocol has a dramatic impact on the mixing rate: If we move a stirring rod back and forth through a fluid, we see no mixing at all because of kinematic reversibility [see Fig.~\ref{fig:viscousflows} in \S\ref{sec:HoneyMicro}], as explained by the scallop theorem \cite{purcell1977life}. Thus, to mix fluids at low \text{Re} we must break time-reversal symmetry, so we change our strategy and stir in circular patterns instead. Now the fluids do mix, but slowly, because they are stretched only linearly. Next, we stir in figure-of-eight patterns \cite{thiffeault2011moving} [see Fig.~\ref{fig:ChaoticAdvection}a], which speeds up the mixing rate dramatically as this strategy yields exponential stretching \cite{meunier2003vortices}. This can also be achieved by rotating two rods, at the same speed but in opposite directions, as in commercial egg beaters \cite{franjione1992symmetry}.

To come up with an optimal mixing protocol is a difficult non-linear task \cite{eggl_schmid_2020} but a number of clever ideas have been proposed. A particularly efficient mixing protocol is the `blinking vortex'\index{blinking vortex}  \cite{aref1984stirring}, where two rotors alternately spin in the same direction. For the first half period, the first rod rotates while the other one is stationary, then vice versa. 
A task that could perhaps be performed by a cooking robot \cite{bollini2013interpreting}.
This canonical and time-periodic blinking vortex is used ubiquitously in chaotic mixing theory to compare the effectiveness of different mixing protocols \cite{aref2017frontiers}.
Similarly, the Arnold-Beltrami-Childress (ABC) flow is often considered the archetypal flow for many studies on chaotic advection in 3D \cite{arnold1998topological}.
Besides expedient cooking, chaotic advection has numerous applications in other disciplines \cite{stroock2002chaotic}. For example, \citet{gilpin2018cryptographic} used a blinking vortex model to create digital hash functions with potential applications in cryptography [Fig.~\ref{fig:ChaoticAdvection}b].

By adding another pair of counter-rotating rods, for example by using two eggbeaters, one obtains the famous `four-roll mill'\index{four-roll mill}. This concept was invented by \citet{taylor1934formation} to study the formation of emulsions [\S\ref{subsec:Emulsions}].
Oil drops immersed in golden syrup (ideal for baking) were placed at the center of the mill, in a stagnation-point flow, which elongates them. 
The drops split when the viscous stresses on their surface exceed the stabilising surface tension, as described by the capillary number,
    \begin{equation}
    \label{eq:CapillaryNumber}
    \text{Ca} 
    \equiv \frac{\text{viscous stresses}}{\text{surface tension}}
    = \frac{\mu U_0}{\gamma}, 
    \end{equation} 
where the characteristic velocity, $U_0=\dot{\gamma} L_0$, can be written as the local shear rate times the droplet size.
The four-roll mill laid the basis for studies of droplet break-up and stability, but fluctuating stagnation points made practical implementation difficult. 
To alleviate this issue, \citet{bentleyLeal1986computerFourRollMill} implemented an image-based feedback loop that controlled the speed of each roller independently. Using their invention, the same group \cite{bentleyLeal1986experimental} validated theoretical limits for drop deformation \cite{barthes1973deformation, hinchAcrivos1979steady} and paired their experiments with theory for studying drop dynamics \cite{stoneLeal1989influence}. Flow fields generated by the automated four-roll mill also pioneered polymer elongational rheometry \cite{fuller1980flow, fuller1981flow}.
More recently, \citet{hudson2004microfluidic} introduced the microfluidic analogue of the four-roll mill,  which has been used extensively to characterise the material properties of biomaterials and single cells by extensional rheometry \cite{haward2016microfluidic}.
And in applications of microfluidic stagnation point flows it has been extended to include substrate patterning \cite{perrault2009microfluidic, juncker2005multipurpose, safavieh2015two} and the trapping of cells by hydrodynamic confinements, allowing new developments in analytical chemistry and in life sciences \cite{brimmo2017stagnation}.

\subsection{Sweetening tea with honey}
\label{section:lowRehighPemixing}

Returning to the mixing of two liquids, we now make a cup of tea sweetened with a drop of honey. By pure dissolution, a viscous drop mixes slowly with the tea, but stirring can help us again. However, since the drop is very viscous, turbulent eddies cannot stretch the drop into thin filaments, as was the case with much less viscous milk drops [see \S\ref{subsec:ChaoticAdvection}]. Instead, the sharp flow velocity gradients around the drop increase the mass transfer by maintaining a correspondingly sharp concentration gradient \cite{leal2007advanced}. 
We seek an estimate of the mixing time.
We consider a drop of honey of size $L_0\sim \SI{1}{\milli\metre}$ and diffusivity $D \sim \SI{e-10}{\metre\squared\per\second}$ in water \cite{fan1967apparent}. 
We also assume a very viscous honey drop, so the mass transport is dominated by advection due to large Schmidt numbers [Eq.~\eqref{eq:SchmidtNumber}].
Using a stirring speed of $U_0 \sim \SI{1}{\milli\metre\per\second}$, the P\'eclet number [Eq.~\eqref{eq:PecletNumber}] is large, $\text{Pe} \sim 10^4$, but the flow close to the drop is still laminar at an intermediate Reynolds number, $\text{Re} \lesssim 1$.
Then, as the drop dissolves, a diffusion layer develops between the pure phases.
\citet{acrivos1965} showed that, in the low Reynolds number and high P\'eclet number limit, this diffusion layer has thickness   
    \begin{equation}
        \delta \sim L_0 \text{Pe}^{-1/3}.
        \label{eq:AcrivosGoddard1965}
    \end{equation}
In our case, at high Pe, the boundary layer is rather thin, $L_0/\delta \sim 20$.
By substituting $\delta$ for $L_0$ in the expression for the diffusion time, $\tau_{D} \sim L_0^2/D$, following \citet{mossige2021dynamics}, we obtain a typical mixing time
    \begin{equation}
        \tau_\text{mix} \sim (L_0^2/D) \text{Pe}^{-2/3}.
        \label{eq:advectiondiffusionTimeScale}
    \end{equation}
By inspecting this expression, we can immediately appreciate the dramatic effect of fluid flow: It can reduce the mixing time by a factor of a thousand or more. 
Putting in the numbers, it takes $\sim\SI{22}{\second}$ to stir the viscous honey droplets (or sugar grains) into our tea. 
This approximation can be improved by accounting for open streamlines and inertial effects \cite{krishnamurthy2018heat1, krishnamurthy2018heat2}.

Instead of stirring, we can also let the honey drop sediment down. If it is sufficiently small, it will remain spherical and sediment at low Reynolds number [\S\ref{subsec:Sedimentation}]. When we substitute the terminal velocity $U_\infty$ [Eq.~\eqref{eq:terminalVelocity}] for $U_0$ in Eq. \eqref{eq:advectiondiffusionTimeScale}, we obtain a characteristic time scale of mixing for the sinking drop,
    \begin{equation}
        \tau_\text{mix, sink} \sim \left(\frac{\mu^2}{\left(\Delta\rho g\right)^2 D}\right)^{1/3}.
        \label{eq:diffusionConvectionTimescale}
    \end{equation}
This timescale also applies to the inverted system of a water drop rising in another viscous, miscible liquid, such as corn syrup, as recently examined both experimentally  \cite{mossige2021dynamics} and numerically \cite{vorobev2020shapes}.



\section{Washing the Dishes: Interfacial Flows}
\label{sec:TheDishes}

After our long meal, from cocktails to coffee and tea [\S\ref{sec:Drinks}--\S\ref{sec:Turbulence}], it is now time to do the dishes.
Food hygiene is paramount [see also \S\ref{subsec:dryingHands}], and the cleaning will not be difficult if everyone helps a little. 
As Johann Wolfgang von Goethe (1749-1832) wrote,
\begin{quote}
    \emph{Let everyone sweep in front of his own door, and the whole world will be clean.}
\end{quote}
Moreover, doing the dishes is much alleviated by the mesmerizing colors of soap bubbles and the startling wave dynamics.
Fun, you might think, but interfacial phenomena have led to exceptional scientific discoveries ranging from cell biology to nanotechnology \cite[see e.g.][]{rosen2012surfactants, myers2020surfactant}.
In this penultimate section, we will pop the bubble of some old misconceptions, and catch the wave of the latest developments concerning interfacial flows.

\subsection{Greasy galleys smooth the waves}
\label{subsec:GreasyGalley}

\begin{figure}[t]
    \includegraphics[width=1\linewidth]{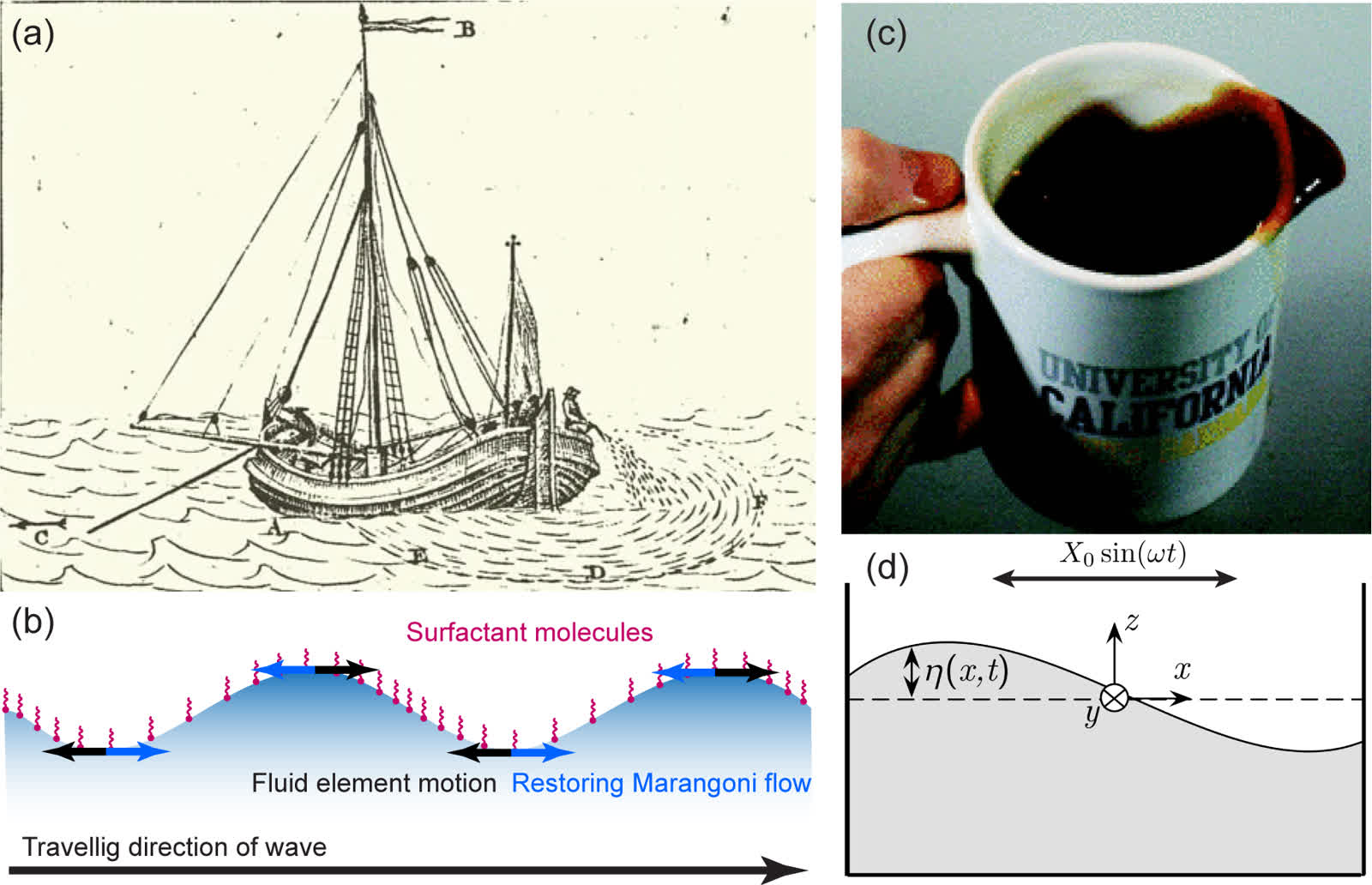}
    \caption{Waves and splashes.
    \textbf{(a)} Protecting a ship by calming the waves with oil. The Dutch fisheman Isak Kalisvaar reported to have conducted this experiment, in a letter to Frans van Lelyveld in 1776, after his ship got into a violent storm. From \citet{mertens2006oil}. 
    \textbf{(b)} Diagram of capillary wave dampening by surfactants. 
    \textbf{(c)} Representative image of coffee spilling. From \citet{mayer2012walking}.
    \textbf{(d)} Schematic of sloshing dynamics in an oscillated container. From \citet{sauret2015damping}. 
    }
    \label{fig:Waves}
\end{figure}

Benjamin Franklin (1706-1790) noticed a remarkable phenomenon during one of his journeys at sea, sailing in a fleet of 96 ships.
\textit{``I observed the wakes of two of the ships to be remarkably smooth, while all the others were ruffled by the wind, which blew fresh''}
\cite{franklin1774stilling}.
Being puzzled with the differing appearance, Franklin at last pointed it out to the captain, and asked him the meaning of it.
The captain's answer may come as a surprise:
\textit{``The cooks, says he, have, I suppose, been just emptying their greasy water.''}
The calming effect of oil on water was common knowledge to seamen at the time, and had indeed been described since ancient Greeks. 
However, legends circulated about a ship that miraculously survived a storm by taming it with olive oil [Fig.~\ref{fig:Waves}a], so Franklin decided to initiate a series of systematic experiments \cite{franklin1774stilling}.
The amusing details of these stories, and the scientific interest that emerged since, are described eloquently by \citet{mertens2006oil} and \citet{tanford2004ben}.

While the dampening of surface waves was known for millennia, its precise cause was a mystery until recently, as described by \citet{henderson1994surface, nicolas2000note, behroozi2007calming, kidambi2011damping} and references therein.
Franklin thought that the oil film stopped the wind from catching the water, but more than a century passed before more progress was made. In her kitchen, Agnes Pockels performed pioneering experiments on the surface tension of oil films [see \S\ref{subsec:SoapFilmDynamics}].
We now know that this surface tension increases when the oil film is stretched thin, for example by the wind.
Because of the Marangoni effect [see \S\ref{subsec:Tears}], the resulting gradients in surface tension then induce flows that oppose the film deformation, thus dampening surface waves [Fig.~\ref{fig:Waves}b]. 
This interfacial restoring force is referred to as the Gibbs surface elasticity, or the Marangoni elasticity \cite{kim2017marangoni}, which is a multiphase flow effect that occurs in many other applications, as reviewed extensively by \citet{brennen2005fundamentals}.

\subsection{Splashing and sloshing}
\label{subsec:Splashes}

No culinary achievement happens without a little mess left behind, be it an accidental spill, or the usual drop of wine from the cook's glass on the kitchen table [see \S\ref{subsec:CoffeeRingEffect}].
Splashes and spills can be dangerous, though.
\citet{carmody2022chicken} recently investigated the health concerns of washing raw chicken, where splashes could contaminate culinary surfaces.
The question of sloshing,\index{sloshing} why liquids spill\index{spill} out of a container under acceleration, has received prior attention in the context of space vehicles and ballistics: Depending on the size of the container, and the type of agitation, large-scale oscillations of the encased fluid can be enhanced to the point of spilling \cite{ibrahim2005liquid, herczynski_weidman_2012}. 
In the academic context, it is known to everyone trying to walk to seminars with their coffee cup [Fig.~\ref{fig:Waves}c]. 
It turns out that spilling results from a combination of excess acceleration for a given coffee level when we start walking, and a complex enhancement of vibrations present in the range of common coffee cups sizes \cite{mayer2012walking}. With some relief came the realisation that beer does not slosh so easily, since the presence of even a few layers of foam\index{foam} bubbles on the free surface introduces strong damping of surface oscillations \cite{sauret2015damping} [Fig.~\ref{fig:Waves}d]. 

We generally want to avoid or control splashing or spreading, especially when mixing and pouring liquids. The impact and breakdown of droplets on a solid or liquid surface is mainly controlled by the Reynolds number [Eq.~\eqref{eq:ReynoldsNumber}] and the Weber number [Eq.~\eqref{eq:WeberNumber}]. Another important factor determining the splashing behaviour is the type of substrate, which regulates the contact angle dynamics of impinging droplets \cite{quetzeri2019role}. The elasticity of the substrate also plays a role \cite{vella2019buffering}, because soft solids absorb kinetic energy from  fluids in motion and noticeably reduce or even eliminate splashing. Estimates and experiments show that the droplet kinetic energy needed to splash on a very soft substrate can be almost twice as large as in the rigid case \cite{howland2016harder}. Droplet spreading and recoil can result in a number of complex fluid dynamics phenomena, when the elongating and stretching drops form jets and sheets which further destabilise into smaller droplets via the Rayleigh-Plateau instability [\S\ref{subsec:RayleighPlateau}]. The possible outcomes of a collision of a droplet with a solid substrate involve deposition, a fervent splash, so-called corona splash\index{corona splash} in which the liquid forms a circular layer which detaches from the wall, and retraction in which the droplet can de-stabilise and break up or rebound (partially or entirely) \cite{richard2000bouncing, richard2002contact, liu2014pancake}. The process is controlled by the wettability of the surface, the parameters of the droplet, and its impact speed. 
At a larger scale, inverted bell structures are formed when a jet impacts a liquid container, as observed during the washing of vials \cite{mohd2022open}.

Before a stream separates into impacting droplets, liquid jets are frequently seen and used in the kitchen \cite{eggers2008physics}. When plating a gourmet meal, the way sauces are spread on a plate is carefully engineered to achieve a variety of shapes and textures. The same questions appear when glazing a cake, where various edible jets and streams are produced on surfaces in an artful manner that manages buckling instabilities. In artistic paining, the understanding of hydrodynamics was crucial to Jackson Pollock, for one, who used a stick to drizzle paint on his canvas in a variety of ways \cite{palacios2019pollock}. The complex fluid dynamics behind different painting effects has only recently been analysed and reviewed by \citet{herczynski2011painting} and \citet{zenit2019fluid}.


\subsection{Dishwashing and soap film dynamics}
\label{subsec:SoapFilmDynamics}
 
\begin{figure}
    \centering
    \includegraphics[width=1\linewidth]{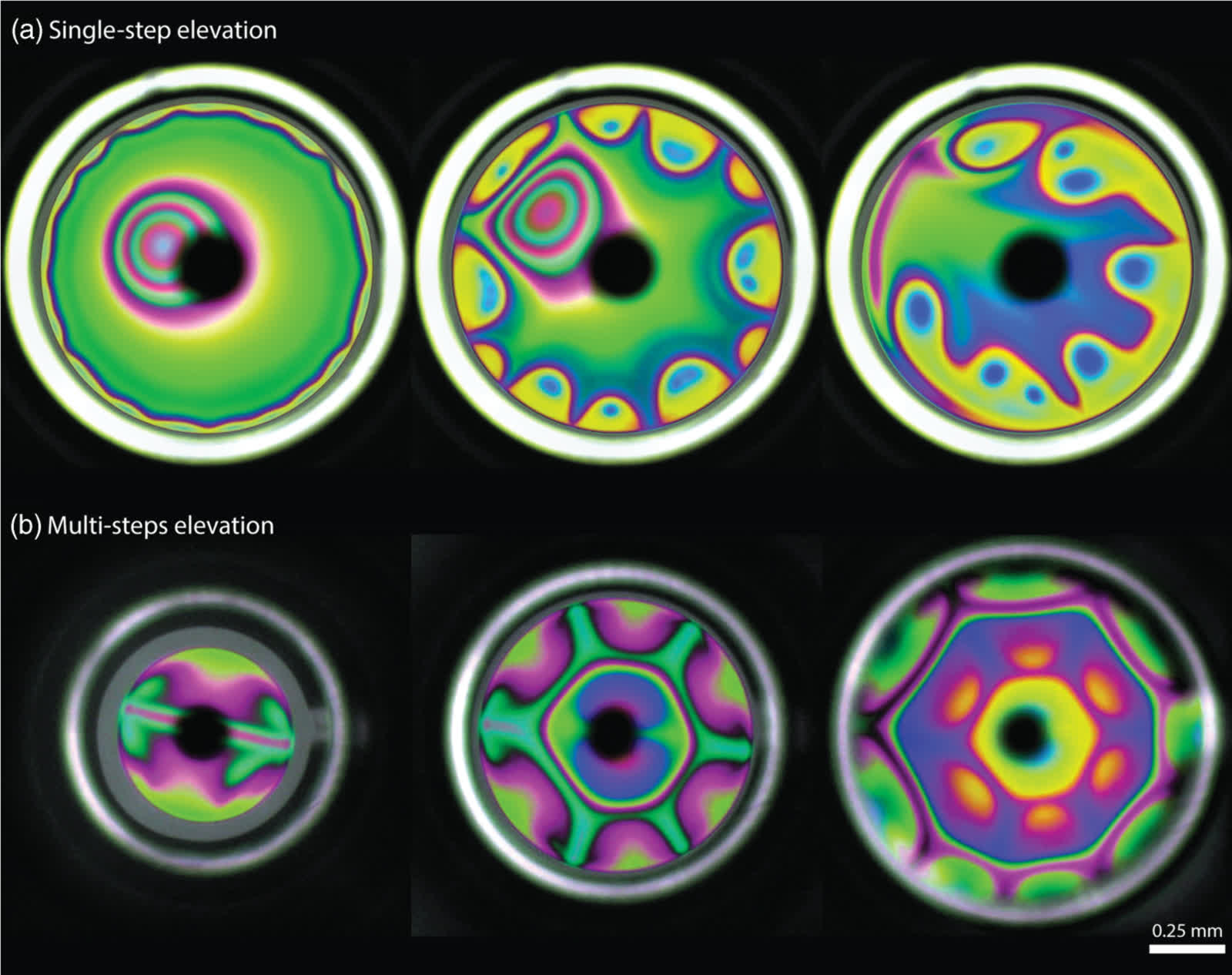}
    \caption{Thin-film interferograms showing the evolution of surfactant driven flows in soap bubbles. 
    \textbf{(a)} When the soap bubble is brought through an air-liquid interface in one single step, an unstable dimple forms at the apex. The dimple is quickly washed away by surfactant plumes rising from the periphery. 
    \textbf{(b)} When the bubble is instead elevated in multiple, small steps, controllable Marangoni instabilities can be utilized to stabilise the bubble and to prolong its life span. 
    From \citet{bhamla2016placing}.
    }
    \label{fig:SoapFilmDynamicsMarangoniArrest}
\end{figure}

The interference patterns on soap bubbles have fascinated physicists for centuries \cite{patsyk2020observation}, which has resulted in pioneering discoveries in optics, statistical mechanics, and in fluid mechanics by \citet{newton1952opticks}, \citet{plateau1873experimental}, and \citet{de2004capillarity}. 
An even more remarkable story is how the self-taught chemist Agnes Pockels (1862-1935) was inspired to study surface tension while doing the dishes. 
Women were not allowed to enter universities, so she did not have a scientific training and could not publish her work in scientific journals \cite{byers2006out}.  
Ten years after her first experiments, she was encouraged to write a letter explaining her findings to Lord Rayleigh, who then forwarded it to Nature \cite{pockels1891surface}. Along with her subsequent papers \cite{pockels1892relative, pockels1893relations, pockels1894spreading, pockels1926measurement}, all in top-level journals, she contributed to the establishment of the field of surface science. Without formal training and without access to a lab, Pockels also used simple kitchen tools to develop the precursor to the now widely celebrated Langmuir trough, which is now used to measure the surface pressure of soap molecules and other surfactants upon compression \cite{fuller2012complex}.

Soap bubbles are comprised of a thin aqueous film that is sandwiched between two surfactant layers, where each color corresponds to a different film thickness.
This film starts to drain immediately due to gravity \cite{bhamla2016placing}. The drainage causes a small deficit in soap concentration at the bubble apex and the formation of a small dimple [Fig.~\ref{fig:SoapFilmDynamicsMarangoniArrest}a, left panel]. This gradient in surfactant concentration sets up a Marangoni flow towards the apex. By replenishing interfacial material, these flows stabilise the bubble against rupture. 
However, such Marangoni flows are short-lived and are quickly destroyed by chaotic flows which de-stabilize the soap film [Fig.~\ref{fig:SoapFilmDynamicsMarangoniArrest}a, right panel]. A simple trick can solve this issue. \citet{bhamla2017interfacial} showed that by elevating the soap bubble in multiple small steps through a soap solution (instead of in one huge step), it is possible to induce a cascade of Marangoni instabilities. Each Marangoni instability arrests the previous one, and this prevents chaotic flows from developing. This method, coined `placing Marangoni instabilities under arrest' by the authors, produces beautiful flow patterns, as displayed in Fig.~\ref{fig:SoapFilmDynamicsMarangoniArrest}b. 

The rate of draining depends on the viscosity of the soap film. Adding glycerol, a natural ingredient in soap, effectively extends the life span of a bubble. Adding corn syrup or honey does the same job, but it might not help to clean your dishes. However, it \emph{will} help to make giant soap bubbles. By retarding film drainage and by reducing the evaporation rate, a bubble stabilized by viscosity has sufficient time to grow before it eventually pops. Another approach was taken by \citet{frazier2020make}. They appreciated the central role of viscoelasticity in stabilizing thin liquid films, and utilized polyethylene glycol (PEG), a long-chained polymer commonly found in hand sanitizers, to create bubbles with surface areas close to $\SI{100}{\metre\squared}$, the area of a badminton field. 

Interfacial fluid mechanics offers exciting avenues for future research.
Recently, looking at the hardness of domestic tap water, \citet{giacomin2021black} used interfacial rheometry to study how this affects thin films floating on black tea, which crack like sea ice.
Revisiting the work by Sir James Dewar after 100 years, \citet{seimiya2021revisiting} shed new light on the emergence of pearl string structures during bubble drainage.
In modern video games, to render bubbles realistically, computer-generated imagery (CGI) techniques are coupled to interfacial flow models that vary soap film thickness~\cite{ishida2020model}.
Moreover, in biology, the nature of the boundary between water and oil is crucial to many nanometre-scale assembly processes, including biological functions such as protein folding and liquid-liquid phase separation~\cite{chandler2007oil, hyman2014liquid}.

Finally, when we are finished with doing the dishes, when we pull the plug from the kitchen sink so that the liquid drains away, we observe a vortex.
When the drain flux is small enough to avoid the formation of a central surface dip, this vortex can be approximated as a
Rankine vortex \cite{tyvand2022viscous}.
The hysteresis in two-liquid whirlpools was investigated by \citet{naumov2022hysteresis}.

\subsection{Ripples and waves}
\label{subsec:Ripples}

Whenever an interface between two fluids is disturbed, ripples and waves\index{wave} emerge and propagate along the surface [also see \S\ref{subsec:LayeredCocktails}]. When a group of waves moves across a pond, we see waves of different wavelengths $\lambda$ propagating at different speeds and, importantly, groups of waves travelling at different speeds than the crests and troughs of individual sinusoidal perturbations. The reason for this is the dispersity of water waves. Dispersity refers to the dependence of the wave propagation speed on the wavelength, with longer waves generally travelling faster. For a wave of frequency $\omega$, the relationship between the {\it wave speed}\index{wave speed} $c$ and the wavenumber $k$ is $c = \omega/k$, where the dependence of the frequency $\omega=\omega(k)$ on the wavenumber is called the {\it dispersion relation}\index{dispersion relation}. In a wave packet, where each crests travels at the speed $c$, the velocity of travel of the whole group is $c_\text{g} = \mathrm{d}\omega/\mathrm{d}k$ and is called the {\it group velocity}\index{group velocity}. 

On deep water, where the dispersion relation reads $\omega^2={gk}$, the wave speed $c=\sqrt{g/k}$ is twice as large as the group propagation speed. This is the reason why in a travelling wave packet, individual crests will seem to continuously appear at the back of the packet, propagate through it towards the front, and eventually vanish there. Here, deep water means that the depth of the layer is much larger than the wavelength, $\lambda=2\pi/k$, in which case the dispersion relation above is obtained from the assumption of a potential flow with a linearised boundary condition at the free surface, which is appropriate when the wave amplitude is small compared to the wavelength \cite{acheson}. Such waves are referred to as {\it gravity waves}. 

However, in many small-scale flows, the surface tension forces at the interface cannot be neglected. Accounting for them leads to a dispersion relation for {\it capillary-gravity waves}\index{capillary-gravity waves}, $\omega^2 = gk + \gamma k^3/\rho$, where the importance of the surface tension parameter is measured by the dimensionless number $\text{S} = \gamma k^2 / \rho g$. 

For very short waves, the capillary term dominates, so $\text{S}\ll1$ and the dispersion relation simplifies to $\omega^2= \gamma k^3/\rho$. Such waves are termed {\it capillary waves}\index{capillary waves}, and for water, the typical cross-over wavelength when $\text{S}=1$ is about 1.7 cm.  Notably, for capillary waves, the group velocity exceeds the wave (or phase) velocity ($c_\text{g}=3c/2$). So crests move backwards in a propagating wave packet. In most small-scale kitchen flows, surface tension has a pronounced effect on the appearance and propagation of waves. A familiar example of this kitchen phenomenon are the waves created by a dripping faucet in a filled sink.

Moreover, in various food science circumstances, we might have to consider waves of wavelength comparable to the depth of the vessel in which they propagate. For such {\it shallow-water waves}\index{shallow-water waves}, the propagation speed depends on the local depth with larger speeds at deeper water. In particular, for gravity waves the dispersion relation becomes in this case $\omega^2 = gk \tanh (kh)$, with $h$ being the water depth. This again holds for wavelengths small as compared to $h$. The general case is much more complex and nonlinear in nature, yet the linear wave theory is often enough to grasp the dominant behaviour. We considered here only free-surface flows but the reasoning can easily generalised to any fluid-fluid interface \cite{lamb1945hydrodynamics}.

\subsection{Rinsing flows: Thin film instabilities}
\label{subsec:RinsingFlows}

Thin fluid films lead to remarkable kitchen flows.
For example, in a wine decanter, thin film instabilities give rise to ripples that enhance wine aeration [Fig.~\ref{fig:wineAeration}b], and similar ripples are seen when rinsing plates or chopping boards.
The stability of falling films\index{liquid film} was the subject of investigation of a father-son team of the Kapitza family, led by the elder Nobel prize winner Pyotr Leonidovich Kapitza, in the 1940s \cite{kapitza1948wave}. After World War II, Kapitza was removed from all his positions, including the directorship of his own Institute for Physical Problems, for refusing to work on nuclear weapons. He was ordered to stay at his country house and, deprived of advanced equipment, devised experiments to work on there, including a famous set of experiments on falling films of liquid \cite{kalliadasis2012}. \citet{kapitzacollection} were the first to experimentally investigate traveling waves\index{waves} on the free surface of a liquid film falling down a smooth plate. The emerging Kapitza instability\index{Kapitza instability} takes form of roll waves \cite{balmforth2004dynamics}, and evolves from a two-dimensional disturbance (i.e., invariant in the spanwise direction) into a fully developed three-dimensional flow \cite{liu1995three}. Since the early works of Kapitza, the dynamics of waves in viscous films over the flat substrates have been reviewed extensively \cite{chang1994wave, oron1997long, craster2009dynamics}. We often encounter such waves after a rainfall, on an inclined asphalt road, or even in flowing mud \cite{balmforth_liu_2004}. Film and rivulet flows at solid surfaces bear importance for gas exchange also in industrial applications, including distillation columns \cite{de1991mechanics},  and in coating processes,  they must be suppressed to obtain a smooth surface without ripples. In the kitchen context, they emerge predominantly in rinsing flows or spreading flows, where the thin film dynamics may be governed either by capillarity or external driving, such as gravity or centrifugal forces \cite{walls2019spreading}. We discussed the viscous spreading phenomenon in \S\ref{subsec:ViscousGravityCurrents}, thus here we focus purely on the waving instability.

In the Kapitza instability, the formation of the roll waves is governed by two dimensionless parameters: the Reynolds number\index{Reynolds number} describing the flow character, and the Kapitza number:\index{Kapitza number} 
\begin{equation}
    \label{KapitzaNumber}
    \text{Ka} = \frac{\text{surface tension}}{\text{inertial forces}} = \frac{\gamma}{\rho g^{1/3} \nu^{1/4}}, 
\end{equation}
where $g$ is the gravitational acceleration driving the flow. The latter is derived as the ratio of capillary to viscous damping forces, $\text{Ka}=(\lambda_c/\lambda_\nu)^2$, where $\lambda_c = (\gamma/\rho g)^{1/2}$ is the capillary length\index{capillary length}, and $\lambda_\nu = (\nu^2/g)^{1/3}$ is the viscous-gravity length scale\index{viscous-gravity length} \cite{kalliadasis2012,mendez2017low}. For a flow down a slope with inclination angle $\beta$, the gravitational acceleration $g$ is replaced by its streamwise component $g\sin\beta$. In the context of thin-film flows down an inclined slope, the formation of roll waves can also be discussed in terms of the Froude number, $\text{Fr}$, defined in Eq.~\eqref{eq:FroudeNumber} in \S\ref{subsec:hydraulicJumps}. For moderate Reynolds numbers, the value of $\text{Fr}\approx 2$ marks the onset of instability in the thin film flow equations \cite{barker2017note}. However, Benjamin has shown in his seminal paper \cite{benjamin1957wave} that such a flow is unstable for all values of $\text{Re}$. He also found that the rates of amplification of unstable waves become very small when $\text{Re}$ is made fairly small, while their wavelengths tend to increase greatly. He proposed a criterion that for an observable instability of flow down a slope, the critical Reynolds number is $\text{Re} = \frac{5}{6} \cot{\beta}$, as later corroborated by \citet{yih1963stability}.\index{Froude number}






\subsection{Dynamics of falling and rising drops}
\label{subsec:immiscibledropsdynamics}
\index{Immiscible Drops}

\subsubsection{Immiscible drops}
\label{subsec:immiscibledropshape}
\index{Immiscible drops}



When a drop of water is released in cooking oil (or vice versa) it falls (or rises) due to gravity. 
During its journey, surface traction from the outer liquid mobilises the fluid-fluid interface and the degree of surface mobility is given by the viscosity ratio between the ambient and drop fluid, $\hat{\mu}/\mu$. For very viscous drops translating through low viscosity liquids (such as a drop of oil rising through water, $\hat{\mu}/\mu\rightarrow 0$), the small surface traction is insufficient to mobilise the interface: this results in the drop translating at the velocity of a rigid Stokes' sphere of the same size and volume [see Eq. \eqref{eq:StokesLaw}], which we discussed in \S\ref{subsec:Sedimentation} about the sedimentation dynamics of coffee grounds. 
The opposite mobility limit is reached when the viscosity ratio is reversed such that $\hat{\mu}/\mu\rightarrow \infty$): the interface is then expected to be completely mobile, which causes a vortex ring to develop within the drop. A completely mobile interface is not able to resist viscous stresses, which reduces the prefactor from 6$\pi$ to 4$\pi$ in the Stokes law \eqref{eq:StokesLaw}. This leads to the terminal drop velocity becoming one and a half times as high as that of a Stokes' sphere of the same size and density [Fig.~\ref{fig:Interfacial}a,b]. The solution to the flow field within a translating, rigid drop at low Reynolds number was worked out simultaneously and independently by the French mathematician Jacques Salomon Hadamard (1865-1963) \cite{hadamard1911mouvement} and the Polish physicist and mathematician Witold Rybczyński (1881-1949) \cite{rybczynski1911uber} as early as in 1911. 
\index{Hadamard-Rybczyński solution}

In reality, most small droplets rise or descend at velocities that lie between the theoretical prediction by Hadamard and Rybczyński and the Stokes prediction for rigid spheres \cite{manikantan2020surfactant}, and this is true even in pure liquids with no surfactants added. The terminal velocity generally depends strongly on size, as reported by \citet{bond1927lxxxii}, who found small water droplets to descend through castor oil at only 1.16 times the Stokes' velocity, while drops exceeding a critical radius of about 0.6 cm descended at 1.4 times the Stokes' velocity. To explain the sudden jump in velocity with drop size, \citet{bond1928lxxxii} postulated that a ratio of buoyancy to surface tension determines the mobility of the interface. \citet{boussinesq1913existence} instead suggested that an increased viscosity at the drop's surface is responsible for slowing the drop. However, without experimental evidence of the flow field within the drop, it is impossible to judge the correctness of these models.  

\index{Stagnant cap model}
Aiming to obtain a better description,  \citet{savic1953circulation} published photographic evidence of the flow streamlines inside water droplets descending through castor oils [Fig.~\ref{fig:Interfacial}c]. His visualizations showed that the streamline patterns of drops exceeding 1 cm in radius are almost indistinguishable from the Hadamard-Rybczynski solution and that the terminal velocities for large drops are in good agreement with theory as well. However, for smaller drops, the vortex rings are shifted forward, and this occurs as a stagnant cap emerges in the rear of the drop. As the drop size is further reduced, the stagnant region covers a larger and larger portion until it envelops the entire drop, with the result of the drop sedimenting as a Stokes sphere.  

To explain his observations, \citet{savic1953circulation} proposed that the interface is immobilized by \emph{surface active} molecules, which are in turn de-stabilized by viscous stresses from the outer fluid. For the smallest drops, the viscous stress is insufficient to distort the surface layer: this leads to a complete immobilization. However, as the drop size increases, the shear stress increases as well, and this leads to a gradual removal of the surface layer until the Hadamard-Rybczynski theory is fully recovered for the largest drops. 

\citet{savic1953circulation} also developed a theory to calculate the drag of a drop from the degree of surface coverage, which he extracts from the flow visualizations. He also attempted to calculate the critical drop size of the transition between a mobile and an immobile no-slip boundary, however the transition occurred at larger radii than predicted, and he suggests this discrepancy to be due to a finite solubility between water and castor oil not accounted for in the theoretical model. Later, \citet{davis1966influence} improved Savic's analysis to obtain better agreement with experiments, and \citet{sadhal1983stokes} extended these results to obtain an exact solution of the drag force on the drop for a given surface coverage. For a droplet sedimenting at a given rate, \citet{sadhal1983stokes} also obtained an analytical expression for the total amount of surfactant adsorbed to the interface. However, the solution to the internal flow field and the corresponding sedimentation rate for a drop of a given size remains an open question.

The literature describing buoyant immiscible drops is vast, and we refer the interested reader to the many excellent reviews and books written on  this topic, see e.g. \citet{harper1972motion} and \citet{leal2007advanced}. 

\subsubsection{Miscible drops}
\label{subsec:miscibledropshape}
\index{Miscible Drops}

\begin{figure}[t]
    \centering
    \includegraphics[width=1.0\linewidth]{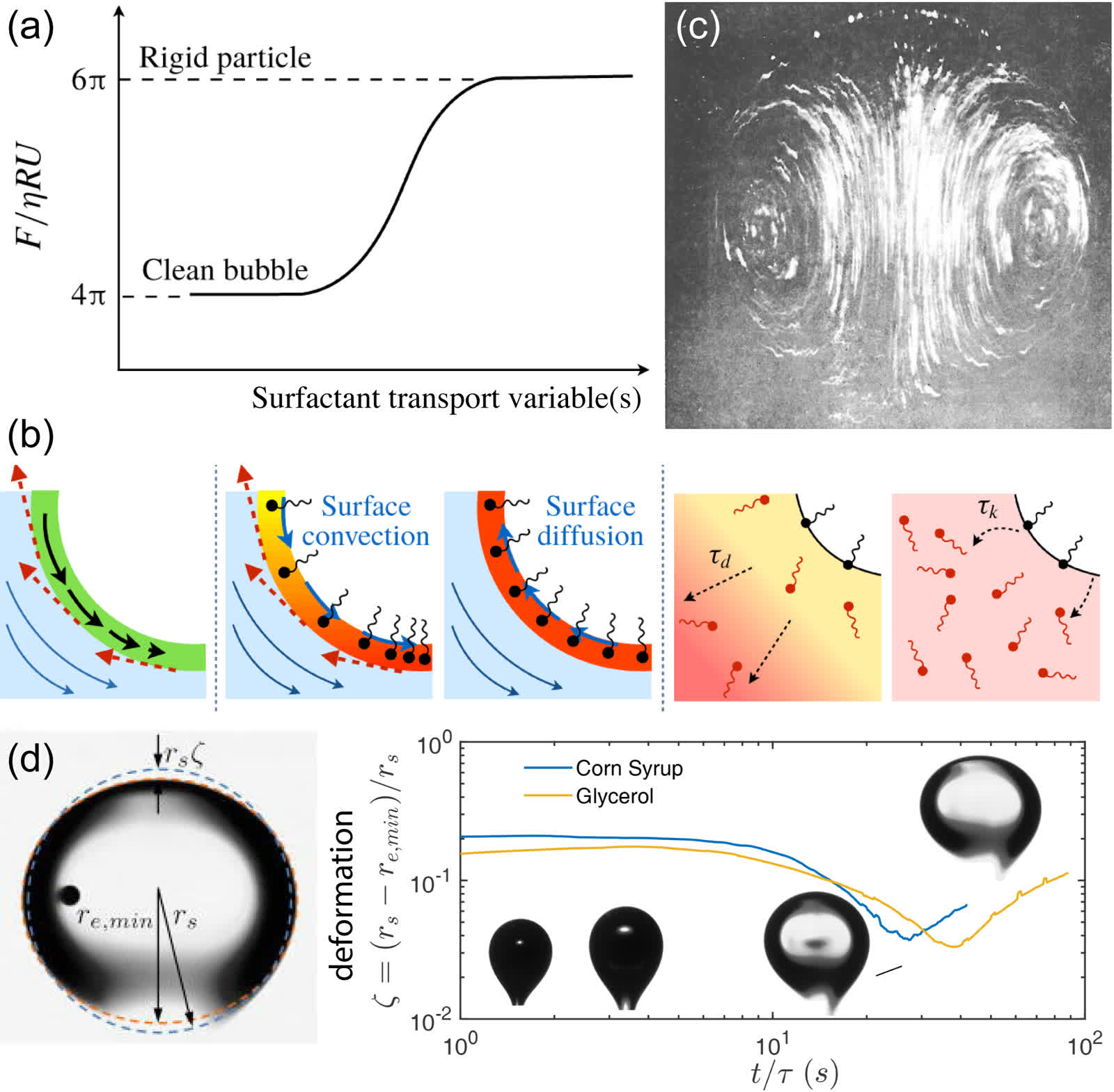}
        \caption{ 
        Interfacial phenomena in  bubbles and drops at low Reynolds number.
        \textbf{(a)} The drag force on a rigid particle  is one and a half times higher than the drag force on a clean bubble having the same size and density. 
        This drag force is affected by `hidden' surfactant transport variables \textbf{(b)} including
        (i) Interfacial viscosity can resist the surface flow. 
        (ii) Surfactant concentration gradients generate Marangoni stresses.
        (iii) Marangoni forces weakened by surface diffusion against the gradient.
        (iv) Diffusive transport of surfactants in the bulk.
        (v) Adsorption and desorption kinetics of soluble surfactants.
        (a-b) From \citet{manikantan2020surfactant}.
        \textbf{(c)} Streakline visualization showing the flow field inside a water drop falling through castor oil. From \citet{savic1953circulation}. 
        \textbf{(d)} Water drops ascending through corn syrup and glycerol undergo shape transformations from prolate to oblate spheroids. The travel time $t$ is rescaled by the characteristic mixing time $\tau$ from Eq. \eqref{eq:advectiondiffusionTimeScale}. From \citet{mossige2021dynamics}.
        } 
    \label{fig:Interfacial}
\end{figure}

As compared to immiscible drops covered in the last section, transport problems involving buoyant miscible drops have enjoyed far less attention, and as a result, their dynamics is far from understood. Or, in the words of Joseph and Renardy, "A basic and basically unsolved problem of fluid dynamics is to determine the evolution of rising bubbles and falling drops of one miscible liquid in another" \cite{joseph1993}. In section \S\ref{section:lowRehighPemixing} we looked at how fluid motion can accelerate the mixing rate between a viscous drop and its surroundings in the low Reynolds number case. In this section, we discuss how finite inertia may influence the shape of falling drops, and we discuss the stabilizing effects of transient tensions between miscible liquids.

When a miscible drop descends in another liquid, it changes shape in response to the viscous drag acting on it, and when it reaches a critical velocity, inertial effects also start to play a role. A simple way to visualise the effect of inertia on the drop shape is to produce a drop of food dye in air and let it fall into a glass of water. Upon impact with the water surface, the central part of the drop gets accelerated upward in a Rayleigh-Taylor instability, and this causes the drop to evolve into an open torus. 
For drops made of honey or corn syrup, this shape transformation is delayed by the high viscosity, but on a long time scale, even the most viscous drops deform into oblate spheroids or donuts.

\citet{kojima1984} developed a theory to explain the shape transitions of miscible drops and validated their theory against experiments with corn syrup drops falling through diluted corn syrup solutions. They showed that when the drop is created in air, immediately above a water surface, the descending drop does not experience inertia in the early stages of its descent. In this case, it is solely deformed by viscous traction forces, causing the drop to develop into an oblate spheroid. However, later in the drop's descent, inertia does play a key role in its shape evolution, and this causes the drop to develop into an open torus. The fact that inertia is relevant at long time scales is intuitive; however, they also had to incorporate a small, but finite tension across the miscible interface to fully explain the material deformation. 
Recently, \citet{mossige2021dynamics} examined the inverted system concerning water drops ascending through corn syrup [Fig.~\ref{fig:Interfacial}d].

The tension existing between miscible liquids is not a surface tension as defined in the classical sense between \emph{immiscible} phases [see \S\ref{subsec:HangingDrop}]. Instead, it is caused by sharp gradients in composition between the pure phases by giving rise to so-called {K}orteweg stresses  \cite{joseph1996} that mimic the effect of a surface tension. These tensions are typically at least two orders of magnitude smaller than the surface tension between immiscible fluids (for example, \SI{0.43}{\milli \newton \per \metre} between glycerol and water \cite{petitjeans1996} as compared to \SI{73}{\milli \newton \per \metre} between water and air) and diminish in time as diffusion smears out the miscible interface; as a result, they are inherently difficult to measure and usually neglected. However, in many situations including miscible displacements in capillary tubes \cite{chen2002} and in Hele-Shaw geometries \cite{chen2008radial}, effective interfacial tensions must be accounted for to accurately describe a deforming, miscible interface, and theoretical and experimental evidence for this is given in Refs.~\cite{pojman2006,pojman2007,lacaze2010, joseph1990,joseph1993,davis1988}. Non-equilibrium stresses are not only of academic interest, but can be tuned to control the morphology of miscible interfaces in modern industrial processes. Notably, \citet{brouzet2019effective} utilized transient tensions to align nanofibrils in microfluidic flow focusing geometries, with implications in the paper production industry and in the development of new, sustainable alternatives to plastics,  and \citet{wylock2014nonmonotonic} explored its potential for controlling gravitational instabilities, with relevance in e.g. carbon sequestration plants.

\section{Discussion}
\label{sec:Discussion}


In this Review, we have presented an overview of culinary fluid mechanics and other currents in food science. 
We discussed that, starting from ancient times, the connection between cooking and fluid mechanics has led to innovations that benefit both. \citet{toussaint2009history} put it more eloquently, 
    \begin{quote}
    \emph{Eating, at first a purely visceral pleasure, became an intellectual process.}
    \end{quote}
We have explored throughout this paper how this connection between science and food grows stronger every day, to the frontier of modern research and gastronomy.
Since kitchen science is so accessible, we can learn an great lot just by observing simple phenomena and see how they are connected. 
Therefore, innovations in fluid mechanics lead to better food, but creativity in cooking can equally generate new knowledge in different areas. 
Indeed, culinary fluid mechanics brings people together from across societies, from chefs to food scientists, physicists and chemical engineers, medical and nutrition specialists, and students across the disciplines.

\subsection{Summary}
\label{subsec:Summary}
 
To make this article accessible to this broad audience, we started our discussion with an overview of kitchen sink fundamentals [\S\ref{sec:kitchenSinkFundamentals}], where we summarised the basics of fluid mechanics in the context of food science.
Beginning the meal with drinks [\S\ref{sec:Drinks}], we reviewed hydrodynamic instabilities in cocktails, Marangoni flows, bubble effervescence and culinary foams.
Getting into the thick of it with a soup for starters [\S\ref{sec:SoupsSauces}], we discussed the rheological properties of viscoelastic food, non-linear sauces, suspensions and emulsions.
Moving on to a hot main course [\S\ref{sec:PotsPans}], we analysed the role of heat in cooking, including the Leidenfrost effect, Rayleigh-B\'enard convection, double-diffusive convection, flames and smoke.
Going for a sticky desert [\S\ref{sec:HoneyMicro}], we described flows at low Reynolds numbers, from Stokes' law to lubrication theory, viscous gravity currents, ice cream and microbial fluid dynamics.
Eager for a postprandial espresso [\S\ref{sec:CoffeeSugar}], we examined the physics of granular matter and porous media flows, different brewing methods, and the coffee ring effect.
Thirsty for another cup of tea [\S\ref{sec:Turbulence}], we delineated the tea leaf paradox and other non-linear flows, succeeded by turbulence and chaos. 
Finally, when doing the dishes [\S\ref{sec:TheDishes}], we explore interfacial phenomena including the Gibbs surface elasticity, soap film dynamics, waves and jets, miscible drops and roll wave instabilities.
Quite a bit to digest, but a place worth coming back to.

\subsection{Learning from kitchen experiments}
\label{subsec:learning}

\begin{figure}[t]
    \centering
    \includegraphics[width=1.0\linewidth]{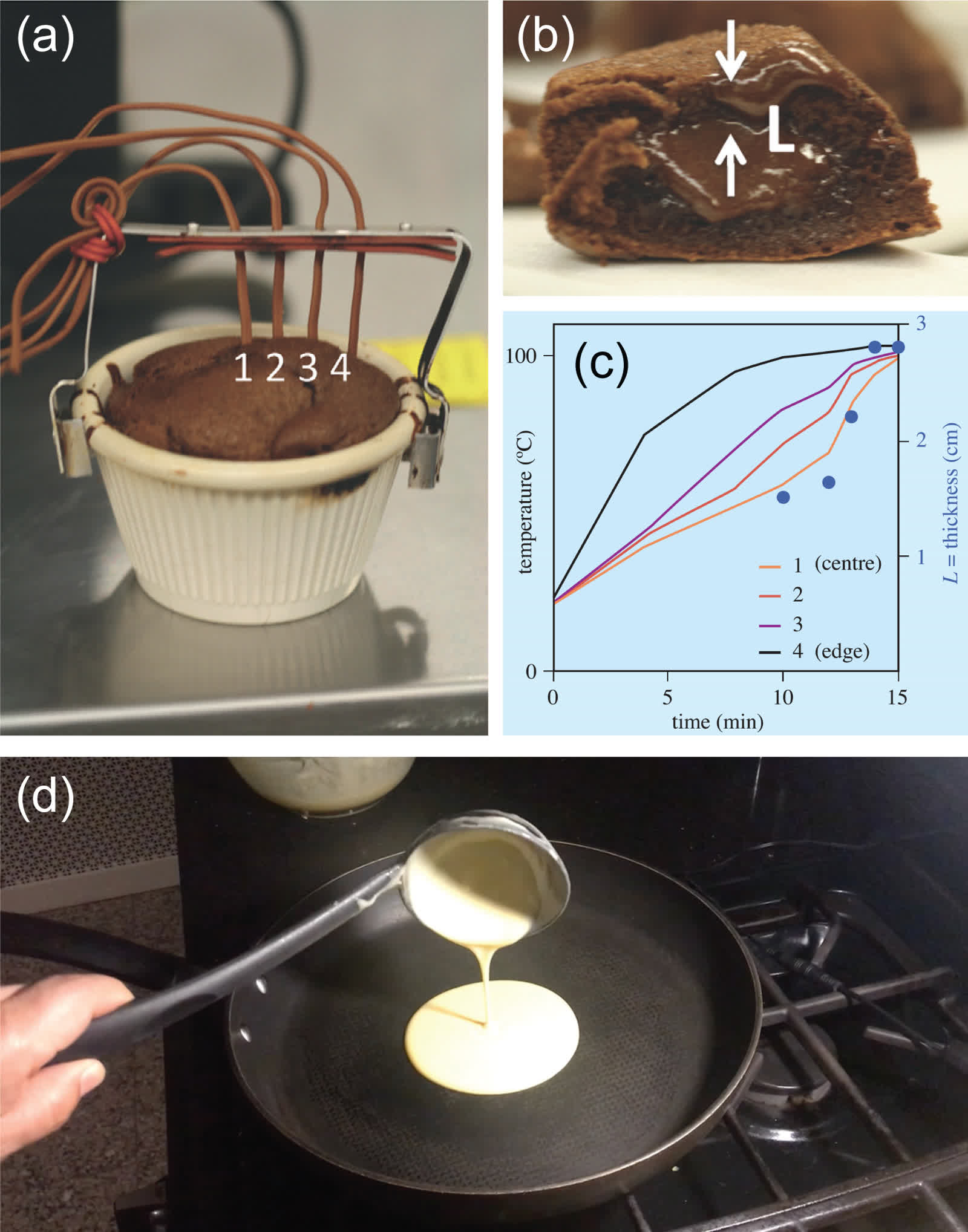}
    \caption{Kitchen-based learning is an affordable and accessible strategy to foster curiosity and intuition for a wide range of physics topics. 
    \textbf{(a-c)} In a science class accessible to non-science majors, cake making was used to demonstrate heat transfer and elasticity. From \citet{rowat2014kitchen}. 
    (a) Thermocouples are used to measure the rise in temperature at different points inside a molten chocolate cake as it bakes in an oven.
    (b) The thickness of the solid crust of the cake $L$ increases over time. 
    (c) Results of experiments.
    \textbf{(d)} Students pour pancake batter to learn about viscous gravity currents. They used cell phones to video-record the spreading rate and fit their data into a theoretical model to back-calculate the viscosity. Image courtesy of Roberto Zenit.}
    \label{fig:zenitPancakePouring}
\end{figure}

Humans are naturally curious. From an infant age, we explore by actively interacting with our surroundings \cite{lindholm2018promoting}. Through touch, smell and taste, we learn about the natural world. 
Becoming a scientist starts with asking questions like ``what?'', ``why?'' or ``how?''.

In physics education, we try to answer these questions the by comparing observations with theoretical descriptions. Traditionally, this knowledge is transferred from teacher to students through in-class lecturing and instructor-made assignments \cite{cagiltay2006students}, but this linear learning protocol is not necessarily compatible with curiosity-driven exploration and observation \cite{kallick2017students}. As a result, students often feel alien to the physics topics taught in class \cite{rowat2014kitchen} and lose the natural intuition and curiosity that is so important for learning \cite{jirout2018curiosity, gruber2019curiosity}. Inevitably, physics has a reputation for being difficult and abstract and with little relevance to students' daily lives, and this disconnection is largely responsible for the relatively poor recruitment to science and education disciplines in higher education \cite{tobias1999will}. To address this issue, it is vitally important to develop effective teaching strategies that foster both intuition, engagement and curiosity. This is best achieved through a hands-on active learning strategy \cite{freeman2014active, deslauriers2011improved}, without creating the perception of learning by ineffective engagement \cite{deslauriers2019measuring}, where experiments that relate to our daily lives have a prominent role. 

The kitchen is an accessible learning environment where simple physics experiments can be performed at home with humble ingredients; 
\citet{nelson2022soft} recently described how rheology can be made more accessible using food materials, and \citet{hossain2022yourself} provide a detailed toolbox for do-it-yourself rheometry.
Morover, \citet{benjamin1999rayleigh} showed that students in elementary physics education can learn about surface tension, mixing and gravity by studying Rayleigh-Taylor instabilities [see \S\ref{subsec:LayeredCocktails}] in their own kitchen. The simple experimental design allowed for a high degree of flexibility and were designed in such a way that they could be performed either individually, or in groups to foster collective accomplishment and collaborative learning. As compared to experiments conducted in school laboratories, kitchen experiments have a higher potential for engagement as we encounter them every single day. and since they require very little equipment, they offer a low-cost `frugal science' alternative that is less susceptible to budget cuts, and more accessible to students from underrepresented socioeconomic backgrounds \cite{whitesides2011frugal, byagathvalli2021frugal}.

Affordable and accessible kitchen experiments can also be utilized to develop intuition for advanced mathematical concepts. Notably, a famous class at Harvard and UCLA teaches general physics concepts such as heat transfer and phase transformations to non-science majors through the lens of cooking [Fig.~\ref{fig:zenitPancakePouring}a-c] \cite{rowat2014kitchen}. In this popular course, top chefs give weekly seminars for further engagement. Kitchen experiments can also be used to learn about more specialized topics in fluid mechanics; For example, take-home experiments such as measuring the flow rate from a hose and estimating the density and the viscosity of household fluids has been used to enhance learning in an introductory fluid mechanics class \cite{teachingFluidMechanics}. In addition, the kitchen can be a gateway to learn about the intrinsic fluid properties that govern these flows. Notably, in a special session on Kitchen Flows at the 73rd Annual Meeting of the APS Division of Fluid Dynamics (APS-DFD), \citet{ZenitPancakes} demonstrated how pancake-making can be used to teach students about fluid viscosity. Instead of extracting the viscosity from a classical sedimenting-sphere experiment \cite{sutterby1973falling}, which is less common in our daily lives, the students were asked to pour pancake batter and other viscous fluids like honey and syrup into frying pans and measure the spreading rate [Fig.~\ref{fig:zenitPancakePouring}d]. By fitting their data to a theoretical prediction \cite{huppert1982propagation}, which is described in \S\ref{subsec:ViscousGravityCurrents}, the students were able to back-calculate the viscosity. In the proposed course of do-it-yourself rheometry, \citet{hossain2022yourself} outlined an efficient way to convey the key notions of rheology to students confined to their homes due to the pandemic. Even twisting an Oreo can be an inspiring physics experiment \cite{owens2022oreology}. Such \emph{in-situ} kitchen measurements can be used for numerous other scientific concepts, as discussed throughout this Review, thus creating direct links between physics and everyday experiences.

In addition, many canonical flows can be generated with simple kitchen tools, including circular hydraulic jumps [see \S\ref{subsec:hydraulicJumps}], and Poiseuille flows [\S\ref{subsec:PoiseuilleFlow}], which can be used to validate theoretical predictions taught in class as a means to develop intuition for advanced mathematical concepts. 
Moreover, \citet{kaye2022overcoming} developed a pedagogical approach with hands-on activities to tackle common misconceptions in fluid mechanics education.
Finally, to further accelerate the learning in fluid mechanics, e-learning tools can be implemented \cite{rahman2017blended} such as the extremely extensive Multimedia Fluid Mechanics Online \cite{homsy2008multimedia}. 

From these examples, it is evident that easy-to-do kitchen experiments can be implemented for enhanced learning and engagement across all ages. They are highly scalable, and can even be taught on an online platform to make learning available for large groups of students, including students who can not afford enrollment in an educational program. Therefore, kitchen-based learning represents a viable strategy to increase the number of competent scientists and engineers in the world, which is necessary to address immediate threats to humankind and ensure a sustainable future \cite{sheppard2008educating}.

\subsection{Curiosity-driven research}
\label{subsec:curiosityDrivenResearch}

As well as being a vehicle for accessible and affordable science education, culinary fluid mechanics is a hotspot for curiosity-driven research \cite{agar20172016}. 
Indeed, Agnes Pockels found inspiration for her breakthrough discoveries in surface science and hydrodynamic instabilities from dish washing [\S\ref{sec:TheDishes}]. 
Valuable data can be extracted relatively quickly from a kitchen-based laboratory, in the spirit of a `Friday afternoon experiment' \cite{smith2015play}.
A minimum of investments of time, training and equipment are needed, which makes this field approachable not only to experimentalists, but also to theorists and researchers in other fields. 
Since many kitchen flows can be described by scaling theories and other analytical techniques, they can serve to validate theoretical models in fluid mechanics and materials science \cite{tao2021morphing}. 
As such, kitchen experiments are attractive to theorists, and by lowering the activation barrier to start a new experiment, they can be combined with mathematical models to solve a large class of problems in science and in engineering. 
Curiosity-driven learning is foundational to human cognition \cite{ten2021humans}, and sometimes the best discoveries are made in a few hours.

Perhaps the most influential fluid mechanicist of all time, Sir G. I. Taylor, was known for his special ability to make groundbreaking discoveries from humble ingredients and to design simple experiments that could be described theoretically \cite{batchelor1996life}. Instead of following the hypes in science and `going with the flow', Taylor was merely driven by his own interest and curiosity, without thinking about specific applications. Outstanding contributions in fundamental science always find useful applications, which is immediately evident when we look at the enormous implication of Taylor's contributions to science and engineering. However, today's funding schemes often require that research should preferentially address a particular problem and have immediate impact \cite{amon2015case}, which leaves little room for curiosity-driven research and scientific investigations for its own sake \cite{woxenius2015consequences}. But, since curiosity is a prerequisite for exploration and discovery, the scientific philosophy of Taylor and his predecessors could serve as inspiration for the modern physicist.

\subsection{Conclusion}
\label{subsec:conclusion}

Culinary fluid mechanics is the study of everything that flows in the food supply chain, covering a wide range of surprising phenomena that can be harnessed for the benefit of gastronomy, food science, and for our planet as a whole.
This field naturally connects practical technologies with basic research, just how fluid mechanics once started.
Culinary flows are accessible to experimentalists and theorists alike: Their intuitive geometry and well-defined conditions are suitable for mathematical modelling, while the relatively low equipment costs reduces the activation energy for pilot investigations, thus catalysing curiosity-driven education [\S\ref{subsec:learning}] and research [\S\ref{subsec:curiosityDrivenResearch}].

Where `kitchen science' papers may initially have been considered occasional or incidental, their breath and depth now constitute a rapidly growing field.
It is a field that this Review can cover only partially because it is so interconnected.
Yet, culinary fluid mechanics is unified by a number of well-defined research directions and goals:
Firstly, it aims to establish a sustainable and fair global food supply \cite{bloemhof2017sustainable}.
Secondly, it has the potential to develop reliable food technologies with a strong fundamental backbone \cite{lopez2015food, knorr2011emerging}.
Thirdly, it can facilitate new discoveries far beyond gastronomy by making science and engineering more accessible \cite{white1998inquiry, lee2003making, tuosto2020making, nelson2022soft}.
Finally, it can advise policy makers on important decisions for our future generations, such as the announced EU ban on PFAS non-stick coatings by 2030 \cite{EUstrategyPFAS} and help the reduction of climate change \cite{dauxois2021confronting, masson2021climate}.
To achieve these goals, scientists from related fields must become even more interconnected.

Indeed, as we discussed, culinary fluid mechanics directly links to other disciplines across the sciences, from molecular gastronomy to biological tissue mechanics and rheology.
Furthermore, it has extensive engineering applications ranging from the stream engine to 3D printing and nanotechnology.  
Not least, there are immediate connections with food safety, microbiology and medicine.
However, unlike many fields in science, kitchen flows create a bond with people who could not have a scientific training. 
People who want to learn more, and people who want to contribute themselves.
People like Agnes Pockels writing to Lord Rayleigh. 
So much talent is lost in this world full of inequality, and we have a responsibility to make science more inclusive and accessible to people from under-represented backgrounds. 
Through science communication, through education, and through research itself.
We hope that more scientists will stand up to this challenge.

\section*{Acknowledgements}

%
We express our gratitude to many colleagues and friends, including but not limited to Anurag Agarwal, Paulo Arratia, Andrea Bertozzi, Michael Brenner, Rajesh Bhagat, Saad Bhamla, Tomas Bohr, Daniel Bonn, John Bush, Otger Campas, Matteo Cantiello, Sam Dehaeck, Damir Dora{\v{c}}i{\'c}, Douglas Durian, Jan Engmann, Pere Castells Esqu\'e, Peter Fischer, Jamie Foster, Gerald G. Fuller, Ion Furjanic, Jeffrey Giacomin, William Gilpin, Nigel Goldenfeld, Elisabeth Guazzelli, Christopher Hendon, Douglas Jerolmack, George Karniadakis, Kiran Kembhavimathada, Rouslan Krechetnikov, G{\'e}rard Liger-Belair, Harishankar Manikantan, Hassan Masoud, Philip Nelson, Enrico Nobile, Crystal Owens, Rosanna Pasquino, Pax Piewpun, David Qu\'er\'e, Ingo Rehberg, Daniel Rosenberg, Amy Rowat, Alban Sauret, Bernhard Scheichl, Gary Settles, Tanu Singla, Pia Sorensen, Todd Squires, Howard Stone, Jean-Luc Thiffeault, Hugo Ulloa, Jan Vermant, Fabian Wadsworth, David Weitz, Tomoaki Watamura, John Wettlaufer, Stuart Williams, Nan Xue, Harunori Yoshikawa, Yonghao Yue, and Roberto Zenit for stimulating conversions, suggesting changes, providing figures, sharing their enthusiasm, and for invaluable support.
A.J.T.M.M. acknowledges funding from the United States Department of Agriculture (USDA-NIFA AFRI grants 2020-67017-30776 and 2020-67015-32330).



\nocite{apsrev41Control}
\bibliographystyle{apsrev4-2} 
\bibliography{references}

\end{document}